\documentclass[a4paper,11pt,final]{article}

\usepackage{graphicx}
\usepackage{lastpage,fancyhdr,graphicx}
\usepackage{epstopdf}
\usepackage{float}
\usepackage{amssymb,amsmath,amsfonts}
\usepackage{hyperref}
\usepackage{tikz-cd}
\usepackage[ruled,vlined,linesnumbered]{algorithm2e}
\usepackage[utf8]{inputenc}

\usetikzlibrary{hobby,decorations.markings,arrows,intersections,shapes}
\usepackage{pgfplots}
\usepgfplotslibrary{fillbetween}
\pgfplotsset{compat=newest}
\usepgfplotslibrary{fillbetween}
\usepackage{verbatim}

\begin{document}

\title{\ \\ \LARGE\bf Spectral dynamics of guided edge removals and identifying transient amplifiers for death-Birth updating}

\author{Hendrik~Richter \\
HTWK Leipzig University of Applied Sciences \\ Faculty of
Electrical Engineering and Information Technology\\
        Postfach 301166, D--04251 Leipzig, Germany. \\ Email: 
hendrik.richter@htwk-leipzig.de. \\
\small{ORCID. (0000-0001-5417-8291)}}

\maketitle

\begin{abstract}

The paper deals with two interrelated topics, identifying transient amplifiers in an iterative process and analyzing the process by its spectral dynamics, which is the change in the graph spectra by edge manipulations. Transient amplifiers are networks representing population structures which shift the balance between natural selection and random drift. Thus, amplifiers are highly relevant for understanding the relationships between spatial structures and evolutionary dynamics. We study an iterative procedure to identify transient amplifiers for death-Birth updating. The algorithm starts with a regular input graph and iteratively removes edges until desired structures are achieved. Thus, a sequence of candidate graphs is obtained. The edge removals are guided by quantities derived from the sequence of candidate graphs. Moreover, we are interested in the Laplacian spectra of the candidate graphs and analyze the iterative process by its spectral dynamics.  The results show that although transient amplifiers for death-Birth updating are rare, a substantial number of them can be obtained by the proposed procedure.  The graphs identified share structural properties and have some similarity to dumbbell and barbell graphs. Also, the spectral dynamics possesses characteristic features useful for deducing links between structural and spectral properties and for distinguishing transient amplifiers among evolutionary graphs in general.

\end{abstract}

\section{Introduction}
Evolution occurs by natural selection and genetic drift. Thus, if a mutant arises in a population of residents, its evolutionary dynamics is affected by the mutant's fitness (as this influences the mutant's chances in natural selection) and drift (which  basically is understood as a random process). 
The balance between natural selection and random drift  may vary over different spatial population structures. 
There are some spatial structures which amplify natural selection, thus shifting the balance towards the influence of fitness. Some other spatial structures suppress natural selection, thus reversing the effect. A mathematical understanding of the relationships between spatial structures and evolutionary dynamics is highly relevant for real biological processes, as for instance shown for
 cancer initiation and progression~\cite{hindersin16,kom03,kom06,nowak03,ver13}, ageing of tissues~\cite{cana16,cana17},  spread of infections~\cite{ott17a,ott17b} and microbial evolution of antibiotic resistance~\cite{krieger20}.

Spatial structures can be interpreted as a network and modelled mathematically as a graph.   Networks amplifying or suppressing selection have been intensively studied in the past decades~\cite{adlam15,alcalde17,allen20,hinder15,hinder19,jam15,mc21,monk18,pav17,pav18,tka20,tka21}.
The ability of a network to amplify (or suppress) selection not only depends on
the spatial structure, but also on other factors. One is where in the network and under what circumstances mutation occurs in the first place. Most previous works~\cite{hinder15,jam15,lieb05,monk14,moell19,pav17} have assumed that heritable mutations mainly occur in adult individuals with the same probability over time. For the network this means mutations equally likely appear 
at all vertices, which is called uniform initialization. As an alternative, we may assume mutations to occur mainly in new offspring, which means in the network mutations appear more likely at vertices more frequently replaced, which is called temperature initialization~\cite{adlam15,allen21,pav18,tka19}. Recently, it has been shown that at least for Birth-death updating amplification properties vary over initialization~\cite{allen21}.  

Another factor is the updating mechanism by which mutants and/or residents propagate on the graph.  Two mechanism frequently studied are Birth-death (Bd) and death-Birth (dB) updating. For Bd updating many graphs have been found which represent population structures with amplification properties. For instance, a comprehensive
 numerical study checking a larger number of random graphs with $N\leq 14$ vertices found a multitude of amplifiers of selection for Bd~\cite{hinder15}. More recently, a systematic study checking all graphs up to $N\leq10$ vertices found that almost all graphs have amplification properties~\cite{allen21}.

For dB updating we find the opposite. The mentioned study checking a larger number of random graphs~\cite{hinder15} found no amplifiers for dB updating. This prompted the assumption that either amplifiers for dB are very rare, at least for graphs with small orders, or even there are none. Meanwhile, theoretical and numerical works have modified and partly corrected this assumption.  On the one hand, it has been shown that for dB updating universal amplification is not possible~\cite{tka20}. At most, an evolutionary graph can be a transient amplifier (meaning the amplification only takes place for a certain range of the fitness).  In addition, a method has been devised which allows computing with polynomial time complexity if for weak selection a graph is a (tangential) amplifier~\cite{allen20}. The method is based on  calculating the coalescence times of random walks~\cite{allen17} on the graph and finding the vertex with the largest remeeting time. If an edge from this vertex is removed, with some likelihood the resulting graph is a transient amplifier.
The method also implies  identifying transient amplifiers for dB updating
by edge removals from regular graphs taken as input to the method. 
Recent results have shown that for a small but non-negligible fraction of all pairwise non-isomorphic regular graphs with certain order and degree transient amplifiers are obtained by such a single edge removal~\cite{rich21}. 

However, the resulting graphs are very close to the  regularity of the input graph and typically only for cubic and quartic regular graphs transient amplifiers have been identified. In this paper we study how transient amplifiers for dB updating can be obtained by multiple edge removals embedded in an iterative algorithmic process. As this gives a larger variety of transient amplifiers, particularly with a stronger perturbation to the regularity of the input graphs, a more profound analysis of  structural and spectral properties of amplification can be done. 
A main tool in this analysis is spectral dynamics, which is concerned with  changes in the graph spectra over graph manipulations~\cite{chen17,zhang09}.
We here study the spectral dynamics of the normalized (and standard) Laplacian spectra over edge removals.
Our main result is that although transient amplifiers for dB updating are rather rare, a significant number of them can be identified by the iterative method. For instance, for graphs on $N=11$  and $N=12$ vertices, for all existing degrees there are regular input graphs which can be disturbed into amplifier by guided edge removals. This also applies for graphs 
on $N=\{14,20,26\}$ vertices with degree $k=N-3$.  
Furthermore, the results show that transient amplifiers for dB updating identified by the iterative process share certain structural properties. They mainly consist of two cliques of highly (frequently completely) connected vertices which are joined by bridges of one or two edges. Thus, these graphs have some similarity to barbell and dumbbell graphs~\cite{ghosh08,wang09}. The analysis of the spectral dynamics 
also reveals shared characteristics. Interlacing results state that the Laplacian spectra generally shrink by edge removals~\cite{atay14,chen04,heu95}. By analyzing the algebraic connectivity as well as the smoothed spectral density of the whole spectrum, it can be shown that edge removal processes leading to transient amplifiers can be distinguished from processes not leading to amplifiers. 
This opens up the possibility to link structural and spectral properties of transient amplifiers and thus adds to answering a fundamental mathematical question in graph theory which is the relationships between the graph spectra and the graph structure.

\section{Methods}

\subsection{Identifying transient amplifiers}

We study evolutionary dynamics on graphs with a population of $N$ individuals on an undirected (and unweighted) graph  $\mathcal{G}=(V,E)$.  Each individual is represented by a vertex $v_i \in V$ and an edge $e_{ij}\in E$ indicates that the individuals placed on $v_i$ and $v_j$ are  (mutually interacting)  neighbors~\cite{allen17,lieb05,ohts07,patt15,rich17}. The graph $\mathcal{G}$ is simple and connected, and each vertex $v_i$ has degree $k_i$, which implies that there is no self-replacement and an individual on $v_i$ has $k_i$ neighbors.

  With mutants and residents there are of two types of individuals. Residents have a constant fitness normalized to unity, while mutants have a fitness $r>0$. An individual may change from mutant to resident (and back) by a fitness-dependent selection process. We consider a death--birth (dB) process, e.g.~\cite{allen17,allen20,patt15}. An individual is chosen uniformly at random and dies, thus vacating the vertex it occupied. One of the neighbors is selected to give birth with a probability depending on its fitness. The neighbor selected transfers its type and thus replaces the dead individual. To indicate that in such an updating process birth is fitness-depended, but death is not, we write dB updating, as suggested by~\cite{hinder15}.  

We consider uniform initialization and define the fixation probability $\varrho_{\mathcal{G}}$ as the expected probability that starting with a single mutant appearing at a vertex uniformly at random all vertices of the graph $\mathcal{G}$ eventually become the mutant type. In particular, we are interested in how this fixation probability $\varrho_{\mathcal{G}}(r)$ compares to the fixation probability $\varrho_{\mathcal{N}}(r)$ of the complete graph with $N$ vertices for varying fitness $r$.  We categorize the graphs as follows~\cite{adlam15,allen20,hinder15,pav17}. A graph $\mathcal{G}$ is called an amplifier of selection if $\varrho_{\mathcal{G}}(r)<\varrho_{\mathcal{N}}(r)$ for $0<r<1$ and $\varrho_{\mathcal{G}}(r)>\varrho_{\mathcal{N}}(r)$ for $r>1$. A suppressor of selection is characterized by $\varrho_{\mathcal{G}}(r)>\varrho_{\mathcal{N}}(r)$ for $0<r<1$ and $\varrho_{\mathcal{G}}(r)<\varrho_{\mathcal{N}}(r)$ for $r>1$. 
Finally, we have a transient amplifier  if
$\varrho_{\mathcal{G}}(r)<\varrho_{\mathcal{N}}(r)$ for $r_{min}<r<1$ and $r>r_{max}$, and additionally there is $\varrho_{\mathcal{G}}(r)>\varrho_{\mathcal{N}}(r)$ for $1<r<r_{max}$ and  some $0<r_{min}<1<r_{max}<\infty$.

The structural and spectral analysis as well as the iterative  procedure is based on three recent results  on amplifiers for dB. First,  it has been shown that for dB updating no universal amplification is possible and  evolutionary graphs can be transient amplifiers at most~\cite{tka20}.  Second, a numerical test has been proposed for weak selection (where $r=1+\delta$ and $\delta \rightarrow 0$), which is executable with polynomial time complexity and allows to detect if a graph $\mathcal{G}$ is a (tangential) amplifier~\cite{allen20}. Third, the test has been applied to check all regular graph up to a certain order and degree. It was shown that a single edge removal produces  transient amplifiers for a small but non-negligible number of cubic and quartic regular graphs. Moreover, a spectral analysis has demonstrated that there is a close relationship between the Laplacian spectra and amplification~\cite{rich21}.

The numerical test identifying transient amplifiers considers coalescing random walks~\cite{allen17} and involves 
calculating 
the effective population size $N_{eff}$ from the relative degree $\pi_i=k_i/\sum_{j \in \mathcal{G}} k_j$ and the remeeting time $\tau_i$ of vertex $v_i$~\cite{allen20}.
The remeeting time $\tau_i$ is obtained by
\begin{equation} \tau_i= 1+ \sum_{j \in \mathcal{G}} p_{ij}\tau_{ij}  \end{equation}
from the coalescence times $\tau_{ij}$ and the step probabilities 
$p_{ij}=e_{ij}/k_i$ (which implies $p_{ij}=1/k_i$, if $e_{ij}=1$ and $p_{ij}=0$, otherwise). Remeeting times observe the condition \begin{equation}\sum_{i \in \mathcal{G}}  \pi_{i} ^2\tau_{i}^{}=1. \label{eq:ident}\end{equation}
The coalescence times $\tau_{ij}$ are computed by solving the system of $\left(\begin{smallmatrix}N\\2\end{smallmatrix}\right)$ linear equations
\begin{equation} \tau_{ij}= \left\{ \begin{array}{cc} 0 & \quad i=j\\ 1+ \frac{1}{2} \sum_{k \in \mathcal{G}} (p_{ik}\tau_{jk}+p_{jk}\tau_{ik}) & \quad i\neq j  \end{array} \right.  .\end{equation}
In~\cite{allen20} it is shown that a graph $\mathcal{G}$ is an amplifier of weak selection if \begin{equation} N_{eff}= \sum_{i \in \mathcal{G}} \pi_i \tau_i >N. \label{eq:neff} \end{equation}
Furthermore, it is argued that an amplifier of weak selection can be identified by the following perturbation method. If the graph 
$\mathcal{G}$  is $k$-regular, then $k_i=k$ and $\pi_i=1/N$ for all $i=1,2,\ldots,N$. Thus, with the identity condition \eqref{eq:ident},  we get from Eq. \eqref{eq:neff}: \begin{equation} N_{eff}= \sum_{i \in \mathcal{G}} \tau_i/N=N \sum_{i \in \mathcal{G}}  \pi_{i} ^2\tau_{i}^{}=N. \label{eq:equal}\end{equation} The equality $N_{eff}=N$ in \eqref{eq:equal} indicates that $k$-regular graphs  cannot be amplifiers of weak selection, which is also a consequence of the isothermal theorem~\cite{lieb05}. But disturbing the regularity may possibly change the equality to $N_{eff}>N$.
Moreover, most promising for such a perturbation is to remove an edge from the vertex $v_i$. This is the vertex of the $k$-regular graph to be tested with the largest remeeting time $\tau_i$, that is $\max(\tau_i)=\underset{i \in \mathcal{G}}{\max} \: \tau_i$. The argument is that if we take a regular graph and induce a small perturbation by removing  an edge, 
the relative degree $\pi_i$ and the remeeting time $\tau_i$ experience  small deviations $\Delta \pi_i$ and $\Delta \tau_i$. For the identity condition  \eqref{eq:ident} it follows \begin{equation} 
 \sum_{i \in \mathcal{G}} \Delta (\pi_{i} ^2\tau_{i}^{}) \approx \sum_{i \in \mathcal{G}} (2 \pi_i \Delta \pi_{i} \tau_{i}^{}+ \pi_{i} ^2 \Delta \tau_{i}^{}) \approx 0. \label{eq:diff_ident}
\end{equation}
As there is $\pi_i=1/N$ for the unperturbed regular graph,  we obtain \begin{equation}  1/N  \sum_{i \in \mathcal{G}} \Delta \tau_{i}^{} \approx   -2 \sum_{i \in \mathcal{G}} \Delta \pi_{i} \tau_{i}^{}. \label{eq:perturb2} \end{equation}
For the effective population size, Eq. \eqref{eq:neff}, the perturbation yields
\begin{equation} \Delta N_{eff}= \sum_{i \in \mathcal{G}} \Delta(\pi_i \tau_i) \approx \sum_{i \in \mathcal{G}} (\Delta \pi_i \tau_i+\pi_i \Delta \tau_i). \label{eq:diff_neff} \end{equation}
By observing $\pi_i=1/N$ for the unperturbed regular graph and inserting Eq. \eqref{eq:perturb2}, we get  
\begin{equation} \Delta N_{eff} \approx -\sum_{i \in \mathcal{G}} \Delta\pi_i \tau_i. \label{eq:approx}\end{equation}
The relationship \eqref{eq:approx} implies that a positive perturbation of the effective population size (and thus the possibility to get $N_{eff}>N$) is obtained if for a large $\tau_i$ the perturbation induces a decrease of the relative degree $\pi_i$, that is a negative $\Delta \pi_i$.
Thus, Eq. \eqref{eq:approx} can be interpreted as a procedure to identify transient amplifiers. We need to find the vertex $v_{n_1}$, $n_1= \arg \underset{i \in \mathcal{G}}{\max} \: \tau_i$, with the largest remeeting time and proceed by removing each of the $k$ edges adjacent to the vertex. Possibly one (or even several) of the perturbed graphs is a transient amplifier, which we can test by  condition \eqref{eq:neff}.

We may repeat this perturbation by removing another edge and thus introducing second (subsequent) deviations $\Delta^2 \pi_i$ and $\Delta^2 \tau_i$ to the relative degree $\pi_i$ and the remeeting time $\tau_i$.  Analogously to Eqs.~\eqref{eq:diff_ident} and \eqref{eq:diff_neff} we obtain for the identify condition \eqref{eq:ident} and the effective population size \eqref{eq:neff} the following second perturbations:
\begin{equation} 
 \sum_{i \in \mathcal{G}} \Delta^2 (\pi_{i} ^2\tau_{i}^{}) \approx \sum_{i \in \mathcal{G}} (2 \pi_i \Delta^2 \pi_{i} \tau_{i}^{}+ 2\left(\Delta \pi_i \right)^2 \tau_i+4\pi_i \Delta \pi_i \Delta \tau_i+\pi_{i} ^2 \Delta^2 \tau_{i}^{}) \approx 0. \label{eq:diff_ident2}
\end{equation}
and
\begin{equation} \Delta^2 N_{eff}= \sum_{i \in \mathcal{G}} \Delta^2(\pi_i \tau_i) \approx \sum_{i \in \mathcal{G}} (\Delta^2 \pi_i \tau_i+2\Delta \pi_i \Delta \tau_i +\pi_i \Delta^2 \tau_i). \label{eq:diff_neff2} \end{equation}
Combining these equations and using $\pi_i=1/N$ yields
\begin{equation} \Delta^2 N_{eff}\approx -\sum_{i \in \mathcal{G}} \left( \left[\Delta^2 \pi_i +2N\left(\Delta \pi_i \right)^2\right]  \tau_i+2\Delta \pi_i \Delta \tau_i\right). \label{eq:approx2} \end{equation}
Also this equation can be interpreted as a calculating instruction to obtain a transient amplifier with a positive $\Delta^2 N_{eff}$, but in addition to the effect of $\tau_i$ (as in Eq. \eqref{eq:approx}), we now also have the influence of $\Delta \tau_i$. In other words, 
we may either remove another edge from the vertex $v_{n_1}$ with the largest $\tau_i$.  But the effect of a negative $\Delta^2 \pi$ is countered by  $2N\left(\Delta \pi_i \right)^2 >0$, which means we need $|\Delta^2 \pi_i| >2N\left(\Delta \pi_i \right)^2$ for the  largest $\tau_i$ to become effective. 
Alternatively, or even additionally,
we may remove an edge from the vertex $v_{n_2}$, $n_2= \arg \underset{i \in \mathcal{G}}{\max} \: \Delta \tau_i$, with the largest $\Delta \tau_i$.
For this procedure repeated another time, we get 
\begin{align} \Delta^3 N_{eff}\approx  -\sum_{i \in \mathcal{G}} & \left( \left[\Delta^3 \pi_i +6N\Delta^2 \pi_i \Delta \pi \right] \tau_i \right. \nonumber  \\   & + \left. \left[3 \Delta^2 \pi_i + 6N\left(\Delta \pi_i \right)^2\right]  \Delta \tau_i   +3\Delta \pi_i \Delta^2 \tau_i \right). \label{eq:approx3} \end{align}
The  instruction for identifying transient amplifiers associated with the third perturbation involves either to remove another edges from $v_{n_1}$ and/or $v_{n_2}$, or to remove an edge from $v_{n_3}$, $n_3= \arg \underset{i \in \mathcal{G}}{\max} \: \Delta^2 \tau_i$, with the largest $\Delta^2 \tau_i$, or any combination of removals from the edge set $(v_{n_1},v_{n_2},v_{n_3})$. 

We may continue  to perturb the graph by further edge removals which suggests a general iterative procedure to identify  transient amplifiers starting with  regular input graphs. The iterative procedure is presented in Algorithm 1. Its basic form is an enumerative, brute-force search.  
Additional steps for an approximative, greedy search are denoted in italics and parenthesis. We next discuss features and properties of the algorithm.
Algorithm 1 has an input as it requires a regular graph $\mathcal{G}_{in}$ with degree $k$. It can be taken from the set of all simple connected pairwise non-isomorphic $k$-regular graphs on  $N$ vertices with degree $k\geq 3$, whose number of known for small $N$, see e.g.~\cite{mer99,rich21,reg_graph}. 
As discussed in the Sec. \ref{sec:results}, not all regular graphs can be disturbed into transient amplifiers.

\begin{algorithm}[t]
\SetAlgoLined
\SetKwInOut{Input}{Input}\SetKwInOut{Output}{Output}
 \Input{Regular graph $\mathcal{G}_{in}$ with degree $k$,  for instance from the set of all non-isomorphic regular graphs;}
 \Output{Set of transient amplifier graphs (if successful);}
 Find vertex $v_{n_1}$ of $\mathcal{G}_{in}$ with largest remeeting time $n_1= \arg \underset{i \in \mathcal{G}_{in}}{\max} \: \tau_i$\;
  Create $k$ graphs by removing each one of the $k$ edges  from vertex $v_{n_1}$\;
  Discard graph if it became disconnected by the edge removal\;
  Store remaining graphs in  $\mathcal{G}(0)$,  $| \mathcal{G}(0) | \leq k$\;
 Set the number of edge removal repetitions $\ell$\; 
 Set
 $j:=1$\;
 \textbf{\textit{(Set upper limit (filter size) $\#_{lim} \gg k$ of graphs to be included in the next edge removal repetition)}}\;
 \While{ $j \leq \ell$ $\lor$  $| \mathcal{G}(j-1) |>0$} { \For{
Each graph of the graph set $\mathcal{G}(j-1)$} {
  Find vertex $v_{n_j}$ with largest remeeting time difference $n_{j}= \arg \underset{i \in \mathcal{G}(j-1)}{\max} \: \Delta^j \tau_i$\;
Create $k$ graphs by removing each one of the $k$ edges  from vertex $v_{n_j}$\;
 Discard graphs which became disconnected by the edge removal\;
  Remove isomorphic graphs\;
  Calculate effective population size $N_{eff}$ by Eq. \eqref{eq:neff} to test amplification\;
 Store graphs in $\mathcal{G}_j$, $| \mathcal{G}_j |\leq k^{j}$\;
 }
\textbf{\textit{(Calculate spectra and other graph measures of all graphs in $\mathcal{G}_j$)}}\;
\textbf{\textit{(Discard graphs exceeding the filter size $\#_{lim}$  depending on a filter criteria derived from spectral graph measures)}}\;
 Set $\mathcal{G}(j)=\mathcal{G}_j$\; Set $j:=j+1$\;
 Remove isomorphic graphs from the graph set  $\{\mathcal{G}(0), \mathcal{G}(1),\ldots,\mathcal{G}(j) \}$\;}
 Return remaining graphs with amplification properties\;

 \caption{Iterative procedure to construct transient amplifiers from regular graphs. \\ Basic form: enumerative, brute-force search. \\
 \textbf{(Additional steps:  approximative, greedy search.)} }
\end{algorithm}

If from a given regular graph we repeat to remove edges, then sooner or later the graph will become disconnected. Thus, an important parameter of the algorithm is  the number of allowed edge removal repetitions $\ell$.
A regular graph has $\frac{kN}{2}$ edges and a connected graph has at least $N-1$ edges, which gives us an upper bound of edge removals:  $1+\frac{(k-2)N}{2}$. Consequently, the number of edge removals  $\ell$ may vary between $1 \leq \ell \leq 1+\frac{(k-2)N}{2}$. This bound is for the total number of edges to be removed from the graph.  From a given vertex at most $k$ edges can be removed before the vertex is no longer connected to the remainder of the graph. If an edge removal disconnects the graph, Algorithm 1 discards the graph. In other words, in order to keep a graph as a potential structure to be perturbed into a transient amplifier, we should not disconnect it by a needless edge removal.  Although suggested by Eqs. \eqref{eq:approx2} and \eqref{eq:approx3} as a possibility, we thus  should sparsely (if at all) remove additional edges from the vertex $v_{n_1}$, which is the vertex with the largest initial remeeting time. In the implementation of Algorithm 1, additional edges from $v_{n_1}$ are only removed if there is $v_{n_j}=v_{n_1}$ for $1<j\leq \ell$.

Algorithm 1 in its basic form is a breadth-first search with a brute-force enumeration of all possible (non-isomorphic) graphs resulting from iterative edge removals. 
It also
relies substantially upon identifying isomorphic graphs. Roughly speaking, isomorphism means that two graphs are structurally alike and merely differ in how the vertices and edges are named.  More precisely,  two graphs are isomorphic if there is a bijective mapping between their vertices which preserves adjacency~\cite{bon08}, pp.~12--14. 
Unfortunately, the computational problem of finding whether or not two finite graphs are isomorphic is not solvable in polynomial time~\cite{arv05,bab18}, which is a major restriction to the applicability of Algorithm 1. Therefore, we substitute detecting isomorphic graphs  by detecting cospectral graphs, which is  computationally less expensive. The rationale of using cospectral as a proxy for isomorphic is that all isomorphic  graphs are cospectral. On the other hand, cospectral graphs can be non-isomorphic. Thus, we might discard graphs which could possibly have been additional sources of transient amplifiers. However, numerical studies suggest that non-isomorphic pairs of graphs with the same spectrum are not very frequent and the effect of mistaking cospectral for non-isomorphic can be minimized by using the spectrum of the normalized Laplacian~\cite{but11}. 

Another limitation of Algorithm 1 in its basic form is the exponential growth of the number of candidate graphs produced by iterative edge removals, which restricts the applicability to small $k$ and $\ell$. However, the transient amplifiers produced for small $k$ and $\ell$ mostly have only a small perturbation to their input regularity and degree distribution. Thus, if we also want to study transient amplifiers with  possibly stronger  perturbations to their regularity  and  more unbalanced degree distributions,  larger $k$ and $\ell$ would be desirable. To counter the growth of the number of candidate graphs and achieve practical computability, we need to modify the basic form of Algorithm 1. Therefore, we set a limit $\#_{\mathcal{G}} \gg k$ to the number of graphs to be included in the next iteration. This restricts the number of graphs taken as an input to the subsequent repetition of  edge removals and thus bounds the exponential growth of the number of candidate graphs. In this paper it is suggested to evaluate spectral graph measures to decide which graphs (if the limit  $\#_{\mathcal{G}}$ is exceeded) 
are included  in the next iteration. In some sense, the modifications to the basic form of the algorithm work like a filter which passes only a limited number of graphs selected by their spectral properties. Thus, we also call the limit $\#_{\mathcal{G}}$ the filter size.  To summarize, the modified Algorithm 1 is a kind of approximative, greedy search for finding transient amplifiers of death-Birth updating. In Algorithm 1 the additional steps augmenting the basic form are given in italics and parenthesis.

\subsection{Graph spectra and edge removals}

An (undirected and unweighted) graph  $\mathcal{G}=(V,E)$ is specified algebraically by a symmetric adjacency matrix $A=\{a_{ij}\}$ with $a_{ij}=a_{ji}=1$ indicating that the vertices $v_i$ and $v_j$ are connected by the edge $e_{ij} \in E$. With the vertex degree $k_i=\sum_{j=1}^{N} a_{ij}$, we additionally specify a degree matrix $D=\rm{diag}$ $(k_1,k_2,\ldots,k_N)$. 
For the  spectral analysis of evolutionary graphs $\mathcal{G}$ we take $A$ and $D$, and consider     the  standard Laplacian $L_{\mathcal{G}}=D-A$  and the normalized Laplacian $\Lambda_{\mathcal{G}}=I-D^{-1/2}AD^{-1/2}$.   The spectrum of the standard Laplacian is denoted by $\mu(\mathcal{G})$
and consists of $N$ eigenvalues $0=\mu_1\leq \mu_2 \leq \ldots \leq \mu_N$, while for the spectrum of the normalized Laplacian we have $\lambda(\mathcal{G})$ with $0=\lambda_1\leq \lambda_2 \leq \ldots \lambda_N\leq 2$. The second smallest eigenvalues $\mu_2$ and $\lambda_2$  are frequently called algebraic connectivity. 
From the Laplacian eigenvalues  a spectral distance $d$ can be defined which is useful for comparing two  families (or classes) of graphs $(\mathcal{G})$ and $(\mathcal{G}')$. For each family only containing a single member, we can also compare two graphs $\mathcal{G}$ and $\mathcal{G}'$.  Therefore, we consider a smoothed spectral density which convolves the eigenvalues $\lambda_i$ with a Gaussian kernel with standard deviation $\sigma$~\cite{ban09,ban12,gu16} 
\begin{equation}
    \varphi_{\mathcal{G}}(x)= \frac{1}{N}\sum_{i=1}^{N} \frac{1}{\sqrt{2 \pi \sigma^2}} \exp{\left(\frac{(x-\lambda_i)^2}{2 \sigma^2} \right)}. \label{eq:density}
\end{equation} We set $\sigma=1/(3N)$.
From this continuous spectral density we can define a pseudometric on graphs by the distance~\cite{gu16} 
\begin{equation}
d(\mathcal{G},\mathcal{G}')= \int_0^2|\varphi_{\mathcal{G}}(x)-\varphi_{\mathcal{G}'}(x)|dx. \label{eq:distance}
 \end{equation}
 Eqn. \eqref{eq:density} and  \eqref{eq:distance} are defined for the normalized Laplacian spectrum $\lambda$, which is contained in the interval $[0,2]$ for any graph order and degree and thus is conveniently comparable over varying order and degree. However, the spectral density and the spectral distance can also be defined for the standard Laplacian spectrum $\mu$ by replacing $\lambda_i$ by $\mu_i$ in Eq. \eqref{eq:density}, albeit with variable upper integration limit in  Eq.  \eqref{eq:distance}.

Suppose we have a graph $\mathcal{G}=(V,E)$ and remove one of its edges. Thus,  the vertex set is preserved but  the edge set is changed. We use $(\mathcal{G}-e_{ij})$ for denoting the graph resulting from the edge $e_{ij} \in E$ being removed from $\mathcal{G}$. As discussed in the previous section, for the greedy algorithm we need spectral measures for deciding which graphs should be included in the subsequent repetition of edge removals. We next review some results about edge removals and spectral characteristics useful for directing the greedy algorithm towards finding transient amplifiers.


Starting point of this discussion are several interlacing results connecting spectra with edge removals~\cite{atay14,chen04,heu95}. For the spectra of the standard Laplacian there is $\mu_{i-1}(\mathcal{G}) \leq \mu_i(\mathcal{G}-e_{ij}) \leq \mu_{i}(\mathcal{G})$, $i=2,3,\ldots,N$. This particularly means for the algebraic connectivity (the second smallest eigenvalue), we have always a positive spectral shift $\mu_2(\mathcal{G})-\mu_{2}(\mathcal{G}-e_{ij})=\alpha$ with $\alpha \geq 0$. 
For the normalized Laplacian, the eigenvalue interlacing is  $\lambda_{i-1}(\mathcal{G}) \leq \lambda_i(\mathcal{G}-e_{ij}) \leq \lambda_{i+1}(\mathcal{G})$, $i=2,3,\ldots,N-1$.
Eigenvalue interlacing differs between the standard Laplacian and the normalized Laplacian.  The eigenvalues of the standard
Laplacian decrease or remain unchanged if an edge is removed, while for the normalized Laplacian the eigenvalues may in fact also increase. This is also known as Braess's paradox~\cite{eld17}. For instance, the algebraic connectivity decreasing or increasing is bounded by $0< \lambda_2(\mathcal{G}-e_{ij}) \leq \lambda_3(\mathcal{G})$.

\section{Results} \label{sec:results}

\subsection{Computational setup}
In the previous section a procedure has been derived and analyzed which identifies transient amplifiers of death-Birth updating by employing an iterative design procedure, see Algorithm 1. We next discuss an application  of the algorithm. 
The algorithm has two variants. It can be either an enumerative, brute-force search or an approximative, greedy search. Due to the exponential growth of the number of candidate graphs produced, the enumerative search is numerically feasible only in exceptional cases of some low values of $N$ and $k$. Therefore, the focus of the numerical investigations is on the approximative search.  
For this variant we need to specify the filter size  $\#_{\mathcal{G}}$ and the filter criteria.  Best results were obtained  with low values of the algebraic connectivity  $\lambda_2$ derived from the normalized Laplacian as filter. We also discuss why  the algebraic connectivity  $\mu_2$ derived from the standard Laplacian is not likely to be a successful option. According to the filter criteria, the graphs with the lowest $\#_{\mathcal{G}}$ values of $\lambda_2$ are included in the next iteration. If there are less than  $\#_{\mathcal{G}}$  graphs, all are taken. Throughout the study we use the filter size $\#_{\mathcal{G}}=500$ as preliminary experiments suggested that such a setting is a good compromise between algorithmic performance and computational effort. 
The algorithm uses the remeeting time difference $\Delta^j \tau_i$ for selecting the vertices from which edges are removed. There are different ways of calculating the remeeting time difference $\Delta^j \tau_i$, for instance, a moving difference
$\Delta^j \tau_i=\tau_i(j)-\tau_i(j-1)$ 
and a forward difference
$\Delta^j \tau_i= \sum_{i=0}^j (-1)^{j-i} \left(\begin{smallmatrix}j\\i\end{smallmatrix}\right) \tau_i(j)$. Preliminary experiments (not shown in figures) have shown that a moving difference gives best results and thus is used.

Studying the algorithm
we have the following performance objectives. Firstly, 
we are interested in the number of transient amplifiers identified.   
 Apart from the actual number of graphs,  it is also relevant how many of these graphs are non-isomorphic, which implies they are structurally different.  
 Additionally, our objective is to identify transient amplifiers
 with small and large $N_{eff}$ as this implies different amplification properties. However, the algorithm is not  explicitly optimizing for large or small $N_{eff}$ by pruning graphs as has been shown by M{\"o}ller et al.~\cite{moell19} using a genetic algorithm (see also an application to finding amplifiers of Bd updating~\cite{allen21}.   Finally, we intend to identify transient amplifiers with  different structural properties, as for instance expressed by the mean degree $\bar{k}$. With respect to the behaviour of the algorithm we are mainly interested in how the spectral dynamics generally relates to  edge removals and how graph evolutions leading to transient amplifiers differ from evolutions not doing so.  
 
 \subsection{Regular input graphs on $N=\{11,12\}$ vertices} \label{sec:11_12}
 
\begin{figure}[htb]

\includegraphics[trim = 60mm 110mm 55mm 110mm,clip, width=3.8cm, height=4cm]{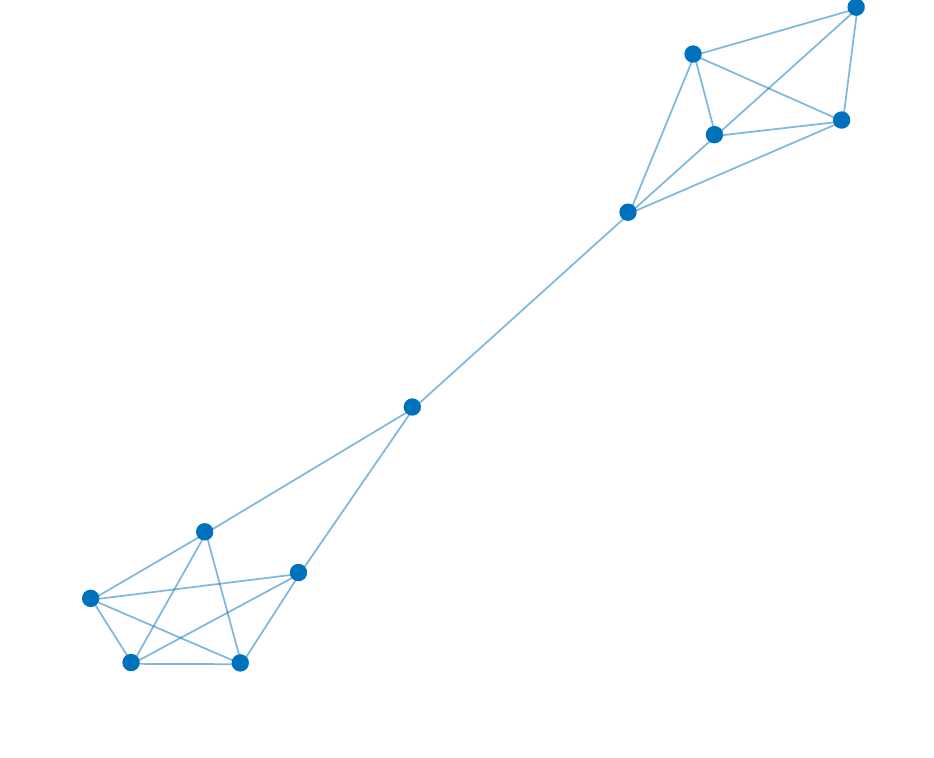}
\includegraphics[trim = 60mm 110mm 55mm 110mm,clip, width=3.8cm, height=4cm]{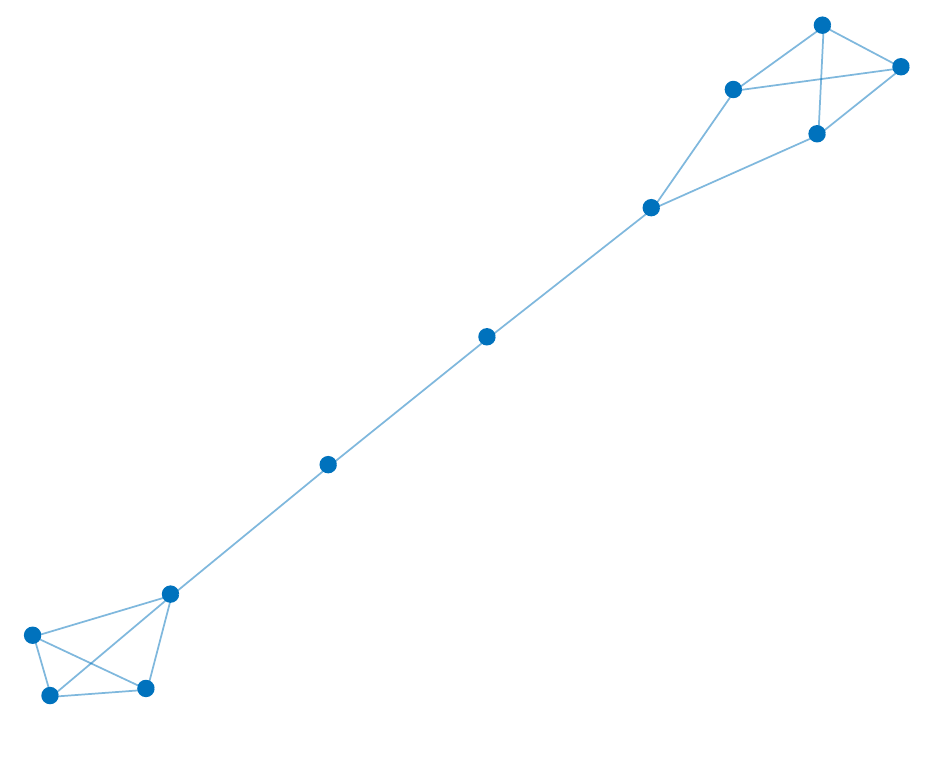}
\includegraphics[trim = 60mm 110mm 55mm 110mm,clip, width=3.8cm, height=4cm]{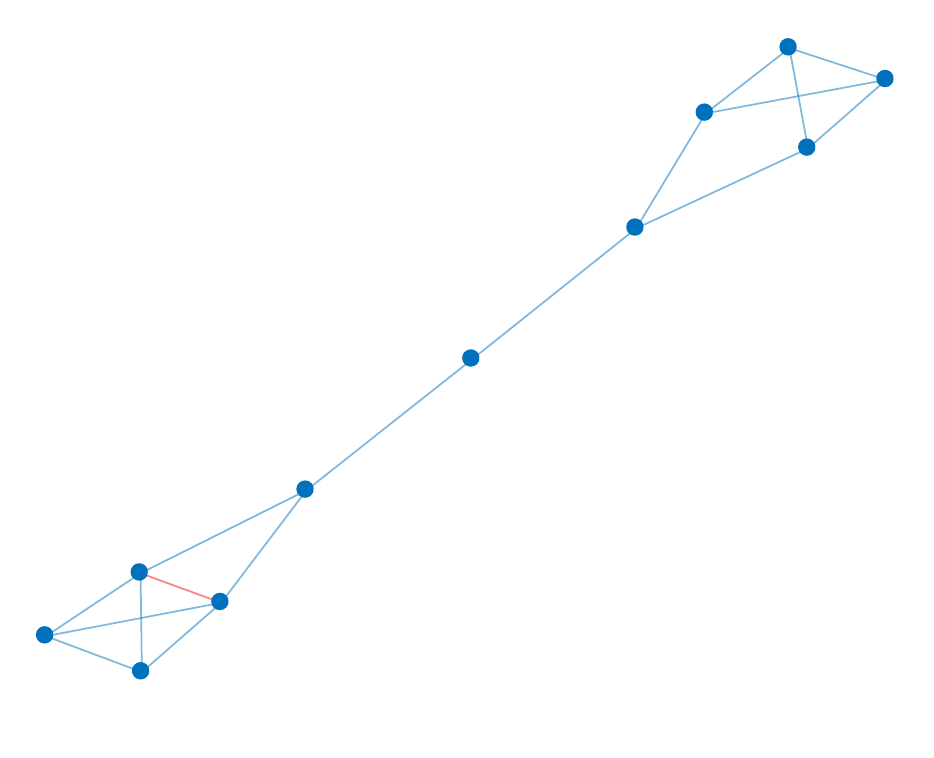}
\includegraphics[angle=5,trim = 58mm 110mm 50mm 110mm,clip, width=3.8cm, height=4cm]{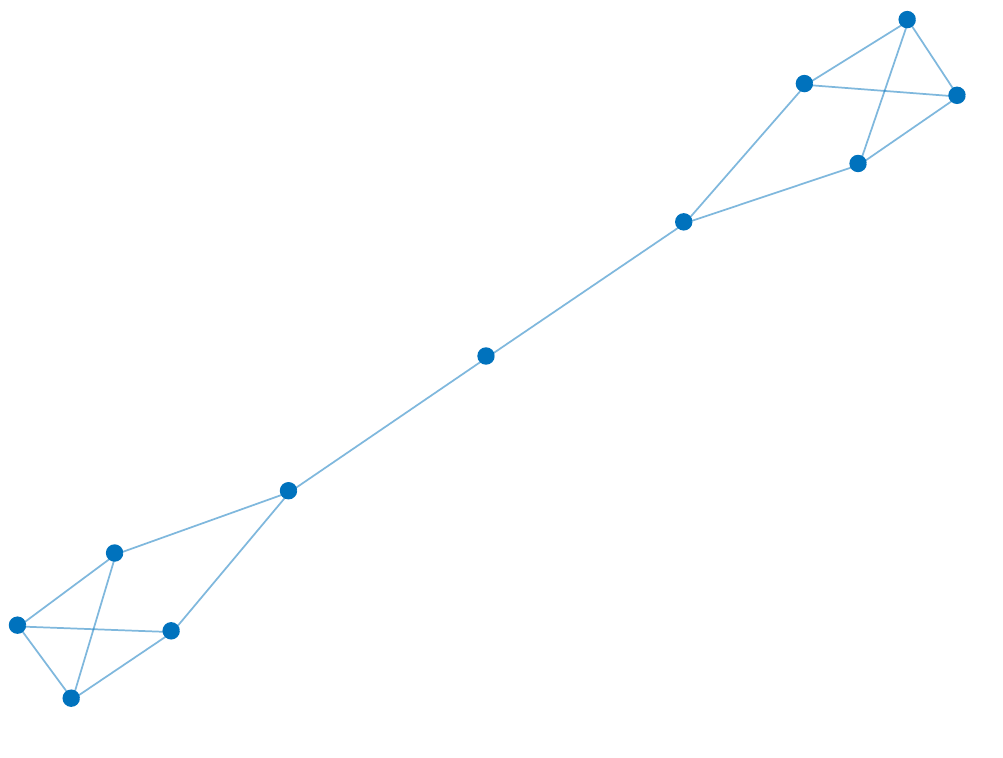}

 (\textbf{a}) \hspace{2.7cm} (\textbf{b})   \hspace{2.7cm} (\textbf{c})  \hspace{2.7cm} (\textbf{d})

\caption{\small{4 non-isomorphic graphs on $N=11$ vertices which are transient amplifiers of death-Birth updating and have maximum degree $\Delta(\mathcal{G})=4$.  (a) $N_{eff}=11.0008$,  $\delta(\mathcal{G})=3$, $\bar{k}=3.8182$, $\lambda_2=0.0567$   (b) $N_{eff}=11.0128$,  $\delta(\mathcal{G})=2$, $\bar{k}=2.9091$, $\lambda_2=0.0399$   (c) $N_{eff}=11.0056$,  $\delta(\mathcal{G})=2$, $\bar{k}=3.0909$, $\lambda_2=0.0451$.      (d) $N_{eff}=11.0952$,  $\delta(\mathcal{G})=2$, $\bar{k}=2.9091$, $\lambda_2=0.0481$. This graph can be obtained by removing the edge depicted in red of the graph in Fig. \ref{fig:graph_11_}(c).      }  }

\label{fig:graph_11_}
\end{figure}

 We start with considering regular graphs on $N=11$ and $N=12$ vertices. For these two graph orders all regular graphs with all occurring degrees have been tested with the numerical resources available in this study. For regular graphs with $N \leq 10$ vertices no transient amplifiers of death-Birth updating were found using the method discussed in this paper. Results of Algorithm 1 (approximative, greedy search) for $N=11$ and $k=(4,6,8)$ are given in  Fig. \ref{fig:graph_11_} and  Tab. \ref{tab:11_graphs}.  

 Fig. \ref{fig:graph_11_} shows 4 non-isomorphic graphs on $N=11$ vertices identified as transient amplifiers of death-Birth updating. The graph in Fig. \ref{fig:graph_11_}(a) has the lowest effective population size  $N_{eff}=11.0008$, the  graph in Fig. \ref{fig:graph_11_}(b) has the highest value $N_{eff}=11.0128$. All 4 graphs have a maximum degree $\Delta(\mathcal{G})=4$; the minimum degree is $\delta(\mathcal{G})=3$ for the graph in Fig. \ref{fig:graph_11_}(a) and    $\delta(\mathcal{G})=2$ for the remaining graphs. Note that the mean degree $\bar{k}$ and the algebraic connectivity $\lambda_2$ scale inversely to the effective population size  $N_{eff}$.
 All graphs share some structural similarities as they all consist of  two cliques of highly connected vertices joined by a bridge, which is a path of one or more  edges. 
 Thus, the graphs can be seen as intermediate forms between a dumbbell graph and a barbell graph~\cite{ghosh08,wang09}, also see the discussion in Sec.~\ref{sec:bar}.

 \begin{table}[htb]
\centering
\caption{\small{Results of Algorithm 1 (approximative, greedy search) for $N=11$ and $k=(4,6,8)$. $\mathcal{L}_k(N)$ is the total number of simple, connected, pairwise nonisomorphic $k$-regular graphs on 11 vertices.   $\mathcal{A}_k(N)$ is the proportion of these regular graphs from which transient amplifiers of death-Birth updating are obtained.  $\#_{tot}$ is the total number of transient amplifiers found,   $\#_{noniso}$ is the number of pairwise non-isomorphic transient amplifiers for each $k$. From all graphs with $N=11$ and  $k=(4,6,8)$
4 non-isomorphic are obtained, see Fig.  \ref{fig:graph_11_}. 
}}
\label{tab:11_graphs}
\begin{tabular}{ccccc}

 $k$ & $\mathcal{L}_k$ & $\mathcal{A}_k$  & $\#_{tot}$ & $\#_{noniso}$   \\ 
 \hline 
 4 & 265 & 5 & 6 &  2  \\
 6 & 266 & 228 & 937 & 4    \\
 8 & 6 &  5 & 15  & 1   \\

\hline

\end{tabular}
\end{table}

 Tab. \ref{tab:11_graphs} gives  results about the algorithmic process for $N=11$ and  $k=(4,6,8)$.  
  A first result is that there are instances of either quartic or sextic or octic regular graphs which can be disturbed into transient amplifiers, but their numbers vary. Whereas for 6-regular graphs $\mathcal{A}_6=228$ graphs out of the  $\mathcal{L}_6(11)=266$  and for 8-regular graphs $\mathcal{A}_8=5$ graphs out of the  $\mathcal{L}_8(11)=6$  produce transient amplifiers, the percentage of 4-regular graphs is much lower ($5$ out of $265$).   Moreover, we know that the result for $k=4$ is not specific for the approximative search with filter size $\#_{\mathcal{G}}=500$ as for the $\mathcal{L}_4=265$ quartic graphs a complete enumeration has been possible and brought exactly the same result. A second finding is that although the total number of transient amplifiers found varies substantially between the graph degrees (there are $937$ for $k=6$, but only $6$ for $k=4$), the number of non-isomorphic graphs is more stable. Overall, 4 non-isomorphic transient amplifier graphs have been found for $N=11$, see Fig. \ref{fig:graph_11_}. All 4 graphs are obtained for $k=6$, but the graph in Fig. \ref{fig:graph_11_}(a) is also identified for $k=4$ and $k=8$ and the graph in Fig. \ref{fig:graph_11_}(b) we also get for $k=4$.
In other words, for all degrees of regular graph with $N=11$ used as an input to Algorithm 1, there is a certain convergence toward graph structures having amplification properties.

\begin{figure}[htb]
\centering
\includegraphics[trim = 25mm 90mm 40mm 80mm,clip, width=6.6cm, height=5.5cm]{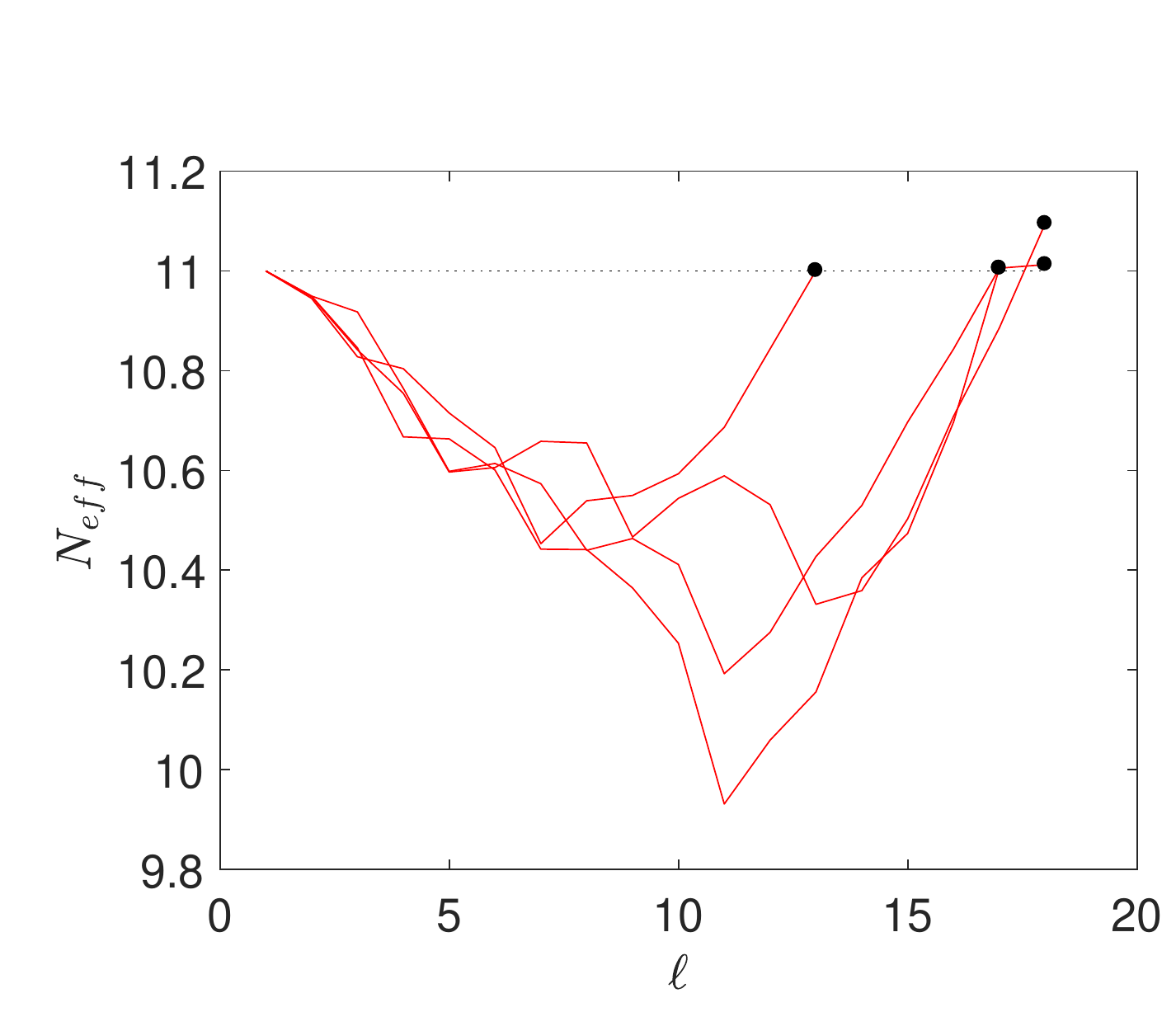}
\includegraphics[trim = 25mm 90mm 40mm 80mm,clip, width=6.6cm, height=5.5cm]{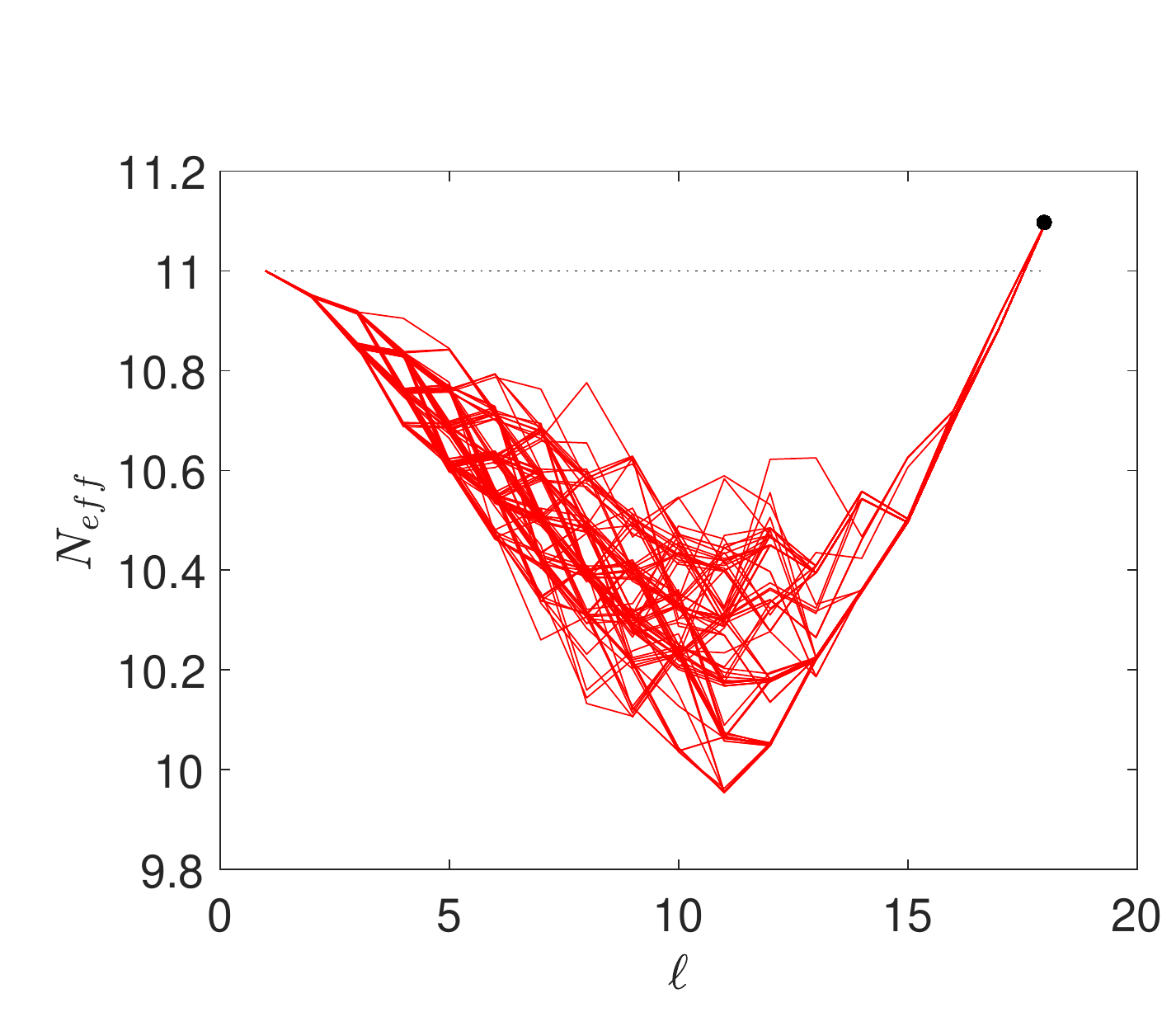}

 (\textbf{a}) \hspace{4.7cm} (\textbf{b})  
 
 \includegraphics[trim = 25mm 90mm 40mm 80mm,clip, width=6.6cm, height=5.5cm]{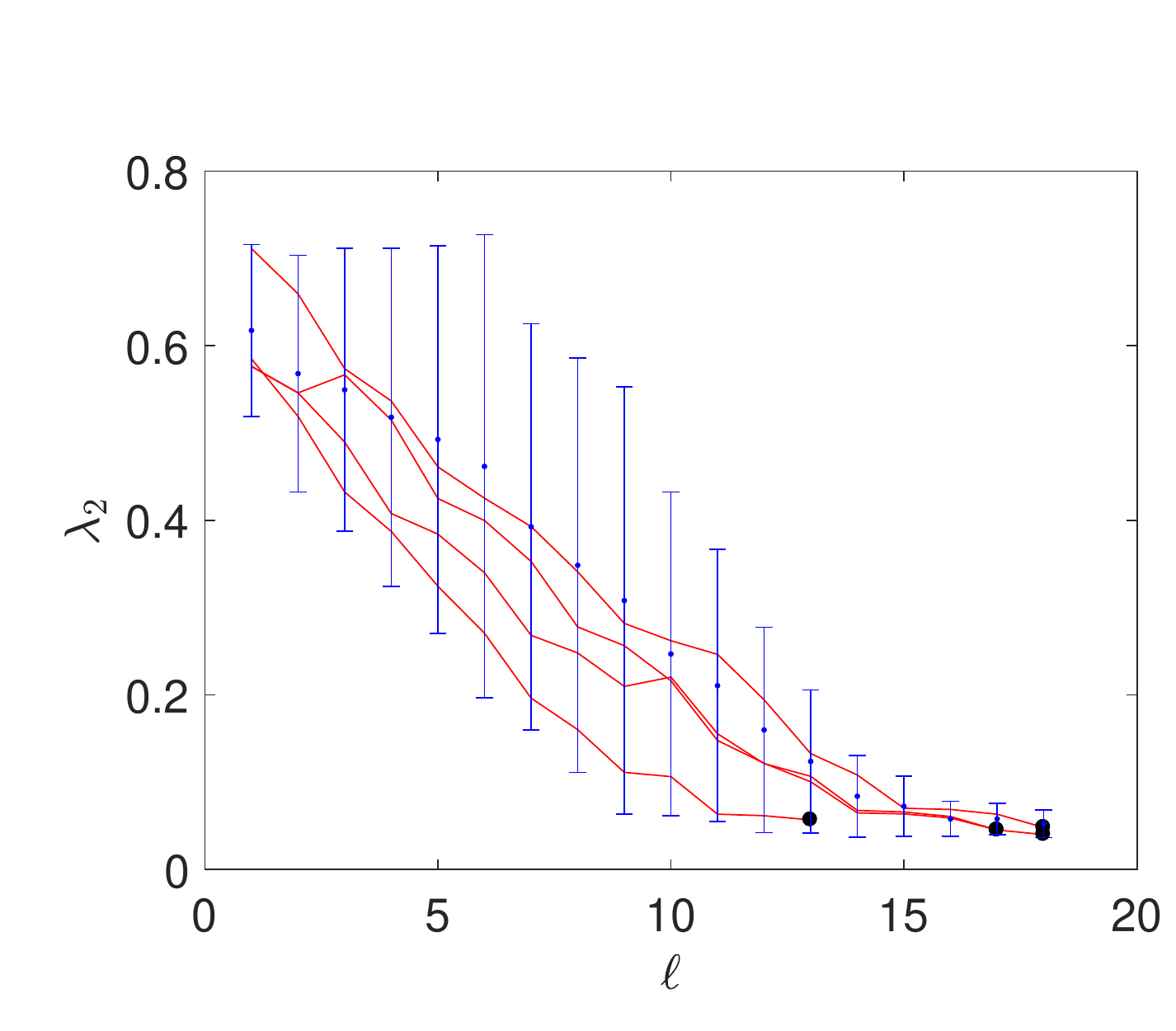}
 \includegraphics[trim = 25mm 90mm 40mm 80mm,clip, width=6.6cm, height=5.5cm]{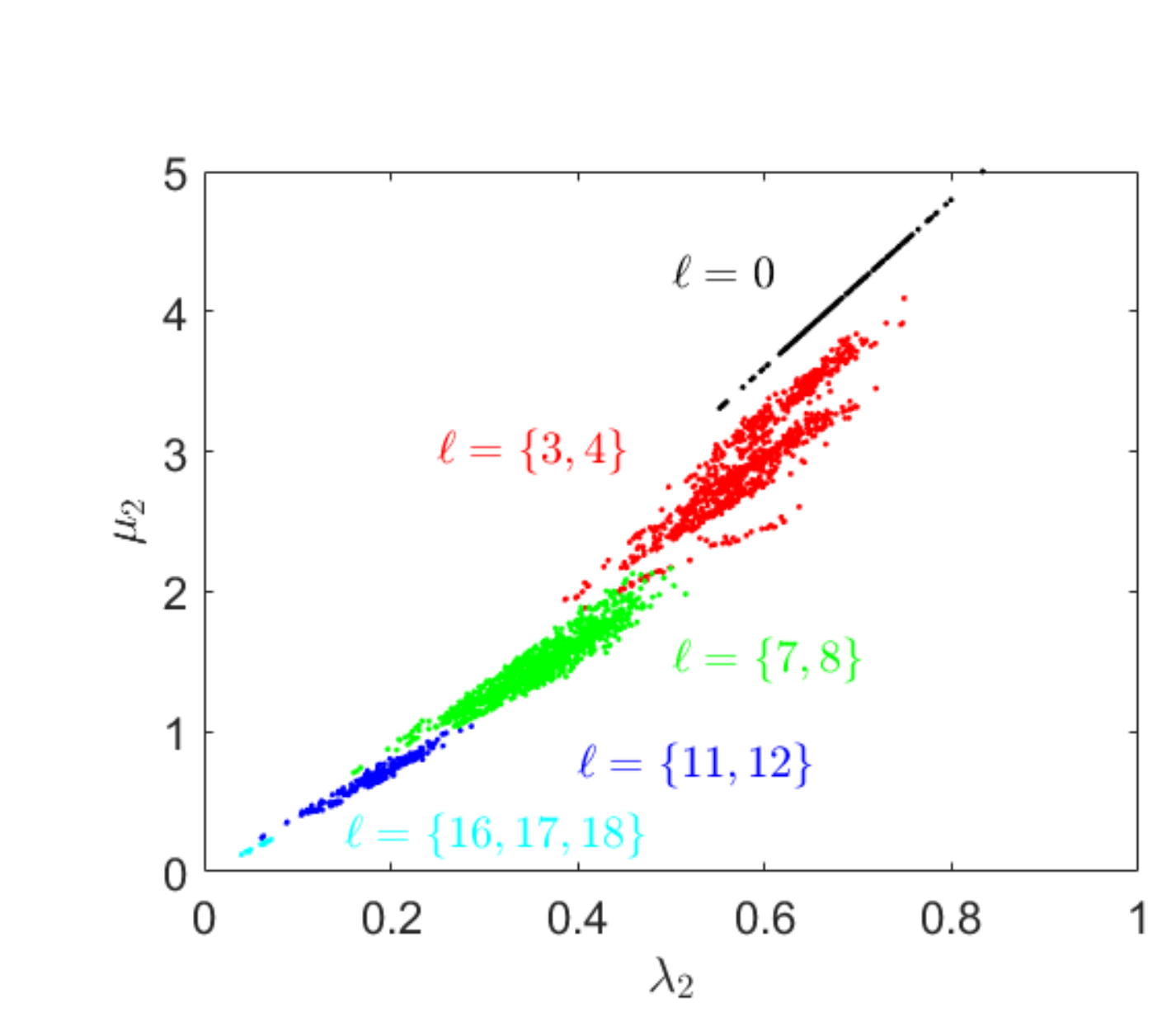}
 
  (\textbf{c}) \hspace{4.7cm} (\textbf{d})  
 
\caption{\small{Behaviour of the approximative search of Algorithm 1 for order $N=11$, degree $k=6$. (a) The effective population size $N_{eff}$ over edge removal repetitions $\ell$ for each of the 4 non-isomorphic transient amplifiers depicted in Fig. \ref{fig:graph_11_}. The black dots mark the final values $N_{eff}>11$.  (b)  $N_{eff}$ over $\ell$ for the graph with the highest final $N_{eff}$, Fig. \ref{fig:graph_11_}(b), for all 33 sextic regular graphs which can be disturbed into this transient amplifier. (c) The algebraic connectivity $\lambda_2$ over $\ell$ for all 4 non-isomorphic transient amplifiers according to Fig. \ref{fig:graph_11_} and produced by taking 6-regular graphs as input.    (d) Scatter plot of algebraic connectivity $\lambda_2$ and $\mu_2$ derived from the normalized and standard Laplacian, respectively, for different edge removal repetitions $\ell$. Black dots $\ell=0$, red dots $\ell=\{3,4\}$, green dots $\ell=\{7,8\}$, blue dots $\ell=\{11,12\}$, cyan dots $\ell=\{16,17,18\}$.}  }
\label{fig:graph_11_6_neff}
\end{figure}

Further insight into the algorithmic process can be obtained
by analyzing the behavior of some quantities connected to finding
transient amplifiers over the run time of the algorithm. Fig. \ref{fig:graph_11_6_neff}(a) shows the effective population size $N_{eff}$, Eq. \eqref{eq:neff}, over edge removal repetitions $\ell$ for all 4 non-isomorphic transient amplifiers according to Fig. \ref{fig:graph_11_} and produced by taking  4 specific 6-regular graphs as input. In other words, each curve in Fig. \ref{fig:graph_11_6_neff}(a) can be interpreted as a trajectory accounting for $N_{eff}$ over graphs experiencing repeated edge removals at iterations $\ell$.   As for regular graph the equality $N_{eff}=N$ applies, the curves start at $N_{eff}=11$, fall to some lower values $10<N_{eff}<11$, before rising up and ending at specific values $N_{eff}>11$. These values are marked by black dots for each of the graphs. The graph in Fig. \ref{fig:graph_11_}(a) with $N_{eff}=11.0008$ is obtained for the lowest $\ell$ as it requires the lowest number of edge removals, and thus has the highest mean degree $\bar{k}=3.8182$. The graphs obtained for the highest $\ell$  (shown in Fig. \ref{fig:graph_11_}(b) and (d))  have the lowest mean degree $\bar{k}=2.0991$. 
The connected black dots of a trajectory between $\ell=17$ and $\ell=18$  indicate that the two transient amplifier graphs differ in just one edge. By removing a single edge (depicted in red) the graph in  Fig. \ref{fig:graph_11_}(c) can be turned  into the graph in  Fig. \ref{fig:graph_11_}(d).

Fig. \ref{fig:graph_11_6_neff}(b) illustrates a different aspect of the same process. 
Here, the effective population size $N_{eff}$ is shown over edge removal repetitions $\ell$ only for the graph with highest $N_{eff}=11.0128$, Fig. \ref{fig:graph_11_}(b), but for all initial sextic regular graphs which can be disturbed into this transient amplifier. We obtain that $33$ out of the $\mathcal{A}_6= 228$ graphs have this property. 
Moreover, we get $99$ different trajectories along $\ell$ as for the $33$ initial 6-regular graphs there are up to $5$ different ways edge removal sequences can lead to the same graph. Whereas basically the same shape of the curves can be observed as in Fig. \ref{fig:graph_11_6_neff}(a), 
it is also worth mentioning that about halfway through the process (about $\ell \approx 10$) a rather large range of $N_{eff}$ can be seen 
which merges in a steep increase of $N_{eff}$ before finally reaching $N_{eff}=11.0128$. This result can be interpreted as starting from the initial regular graphs, after 2 or 3 edge removal repetitions there emerges a substantial structural diversity of graphs by the edge removal process which finally converge to the transient amplifier. 

The occurrence of structural diversity of graphs is supported by the results given in Fig.  \ref{fig:graph_11_6_neff}(c) which shows an aspect of spectral dynamics  with the behaviour of the algebraic connectivity $\lambda_2$ over $\ell$ for all 4 non-isomorphic transient amplifiers according to Fig. \ref{fig:graph_11_} and produced by taking 6-regular graphs as input.  
Again, the values obtained for the final transient amplifiers are marked by black dots. The blue error bars indicate the range between largest and smallest $\lambda_2$ in the whole ensemble of candidate graphs at iteration $\ell$. We see that the values of $\lambda_2$ leading to transient amplifiers mostly are below the mean of all $\lambda_2$ of candidate graphs, but not the smallest. The values of $\lambda_2$ are mostly falling for $\ell$ getting larger. The interlacing result of normalized Laplacian is $0< \lambda_2(\mathcal{G}-e_{ij}) \leq \lambda_3(\mathcal{G})$ which implies that $\lambda_2$ may also increase if an edge is removed. In fact, this can observed, albeit rarely, for instance at $\ell=9$. However, for most edge removals,  $\lambda_2$ is falling or stays constant. 
Also note that the range of $\lambda_2$ as indicated by the error bars initially increases and reaches the largest range at $5 \leq \ell \leq 10$ before shrinking for the final edge removals before a transient amplifier is reached. The transient amplifiers reached at the end of the process have similar $\lambda_2$ and these values are also taken if the number of edge removals required is smaller as to be seen for the graph in Fig.   
\ref{fig:graph_11_}(a) which is reached for $\ell=11$. Finally, we analyze how the algebraic connectivity $\lambda_2$ and $\mu_2$ derived from the normalized and the standard Laplacian, respectively,  evolve for the edge removal process, see Fig. \ref{fig:graph_11_6_neff}(d). The setting is the same as for Fig. \ref{fig:graph_11_6_neff}(c), that is for all 4 non-isomorphic transient amplifiers according to Fig. \ref{fig:graph_11_} and produced by taking 6-regular graphs as input. Fig. \ref{fig:graph_11_6_neff}(d) shows a scatter plot of $\lambda_2$ and $\mu_2$ over edge removal repetitions $\ell$, where black dots are for $\ell=0$, red dots for $\ell=\{3,4\}$, green dots for  $\ell=\{7,8\}$, blue dots for $\ell=\{11,12\}$ and  cyan dots for $\ell=\{16,17,18\}$. This means in addition to the edge removals also the relationship between $\lambda_2$ and $\mu_2$ for the sextic regular input graphs is shown (as black dots). As for regular graphs the normalized Laplacian is $\Lambda_{\mathcal{G}}=I-1/k \cdot A$ and the standard Laplacian is 
$L_{\mathcal{G}}=kI-A$, we have a linear relation $k \lambda_2=\mu_2$
for $\ell=0$, which can be  seen as the line of black dots in Fig. \ref{fig:graph_11_6_neff}(d). However, for $\ell$ getting larger, edges getting removed and the regularity of the candidate graphs being disturbed, the linear relationship collapses. This is particularly visible for   $\ell=\{3,4\}$, see the cloud of red dots in Fig.~\ref{fig:graph_11_6_neff}(d). This effect gets less profound for $\ell$ getting larger and almost vanishes for  $\ell\geq 16$. The result can be interpreted as particularly in the initial phase of the edge removals both types of algebraic connectivity $\lambda_2$ and $\mu_2$ account for different aspects of the graph structure, and thus might be differently suitable for guiding the search process. The spectrum of the  normalized Laplacian capturing  geometric and structural properties differently to the spectra of the standard Laplacian or the adjacency matrix has been already noted~\cite{ban08,ban09,ban12,gu16}. Below we come back to this property and discuss more details.  

\begin{figure}[htb]

\includegraphics[angle=5,trim = 58mm 110mm 55mm 110mm,clip, width=3.8cm, height=4cm]{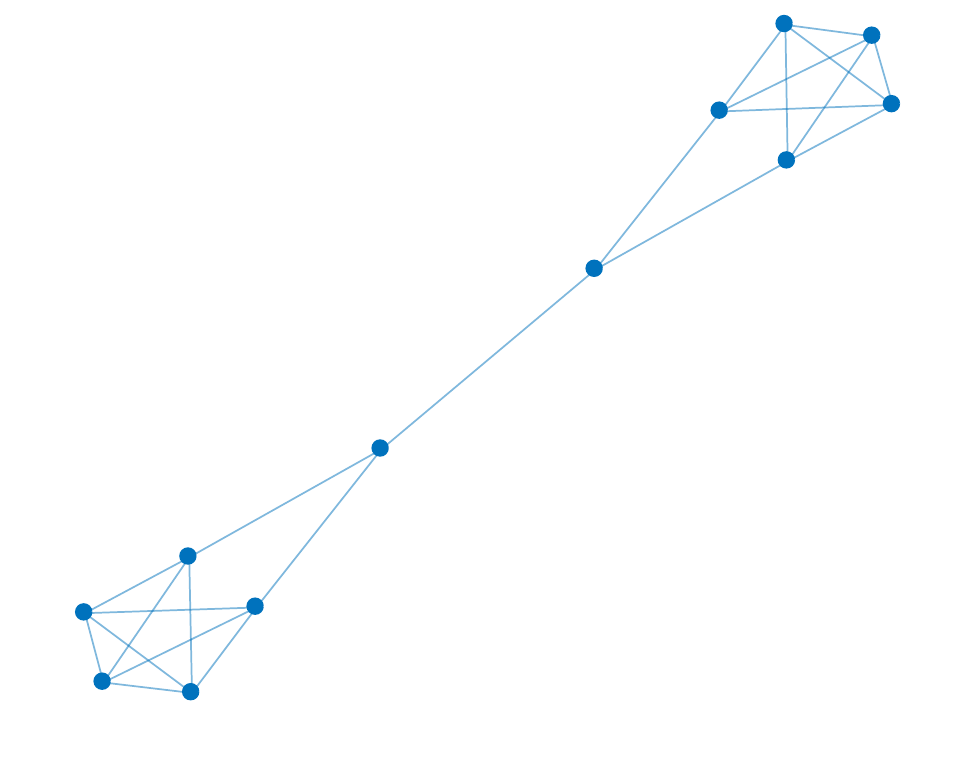}
\includegraphics[trim = 60mm 110mm 55mm 110mm,clip, width=3.8cm, height=4cm]{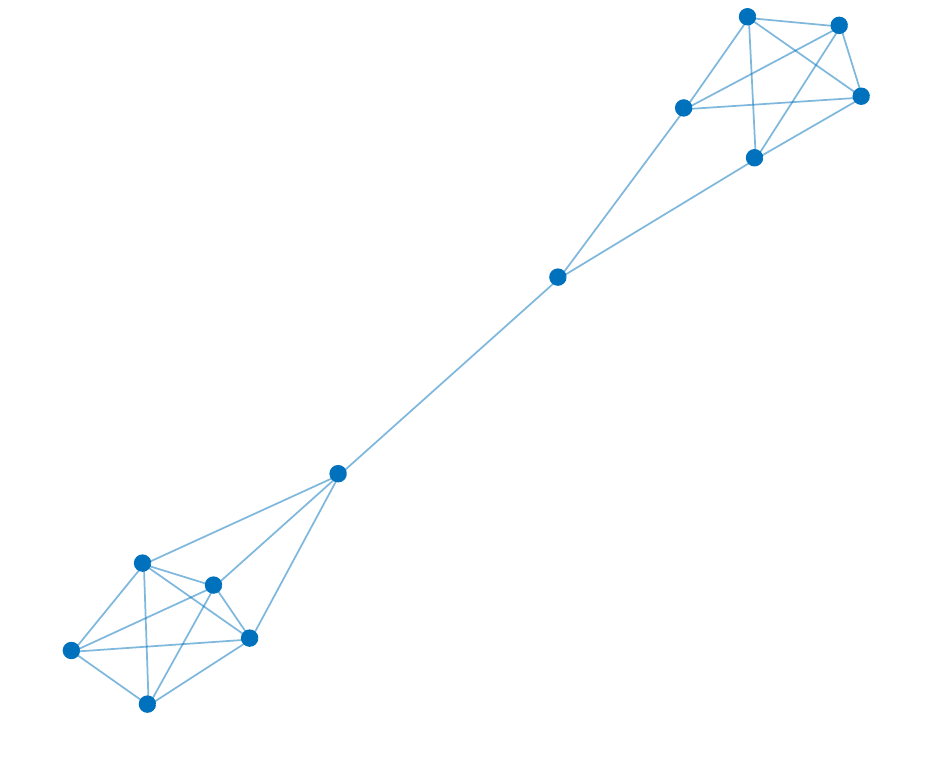}
\includegraphics[trim = 60mm 110mm 55mm 110mm,clip, width=3.8cm, height=4cm]{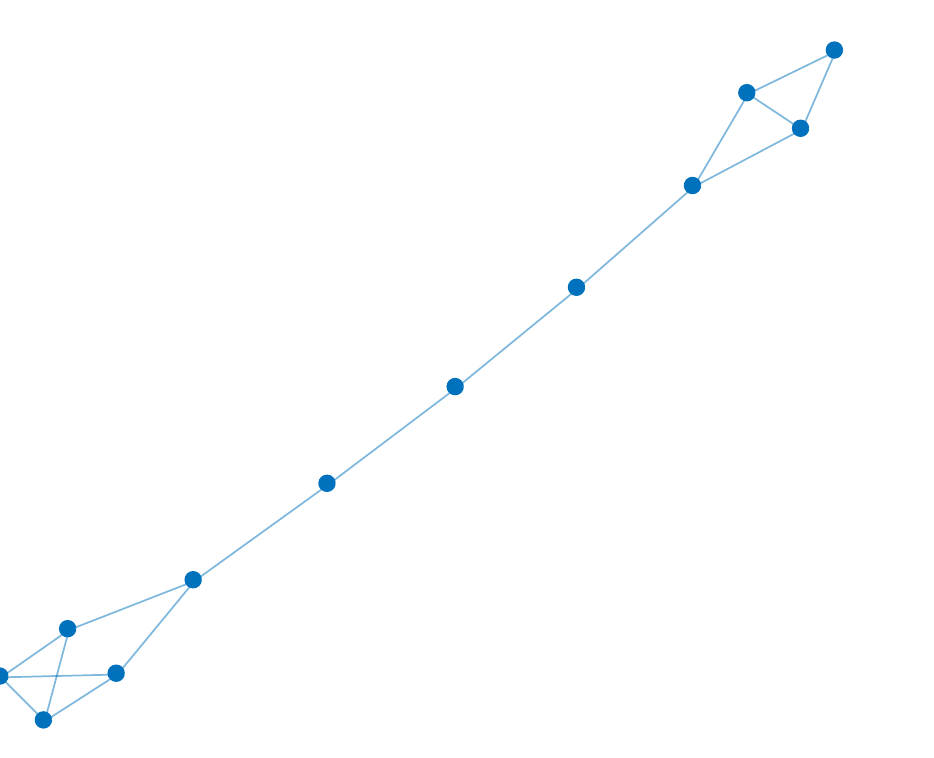}
\includegraphics[angle=5,trim = 58mm 110mm 50mm 110mm,clip, width=3.8cm, height=4cm]{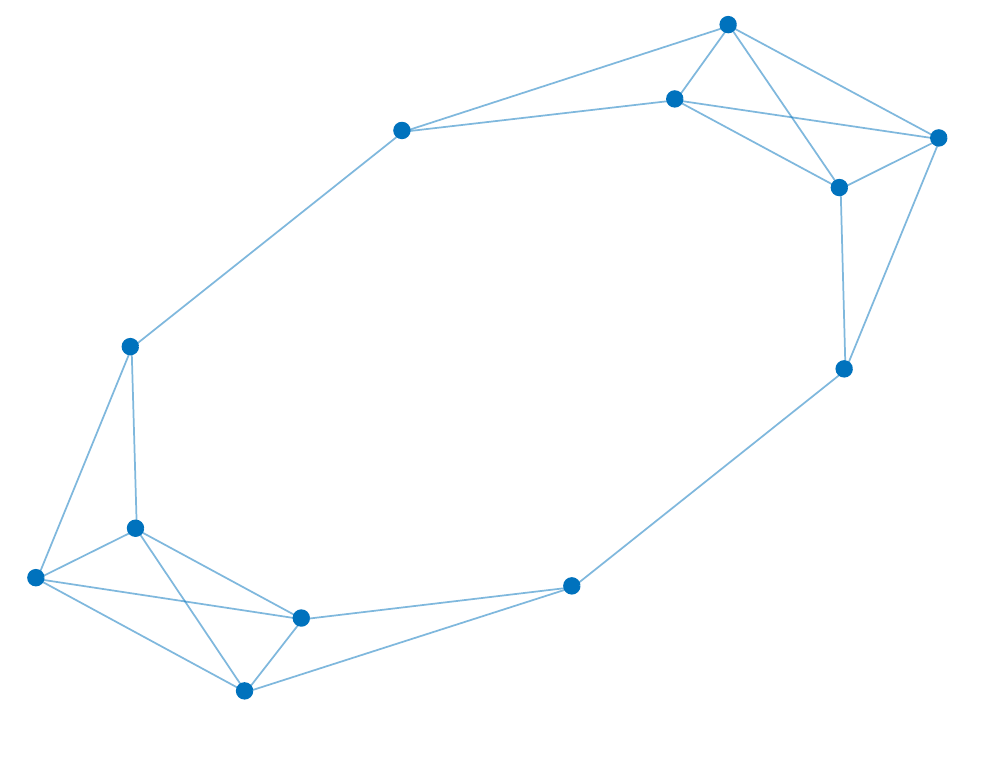}

 (\textbf{a}) \hspace{2.7cm} (\textbf{b})   \hspace{2.7cm} (\textbf{c})  \hspace{2.7cm} (\textbf{d})

\caption{\small{4 examples of 39 non-isomorphic graphs on $N=12$ vertices which are transient amplifiers of death-Birth updating.  (a) $N_{eff}=12.2209$, $\Delta(\mathcal{G})=4$, $\delta(\mathcal{G})=3$, $\bar{k}=3.8333$, $\lambda_2=0.0455$   (b) $N_{eff}=12.0390$, $\Delta(\mathcal{G})=5$, $\delta(\mathcal{G})=3$, $\bar{k}=4.1666$, $\lambda_2=0.0466$   (c) $N_{eff}=12.0666$,  $\Delta(\mathcal{G})=3$, $\delta(\mathcal{G})=2$, $\bar{k}=2.6666$, $\lambda_2=0.0326$   (d) $N_{eff}=12.0028$, $\Delta(\mathcal{G})=4$,  $\delta(\mathcal{G})=3$, $\bar{k}=3.6666$, $\lambda_2=0.1134$.     }  }

\label{fig:graph_12_}
\end{figure}

The next set of results is for $N=12$, see Fig. \ref{fig:graph_12_} and Tab. \ref{tab:12_graphs}.
We obtain 39 non-isomorphic transient amplifier graphs on $N=12$ vertices.  Fig. \ref{fig:graph_12_} shows 4 examples selected by the largest and smallest effective population size ($N_{eff}=12.2209$ for the graph in Fig. \ref{fig:graph_12_}(a) and $N_{eff}=12.0028$ for the graph in Fig. \ref{fig:graph_12_}(d)), and the largest and smallest mean degree ($\bar{k}=4.1666$ for the graph in Fig. \ref{fig:graph_12_}(b) and $\bar{k}=2.6666$ for the graph in Fig. \ref{fig:graph_12_}(c)). Again most of the transient amplifier graphs on $N=12$ vertices have structures where two cliques are connected by a single bridge (also compare to the graphs on $N=11$ vertices, see Fig. \ref{fig:graph_11_}), but there are also 4 graphs where the cliques are connected by two bridges, see an example in  Fig. \ref{fig:graph_12_}(d).
For this graph structure the algebraic connectivity is $\lambda_2=0.1134$, which is substantially higher than the value of the graphs in  Fig. \ref{fig:graph_12_}(a)-(c). In fact, the mean value for the 35 transient amplifier graphs with just one bridge is $\bar{\lambda}_2=0.0405$, while for the 4 graphs with two bridges it is $\bar{\lambda}_2=0.1088$. The values of the algebraic connectivity $\lambda_2$ are generally known to express some structural properties.       Small values of $\lambda_2$ point to large mixing times, bottlenecks, clusters and low conductance~\cite{ban08,ban09,hoff19,wills20}. Additionally, a low algebraic connectivity indicates path-like graphs which can rather easily be divided into disjointed subgraphs by removing edges or vertices. These are exactly the characteristics we see in the graphs in  Fig. \ref{fig:graph_12_}. The graphs which can be disconnected by removing a single edge (one bridge, Fig. \ref{fig:graph_12_}(a)-(c)) have even lower values of $\lambda_2$ than the graphs where two edges must be removed (two bridges, Fig. \ref{fig:graph_12_}(d)).

 \begin{table}[htb]
\centering
\caption{\small{Results of Algorithm 1 (approximative, greedy search) for $N=12$ and $k=(3,4,\ldots,9)$. $\mathcal{L}_k(N)$ is the total number of simple, connected, pairwise nonisomorphic $k$-regular graphs on 11 vertices.   $\mathcal{A}_k(N)$ is the proportion of these regular graphs from which transient amplifiers of death-Birth updating are obtained.  $\#_{tot}$ is the total number of transient amplifiers found,   $\#_{noniso}$ is the number of pairwise non-isomorphic transient amplifiers for each $k$, and $\bar{\bar{k}}$ is the mean degree averaged over these non-isomorphic amplifiers for each $k$. From all graphs with $N=12$ and  $k=(3,4,\ldots,9)$ overall
39 structurally different transient amplifiers are obtained, see Fig.  \ref{fig:graph_12_} for examples. 
}}
\label{tab:12_graphs}
\center
\begin{tabular}{cccccc}

 $k$ & $\mathcal{L}_k$ & $\mathcal{A}_k$  & $\#_{tot}$ & $\#_{noniso}$  & $\bar{\bar{k}}$ \\ 
 \hline 
 3 & 85 & 1 & 1 &  1 & 2.8333 \\
 4 & 1.544 & 226 & 303 & 26  & 3.3141  \\
 5 & 7.848 &  7.473 & 43.974  & 29  & 3.2931 \\
 6 & 7.849& 6.376 & 63.693  &  37 & 3.4595\\
 7 & 1.547 & 935 &11.989 & 21 & 3.6190\\
 8 & 94& 79 & 557 & 4 & 4.0000\\
 9 & 9 & 3 & 23 & 1 & 4.1666\\

\hline

\end{tabular}
\end{table}

Tab. \ref{tab:12_graphs} summarizes further results about identifying 39 non-isomorphic transient amplifier graphs on $N=12$ vertices. As for regular graphs on $N=11$ vertices, compare Tab. \ref{tab:11_graphs}, for all degrees $k=(3,4,\ldots,9)$ instances of graphs can be perturbed into transient amplifiers, but again the number of amplifiers differs substantially.  Particularly for $k=5$ and $k=6$ a substantial number of $k$-regular input graphs have the property to produce amplifiers. For example, of the $39$ structurally different graphs identified, $37$ are associated with  degree $k=6$. Of the remaining $2$, one can be obtained from $k=\{3,4,5\}$ and the other just from $k=\{4,5\}$. Comparing the results for $N=11$ and $N=12$, we see that for a middle range of degrees $k \approx N/2$ the percentage of non-isomorphic (i.e. structurally different) transient amplifiers falls by one order of magnitude. While for $N=11$ and $k=6$ we have $\frac{\#_{noniso}}{\#_{tot}}=\frac{4}{937}=4.3 \cdot 10^{-3}$, for $N=12$ there is $\frac{\#_{noniso}}{\#_{tot}}=\frac{29}{43.974}=6.6 \cdot 10^{-4}$ for $k=5$ and $\frac{\#_{noniso}}{\#_{tot}}=\frac{37}{63.693}=5.8 \cdot 10^{-4}$ for $k=6$. It suggests that for increasing the order from $N=11$ to $N=12$, the algorithmic process constructs roughly  10 times more transient amplifiers which are structurally alike. For several reasons this might appear to be surprising. The total number of pairwise non-isomorphic regular graphs $\mathcal{L}_k$ increases by more than a magnitude, for instance for $k=6$ from $\mathcal{L}_6(11)=266$ for $N=11$ to $\mathcal{L}_6(12)=7.849$ for $N=12$. This would suggest an increased structural diversity of input graphs from which amplifier graphs could emerge. At the same time the number of edges of a regular graph ($kN/2$) increases linearly with $N$, which additionally broadens the possibilities to remove edges and thus to obtain different structures. At least in principle these possibilities should induce divergence in edge removing trajectories and thus potentially enhance structural diversity in transient amplifiers. However, the results show the contrary. The number of structurally different transient amplifiers increases just about linearly. 

There are several possible explanations. A first is that although the search space of possible graph structures increases massively with the graph order $N$ rising,  transient amplifiers of dB updating are most likely subject to severe structural restrictions which to some extent constrain the feasible search space of candidate graphs, see also the discussion about barbell and dumbbell graphs in Sec.~\ref{sec:bar}. Although the algorithmic search discussed in this paper identifies transient amplifiers, they are still very rare.
Another possibility is that the search process guided by the spectral measure algebraic connectivity $\lambda_2$ actually narrows the search to just a subsection of the overall search space. This certainly is plausible and suggests possible directions for future work on algorithmically identifying transient amplifiers by edge removing procedures, see the discussion in Sec.~\ref{sec:dis}.  A third possibility is that the approximative search, and particularly the setting of the filter size $\#_\mathcal{G}$, is responsible for the solely linear increase and a higher number of $\#_\mathcal{G}$ would yield more amplifiers. However, additional experiments with varying   $\#_\mathcal{G}$ showed that there is no clear relationship between increasing $\#_\mathcal{G}$ and performance and a higher filter size sometimes even gets worse results.  
This algorithmic behaviour is a consequences of the iterative process and the fact that small values of $\lambda_2$ point to amplification properties but strictly pursuing only smallest values  is not the best option.

\begin{figure}[htb]
\centering
\includegraphics[trim = 25mm 90mm 40mm 80mm,clip, width=6.6cm, height=5.5cm]{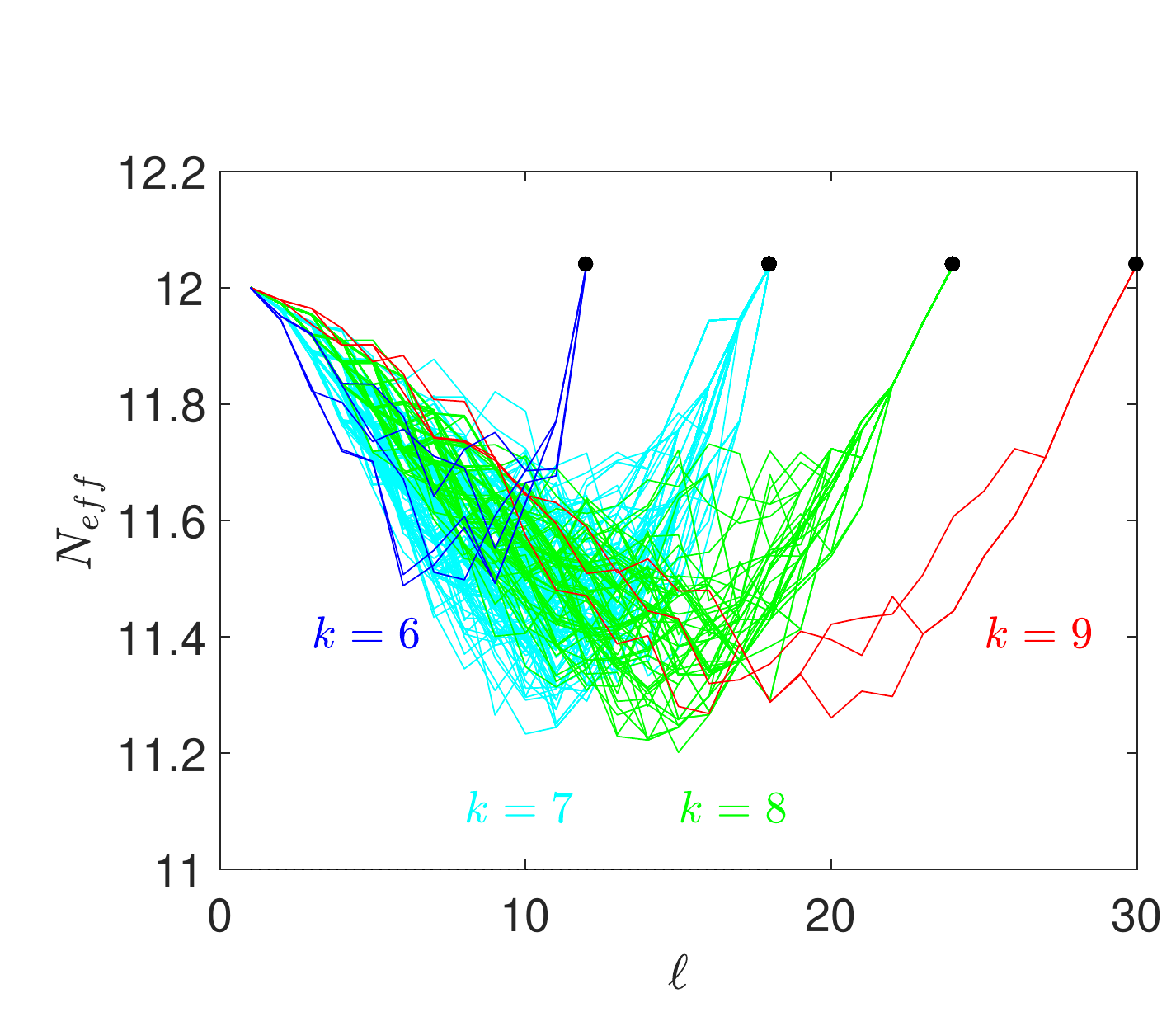}
\includegraphics[trim = 25mm 90mm 40mm 80mm,clip, width=6.6cm, height=5.5cm]{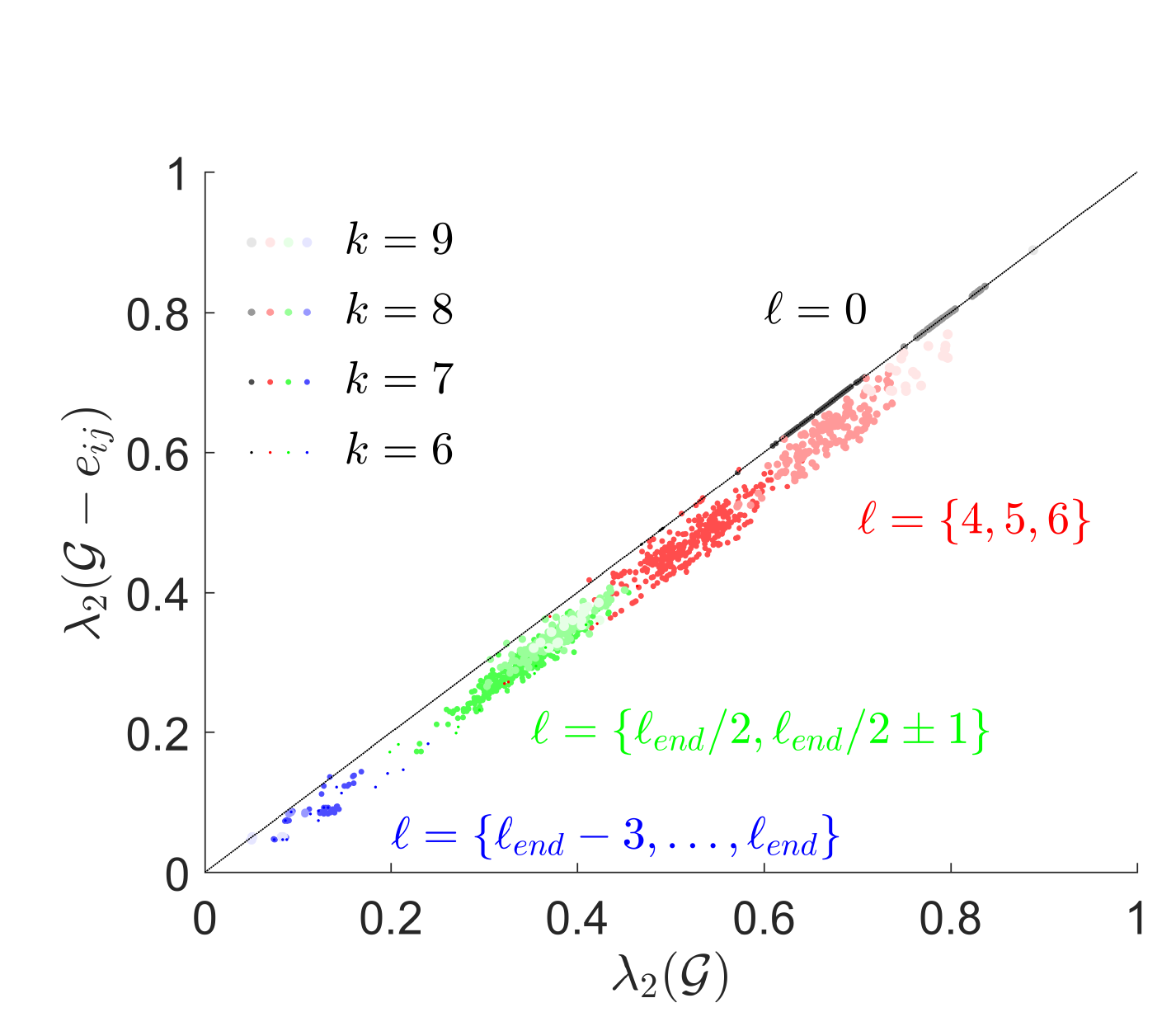}

 (\textbf{a}) \hspace{4.7cm} (\textbf{b})

\caption{\small{Behaviour of the approximative search of Algorithm 1 for input graphs with order $N=12$ yielding the amplifier with $N_{eff}=12.0390$ depicted in Fig. \ref{fig:graph_12_} and all degrees $k=\{6,7,8,9\}$. (a) The effective population size $N_{eff}$ over edge removal repetitions $\ell$. (b) Spectral shift in algebraic connectivity $\lambda_2$ induced by removing an edge $e_{ij}$ depicted as scatter plot of $\lambda_2(\mathcal{G}-e_{ij})$ over $\lambda_2(\mathcal{G})$.  }  }
\label{fig:graph_12_X_neff}
\end{figure}

Also Fig. \ref{fig:graph_12_X_neff} illustrates aspects of the algorithmic process. Fig. \ref{fig:graph_12_X_neff}(a) shows
the effective population size $N_{eff}$ over edge removal repetitions $\ell$ for the amplifier depicted in Fig. \ref{fig:graph_12_}(b). This amplifier can be obtained by taking input  graphs with 4 different degrees $k$, which is the maximal range of input degrees obtained in the experiments with $k$-regular graphs with $N=12$ and $\#_\mathcal{G}=500$. By taking input graphs with any degree from $k=\{6,7,8,9\}$, we see that for each input degree the final value of $N_{eff}=12.0390$ is obtained after a specific number of edge removals $\ell=\ell_{end}$ with $k=6$ needing the smallest $\ell_{end}$ and $k=9$ needing the largest.    The curves for each $k$ resemble each other with setting out at $N_{eff}=12$ and experiencing a prolonged decline to values $N_{eff}<12$. Afterwards, they spread out to a larger range of $N_{eff}$ reflecting structural diversity, 
before the different paths of edge removal trajectories sharply rise and merge before ending at $N_{eff}=12.0390>12$.  The actual amount of edge removals required for different input degrees $k$ mainly influences the length of the curves, but not their shape.  

These shape similarities  point at underlying similarities in the way edges are removed from the input graph. They also becomes noticeable in the graph spectra, see 
Fig. \ref{fig:graph_12_X_neff}(b) which shows the spectral shift over edge removals. The spectral shift is depicted as a scatter plot of the algebraic connectivity $\lambda_2(\mathcal{G}-e_{ij})$ over $\lambda_2(\mathcal{G})$ for the candidate graphs $\mathcal{G}$ before and after an edge $e_{ij}$ is deleted.   The different colors of the dots indicate different edge removal repetitions $\ell$. The different sizes and lightness of the dots label different input degrees $k$. To compensate for the different $\ell_{end}$ for each $k=\{6,7,8,9\}$, the plot gives the spectral shift for different phases in the edge removing process. The plot in Fig. \ref{fig:graph_12_X_neff}(b) shows the spectral shift from the regular input graphs  experiencing their first edge removal ($\ell=0$, black dots), an initial phase  ($\ell=\{4,5,6\}$, red dots), an intermediate phase ($\ell=\{\ell_{end}/2,\ell_{end}/2 \pm1\}$, green dots) and a final phase ($\ell=\{\ell_{end}-3,\ldots,\ell_{end}\}$, blue dots). We particularly see that for the initial edge removal ($\ell=0$, depicted as black dots) almost all value lie on the diagonal
$\lambda_2(\mathcal{G}-e_{ij})=\lambda_2(\mathcal{G})$. In other words, there is hardly any spectral shift.  This is interesting as the algorithmic search sets $\#_{lim} \gg k$, and with just $k$ possibilities to remove a first edge from a $k$-regular graph, all graphs resulting from the first edge removal are kept as candidate graphs. The filter of the approximative search has no influence on the first step. Therefore, all first edge removals on the trajectory to a transient amplifiers cause no or just a tiny spectral shift. For the number of edge removals $\ell$ increasing this ceases to be the case, although there are still instances of a small or zero spectral shift. There are even rare instances where  $\lambda_2(\mathcal{G}-e_{ij})>\lambda_2(\mathcal{G})$, which is known as Braess's paradox~\cite{eld17}.  But mostly we have  spectral shifts which decrease the algebraic connectivity $\lambda_2$. The majority of values are  along a band below 
the diagonal. The band becomes slightly smaller for $\ell$ getting larger which indicates that the magnitude of the spectral shift lowers. With respect to the different degrees $k$ of the input graphs we see that only for the initial phase  ($\ell=\{4,5,6\}$, red dots) clearly separable clouds of dots  occur while for rising $\ell$ the values are more overlapping. 
These results support the notion that the edge removing trajectories resemble each other in shape even if their duration differs. 
To conclude the spectral shift along the edge removing process leading to the transient amplifier depicted in Fig. \ref{fig:graph_12_}(b) with $N_{eff}=12.0390$ follows some characteristic patterns. These patterns can similarly be found for the other amplifiers with $N=12$ and analogously for other graph orders $N$ as well. The notion of the spectral shift of $\lambda_2$ following characteristic patterns appears to be rather self-evident, giving the fact that the approximative search explicitly selects for graphs with  small $\lambda_2$. We next generalize the notion of spectral shifts in three directions, thus studying the spectral dynamics of edge removals. First, we now consider
all edge removal repetitions $\ell$ and not only some selected phases, which is expressed as a sum of edge removal $\sum e_{ij}$.
Second, all trajectories leading to transient amplifiers are recorded and not just those leading to selected amplifiers, and third, we not only account for the algebraic connectivity $\lambda_2$ but for the whole spectrum $\Lambda_\mathcal{G}=\{\lambda_i(\mathcal{G}-\sum e_{ij})\}$.

\begin{figure}[htb]
\centering
\includegraphics[trim = 5mm 1mm 6mm 20mm,clip, width=6.6cm, height=5.5cm]{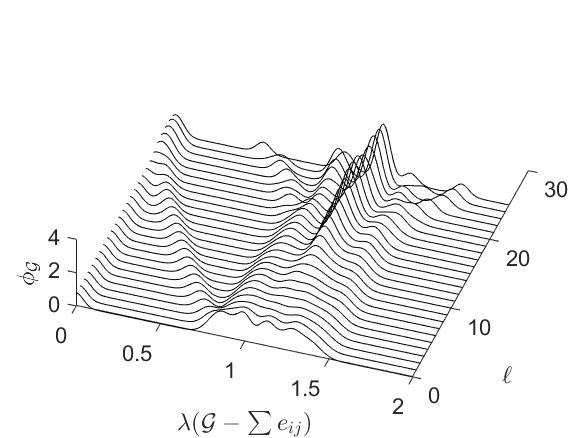}
\includegraphics[trim = 5mm 1mm 6mm 20mm,clip, width=6.6cm, height=5.5cm]{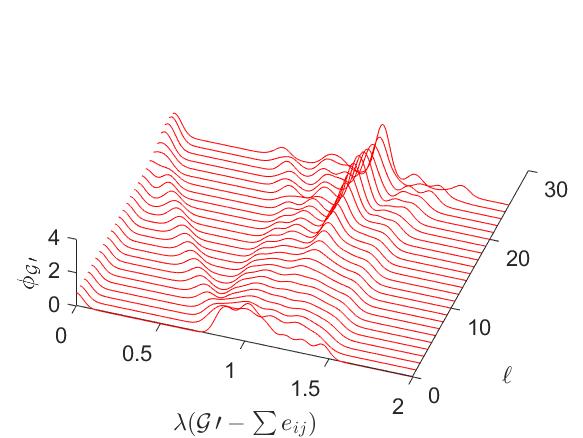}

 (\textbf{a}) \hspace{4.7cm} (\textbf{b})  
 
 \includegraphics[trim = 5mm 1mm 6mm 20mm,clip, width=6.6cm, height=5.5cm]{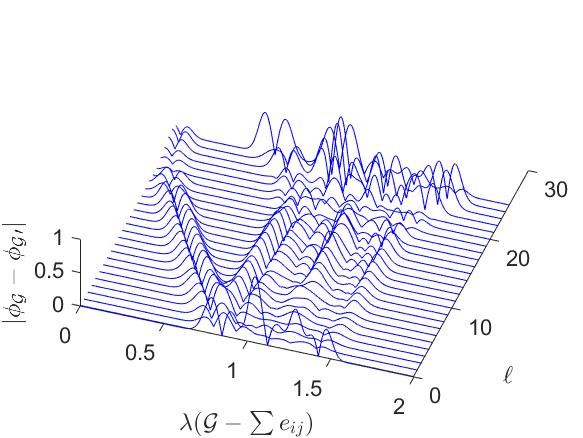}
  \includegraphics[trim = 25mm 90mm 40mm 80mm,clip, width=6.6cm, height=5.5cm]{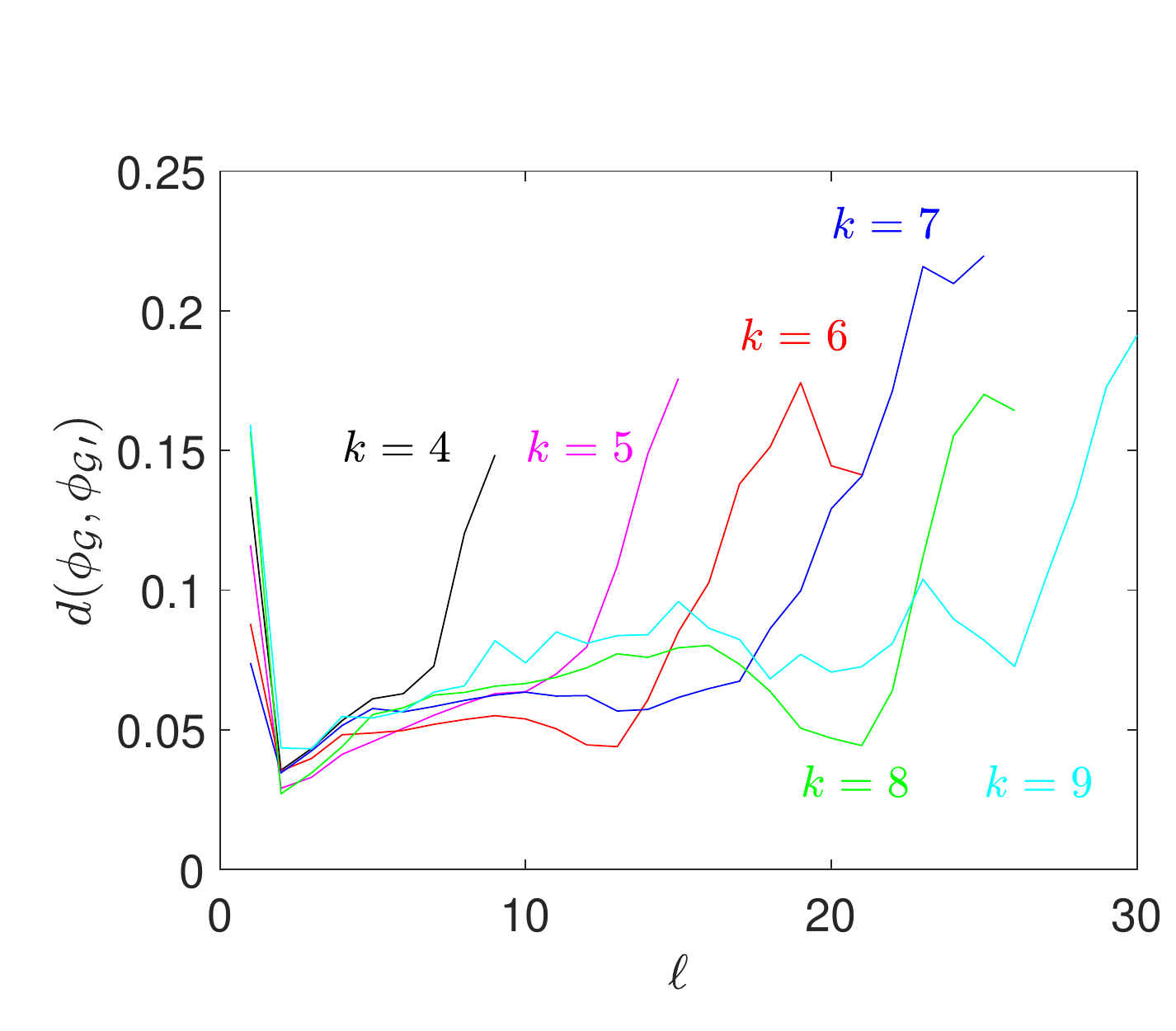}

  (\textbf{c}) \hspace{4.7cm} (\textbf{d})  

\caption{\small{Spectral dynamics of the approximative search of Algorithm 1 for order $N=12$, degree $k=8$. (a) The spectral density $\phi_\mathcal{G}$ describing graph evolutions leading to transient amplifiers. (b) The spectral density $\phi_\mathcal{G}\prime$ describing graph evolutions not leading to transient amplifiers.  (c) The quantity $|\phi_\mathcal{G}-\phi_\mathcal{G}\prime|$  describing the difference. (d) The spectral distance $d(\mathcal{G},\mathcal{G}\prime)$, Eq.~\eqref{eq:distance} for any $k=\{3,4,\ldots,9\}$ for which input graphs on $N=12$ vertices produce transient amplifiers. See Figs. \ref{fig:graph_12_xx_dense}-\ref{fig:graph_12_xxxx_dense} for $\phi_\mathcal{G}$, $\phi_\mathcal{G}\prime$ and $|\phi_\mathcal{G}-\phi_\mathcal{G}\prime|$  of the remaining $k$. }  }
\label{fig:graph_12_X_dense}
\end{figure}

Thus, with Fig. \ref{fig:graph_12_X_dense}  we take a broader and more global look at the algorithmic process and examine the dynamics of the whole Laplacian spectra over edge removals from  regular input graphs. 
We consider the spectral density $\phi_\mathcal{G}$ as defined by Eq. \eqref{eq:density} which convolves all eigenvalues $\lambda_i$, $i=1,2,\ldots,N$, with a Gaussian kernel. Thus, the spectral density $\phi_\mathcal{G}$  can be seen as a smoothed curve over the eigenvalue distribution. 
Furthermore, $\phi_\mathcal{G}$ averages  for each $\ell$ over the graph set yielding all transient amplifiers. For instance, for $\ell=0$ we average over all input graphs  which finally lead to a transient amplifier. Note that for $\ell=0$ the number of graphs in the graph set is explicitly specified by $\mathcal{A}_k$, see Tabs.~\ref{tab:11_graphs} or~\ref{tab:12_graphs}. For $\ell=1$, the graph set comprises of all graphs after the first edge removal which subsequently yield a transient amplifier, and so on for $\ell>1$.      
Fig. \ref{fig:graph_12_X_dense}(a) shows $\phi_\mathcal{G}$ for input graphs with $k=8$ for all trajectories leading to transient amplifiers. Thus, for $\ell=0$ the graph set consist of $\mathcal{A}_8=79$ out of the $\mathcal{L}_8=94$ input graphs. The results for the other degrees $k$ are depicted in the Appendix, see Fig. \ref{fig:graph_12_xx_dense}. 

There are two interesting features in the spectral dynamics shown in Fig. \ref{fig:graph_12_X_dense}(a). The first is that the algebraic connectivity $\lambda_2$ getting progressively smaller and smaller can be seen as a kind of single travelling peak setting out at $\lambda(\mathcal{G}-\sum e_{ij}) \approx 0.9$ for $\ell=0$ and ending at    $\lambda(\mathcal{G}-\sum e_{ij}) \approx 0$ for  $\ell=25$. Over all graphs  the decrease in $\lambda_2$ caused by repeated edge removals is narrowly bounded and almost continuous. This is in contrast to random edge removals which do not exhibit such a behavior. The second important feature is a kind of standing peak at  $\lambda(\mathcal{G}-\sum e_{ij}) \approx 1.1$ which becomes prominent at $\ell \approx 10$ and continuously increases thereafter for $\ell \leq 25$. Such an increase in $\phi_\mathcal{G}$ indicates a multiplicity of eigenvalues which additionally points at doubling of motifs in the graph~\cite{meha15}. These results can be contrasted with the spectral density of graph evolutions which do not lead to transient amplifiers.   Fig. \ref{fig:graph_12_X_dense}(b)
shows the spectral density $\phi_\mathcal{G}\prime$  of the graph set $\mathcal{G\prime}$ over $\ell$ and $\lambda(\mathcal{G}\prime-\sum e_{ij})$. The graph set $\mathcal{G}\prime$ consists of all candidate graphs produced on the edge removing trajectory which are not finally leading to a transient amplifier. Thus, $\mathcal{G}\prime$ can be seen as complementary to $\mathcal{G}$.  For instance, for $\ell=0$ is comprises of remaining 
 $\mathcal{L}_k-\mathcal{A}_k$ input graphs, and so on. Comparing the spectral density $\phi_\mathcal{G}$ of the graph set leading to transient amplifiers, Fig. \ref{fig:graph_12_X_dense}(a), with $\phi_\mathcal{G}\prime$ not leading to transient amplifiers, general similarities can be noted.  Also in Fig. \ref{fig:graph_12_X_dense}(b) we see the travelling peak indicating decreasing $\lambda_2$ over $\ell$ and the standing peak indicating increasing eigenvalue multiplicity. However, the difference  $|\phi_\mathcal{G}-\phi_\mathcal{G}\prime|$, see Fig. \ref{fig:graph_12_X_dense}(c), also reveals significant differences in the graph sets. 
 
 A first is the difference  $|\phi_\mathcal{G}-\phi_\mathcal{G}\prime|$ showing that the travelling peak has a kind of notch which makes the peak appear split and twofold. The geometrical interpretation is that the travelling peak of $\phi_\mathcal{G}$ is more narrow than the one of $\phi_\mathcal{G}\prime$. In other words, the  range of progressively decreasing values of $\lambda_2$   is smaller for graphs evolving towards transient amplifiers than for graphs which do not lead to amplifiers. A second is that while in the initial and intermediate phase of edge removals the differences remain within a certain range of $\lambda(\mathcal{G}-\sum e_{ij})$, they spread in the final phase, particularly for $\ell>20$. This means that in the final phase of edge removals the $\lambda_i$ for graphs evolving towards transient amplifiers are more dispersed than those for graphs not doing so.
 This most likely indicates that on average graphs evolving towards transient amplifiers  build up characteristic structural features which entail certain values and multiplicities in $\lambda_i$. These features become visible in the spectral density $\phi_\mathcal{G}$.  Candidate graphs which do not evolve towards amplification properties do not  specifically possess these features.  In the spectral density $\phi_\mathcal{G}\prime$ the resulting variety of structural features cancels off, leading to differences as compared to $\phi_\mathcal{G}$. For the other degrees $k$, we find similar characteristics, see Figs.  \ref{fig:graph_12_xx_dense}-\ref{fig:graph_12_xxxx_dense} in the Appendix. An exception is $k=3$ for which there is only a single edge removal from input graph to transient amplifier and not the same characteristic curves. However, to some extend for $k=4$ and clearly for $k>4$ the features described become visible. 
 
 For an overall comparison between the spectral densities, the spectral distance $d(\mathcal{G},\mathcal{G}\prime)$, Eq.~\eqref{eq:distance} can be used. Fig. \ref{fig:graph_12_X_dense}(d) shows this quantity for any $k=\{4,\ldots,9\}$ for which input graphs on $N=12$ vertices produce transient amplifiers. The degree $k=3$ is omitted as there is only a single edge removal from input graph to transient amplifier and thus no meaningful comparison over $\ell$.  We see that although the number of required edge removals varies for different $k$, the spectral distance starts at large values for $\ell=0$, before dropping for a certain amount of time but increasing again for the graph evolution about to finish towards transient amplifiers. In other word,    at the beginning and at the end of the edge removal process, the graph set connected with amplifiers and the graph set not connected with amplifiers have clear differences in their normalized Laplacian spectra.  

 \subsection{Regular input graphs on $N=\{14,20,26\}$ vertices} \label{sec:14_20_26}
\begin{figure}[htb]

\includegraphics[angle=5,trim = 58mm 110mm 55mm 110mm,clip, width=3.8cm, height=3.0cm]{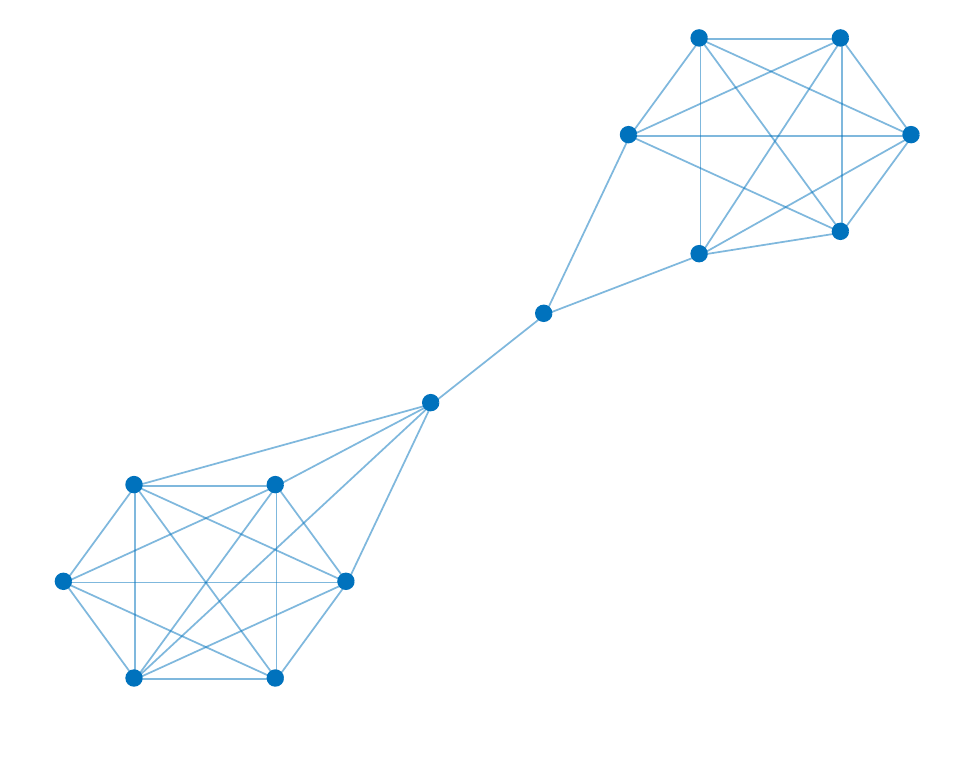}
\includegraphics[angle=5,trim = 60mm 110mm 55mm 110mm,clip, width=3.8cm, height=3.0cm]{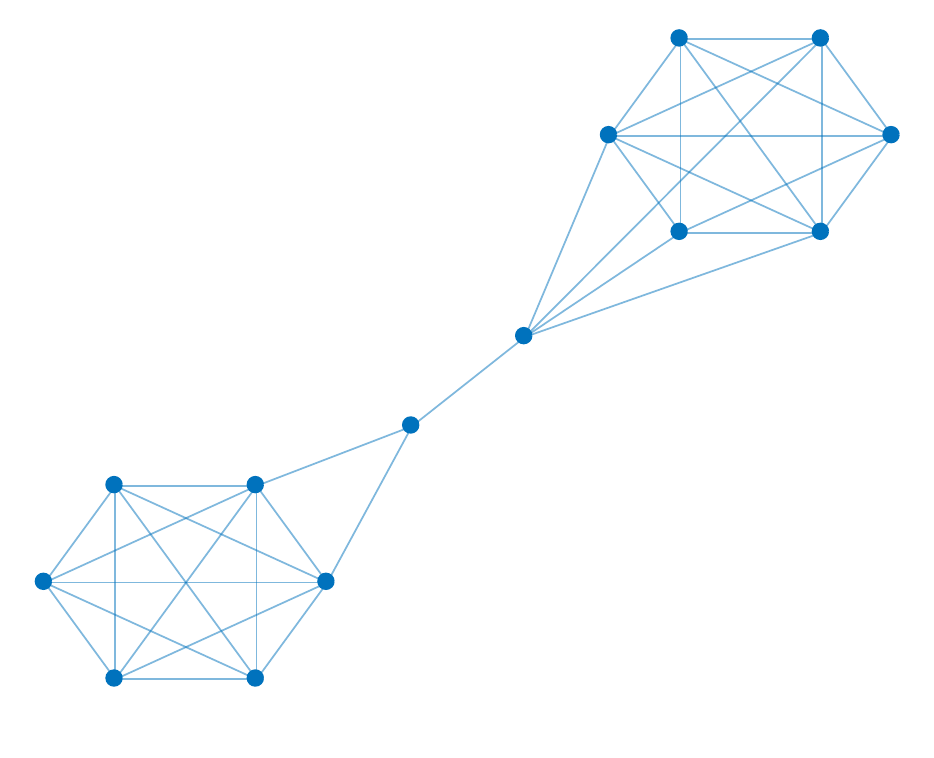}
\includegraphics[angle=5,trim = 60mm 110mm 55mm 110mm,clip, width=3.8cm, height=3.0cm]{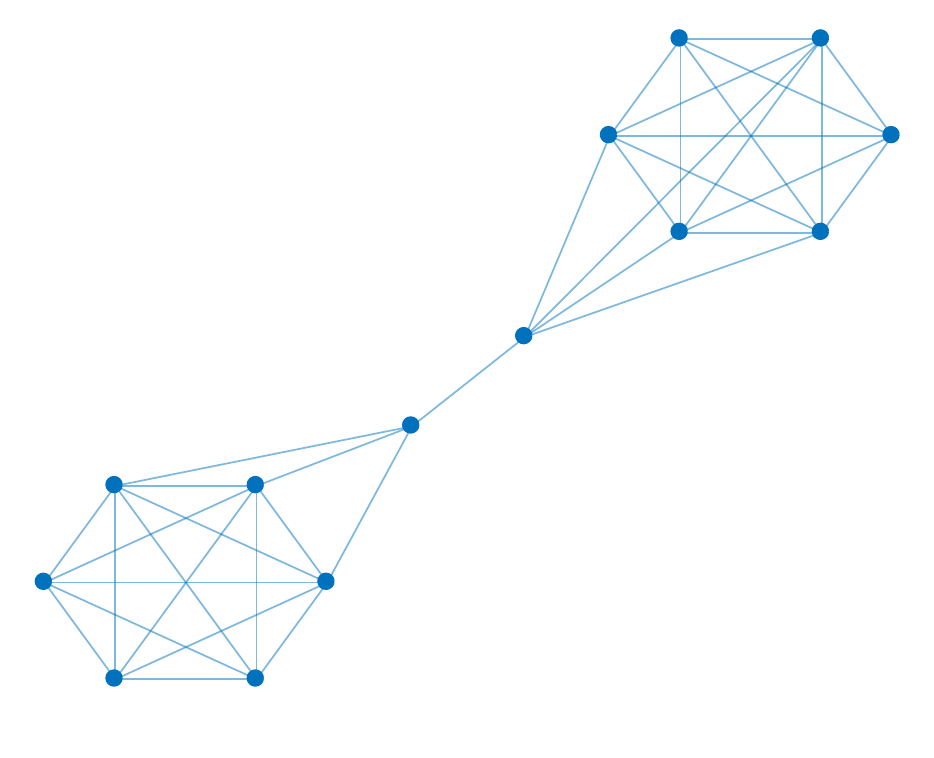}
\includegraphics[angle=5,trim = 58mm 110mm 50mm 110mm,clip, width=3.8cm, height=3.0cm]{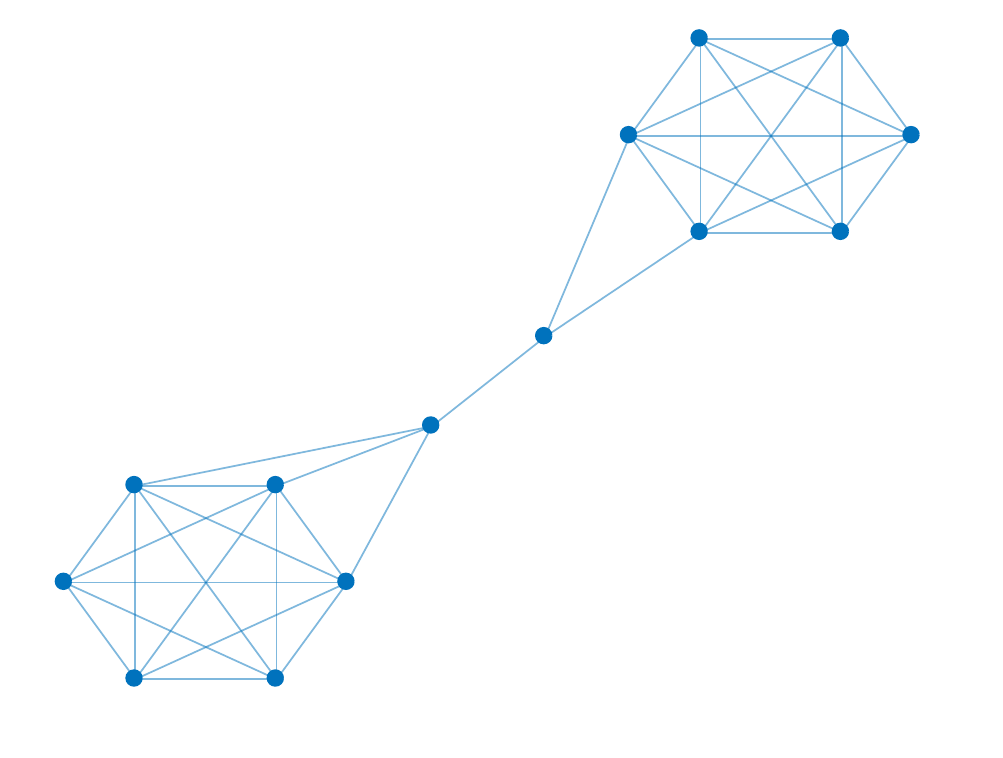}

 (\textbf{a}) \hspace{2.7cm} (\textbf{b})   \hspace{2.7cm} (\textbf{c})  \hspace{2.7cm} (\textbf{d})

\includegraphics[angle=5,trim = 58mm 110mm 55mm 105mm,clip, width=3.8cm, height=3.0cm]{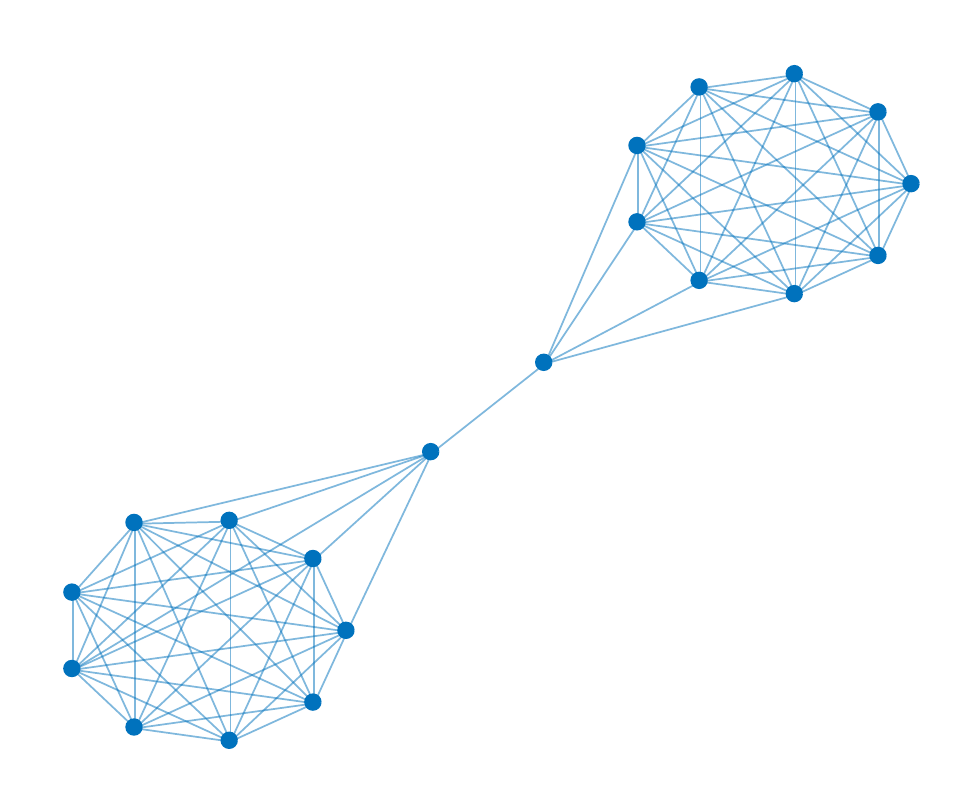}
\includegraphics[angle=5,trim = 60mm 110mm 55mm 105mm,clip, width=3.8cm, height=3.0cm]{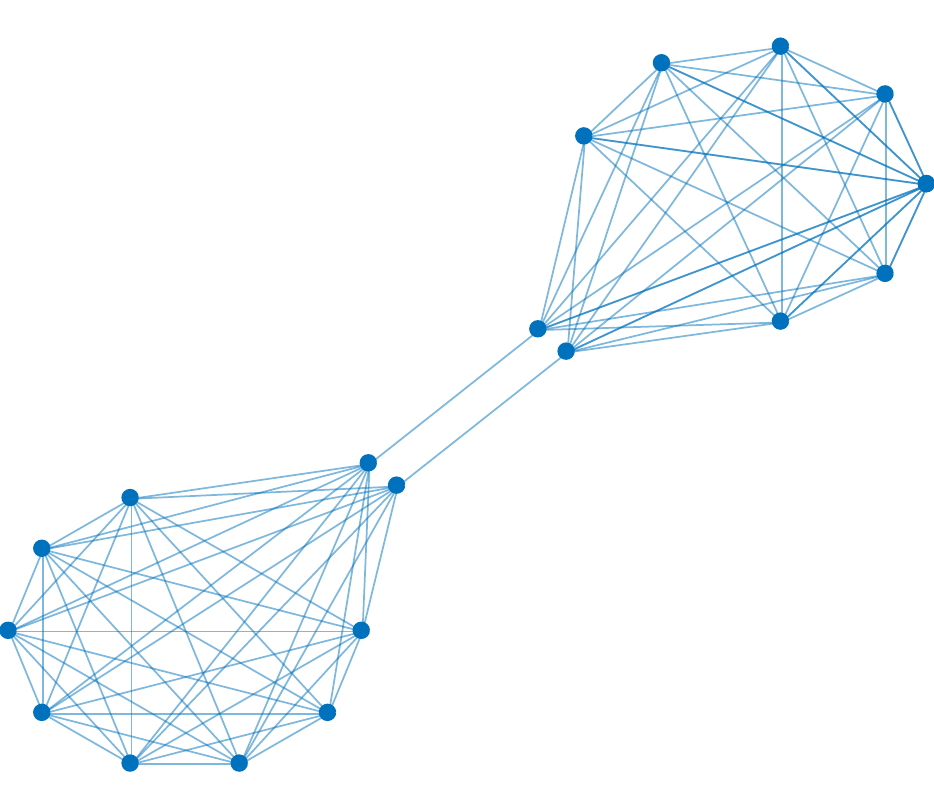}
\includegraphics[angle=5,trim = 60mm 110mm 55mm 105mm,clip, width=3.8cm, height=3.0cm]{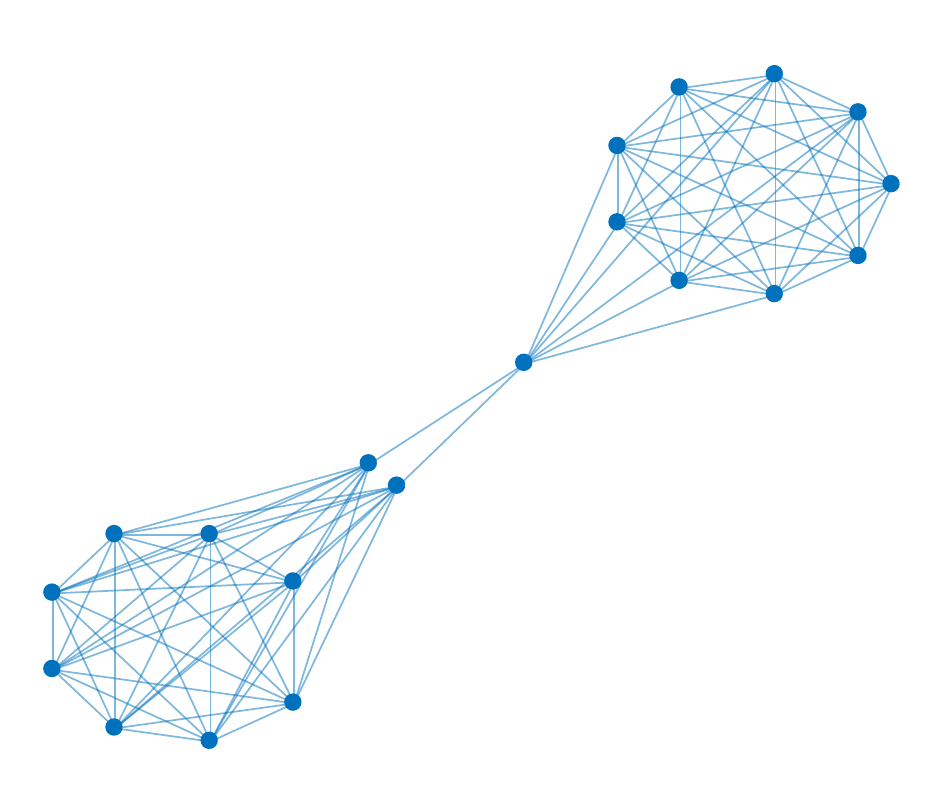}
\includegraphics[angle=5,trim = 58mm 110mm 50mm 105mm,clip, width=3.8cm, height=3.0cm]{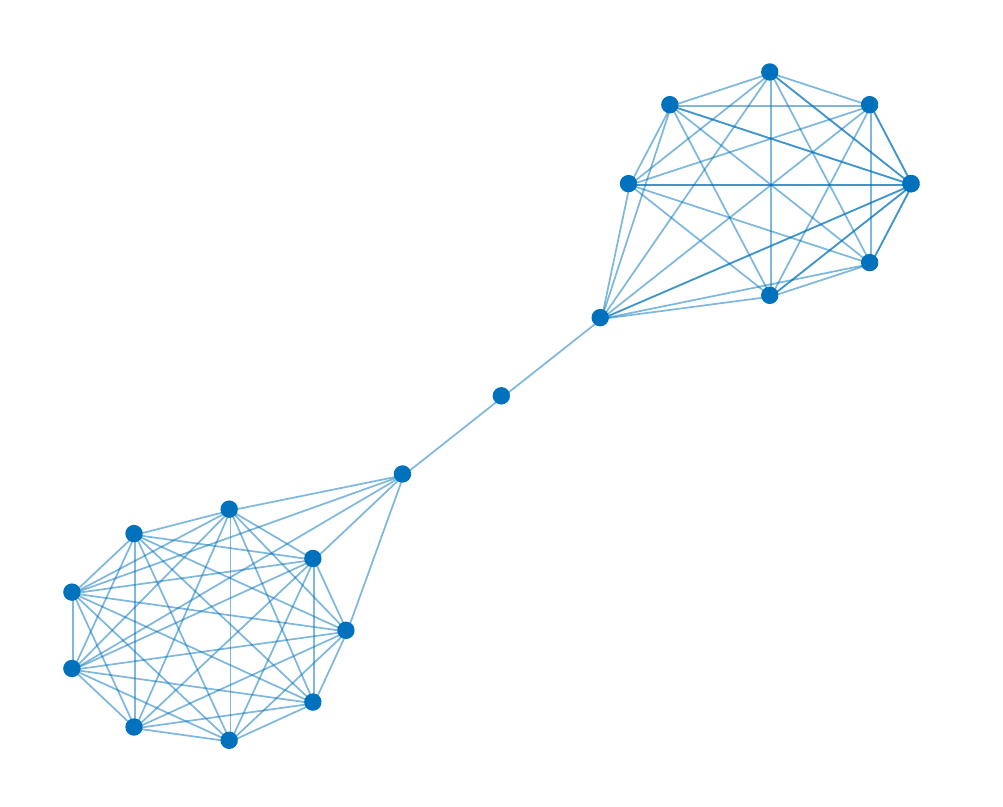}

 (\textbf{e}) \hspace{2.7cm} (\textbf{f})   \hspace{2.7cm} (\textbf{g})  \hspace{2.7cm} (\textbf{h})
 
 \includegraphics[angle=5,trim = 58mm 110mm 55mm 105mm,clip, width=3.8cm, height=3.0cm]{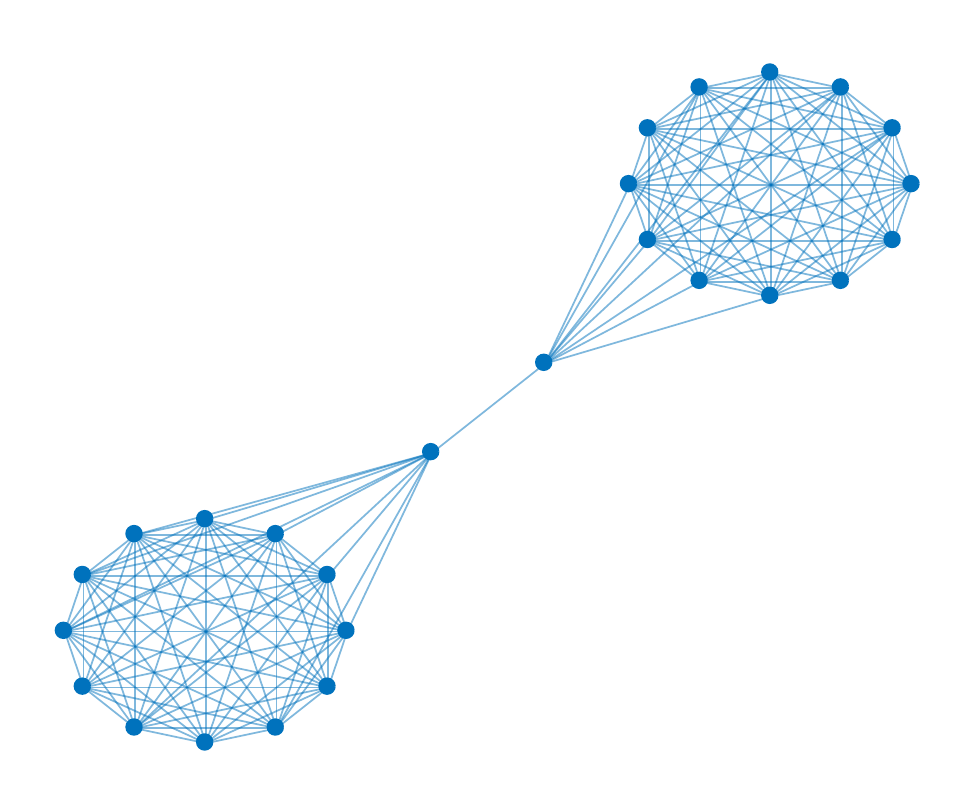}
\includegraphics[angle=5,trim = 60mm 110mm 55mm 105mm,clip, width=3.8cm, height=3.0cm]{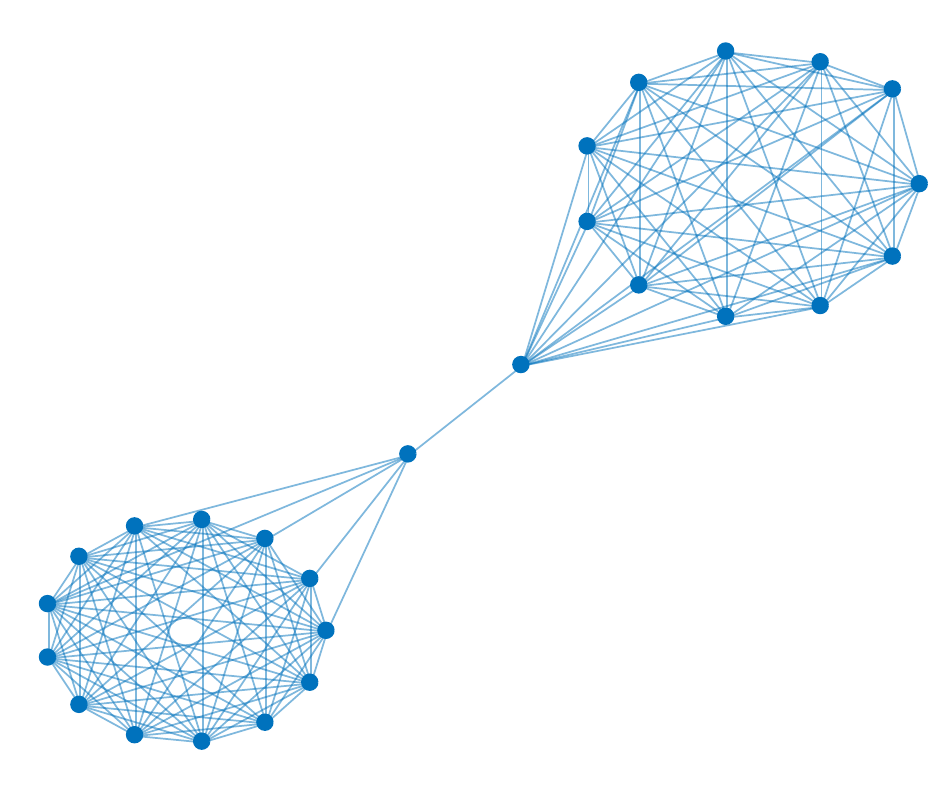}
\includegraphics[angle=5,trim = 60mm 110mm 55mm 105mm,clip, width=3.8cm, height=3.0cm]{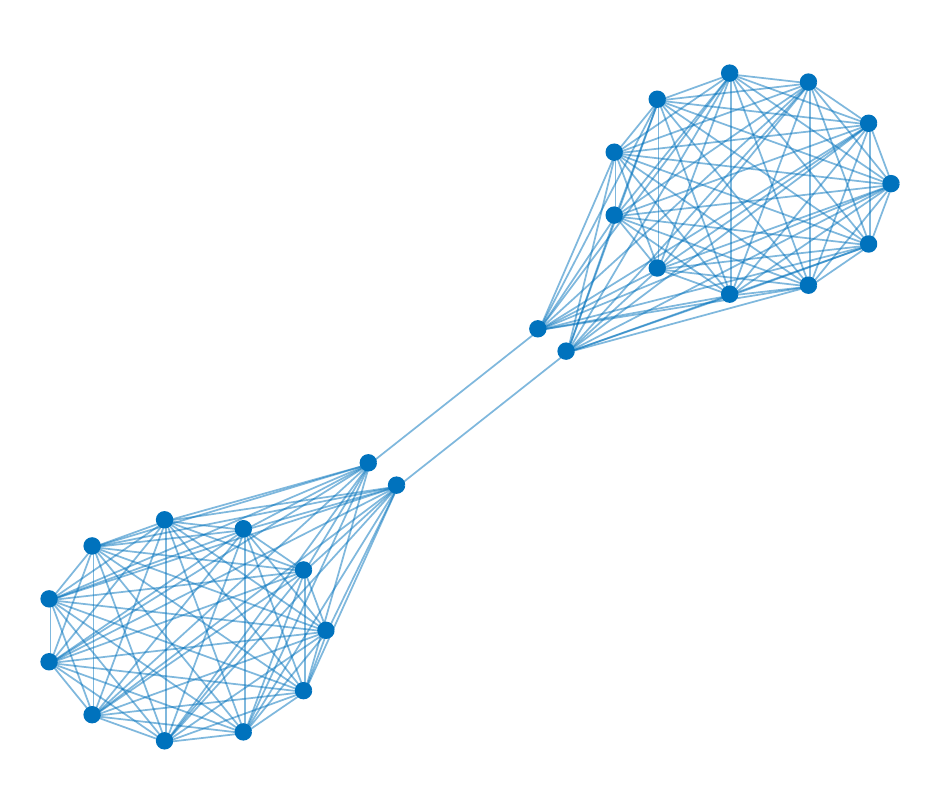}
\includegraphics[angle=5,trim = 58mm 110mm 50mm 105mm,clip, width=3.8cm, height=3.0cm]{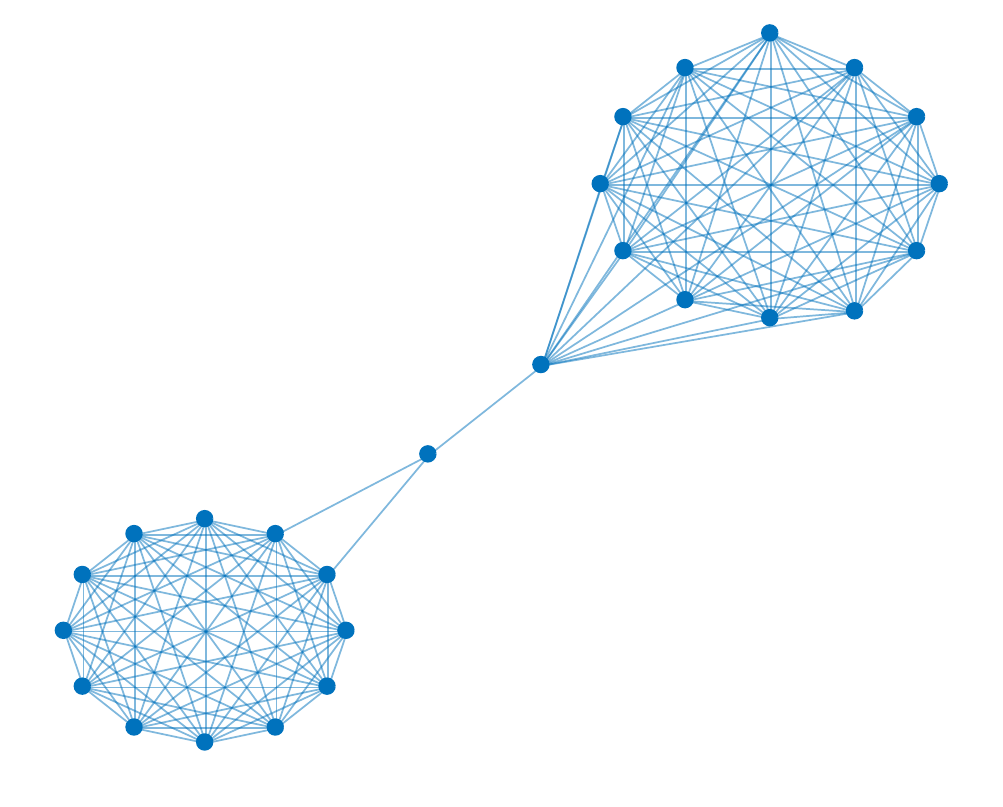}

 (\textbf{i}) \hspace{2.7cm} (\textbf{j})   \hspace{2.7cm} (\textbf{k})  \hspace{2.7cm} (\textbf{l})

\caption{\small{Examples of non-isomorphic graphs on $N$ vertices which are transient amplifiers of death-Birth updating. (a)-(d): $N=14$, which have maximum degree $\Delta(\mathcal{G})=6$. (e)-(h): $N=20$, which have maximum degree $\Delta(\mathcal{G})=9$. (i)-(l): $N=26$, which have maximum degree $\Delta(\mathcal{G})=12$ except (j), which has $\Delta(\mathcal{G})=13$. The values of the effective population size $N_{eff}$, the minimum degree $\delta(\mathcal{G})$, the mean degree $\bar{k}$ and the algebraic connectivity $\lambda_2$ are   
}  }
\label{fig:graph_14_26}
\small
\begin{tabular}{ccccccccc} \hline
 & (a) & (b) & (c) & (d)  & (e) & (f) & (g) & (h)  \\

$N_{eff}$  & $14.0986$ &$14.0073$&$14.0247$&$14.0533$&$20.1978$&$20.0013$ &$20.0021$&$20.0831$ \\
$\delta(\mathcal{G})$ & 3 & 3 & 4 & 3 & 5 & 8 & 8 & 2 \\
$\bar{k}$ & 5.1429 & 5.1429 & 5.4286 & 5.1429& 8.2000 & 8.9000 & 8.8000 & 7.9000\\
$\lambda_2$ & 0.0331 & 0.0329 & 0.0350 & 0.0313 & 0.0172& 0.0383 & 0.0330 & 0.0113\\
 \hline 
& (i) & (j) & (k)& (l) \\ 

$N_{eff}$ &$26.1747$&$26.0019$&
$26.0092$& $26.0170$ \\
$\delta(\mathcal{G})$  & 9 & 6 & 11 & 3\\
$\bar{k}$  & 11.5385 & 11.3077 & 11.9231& 11.2308\\
$\lambda_2$  & 0.0110 & 0.0108& 0.0227 & 0.0086\\

\hline

\end{tabular}
\end{figure}

As shown in the previous section, for regular graphs on $N=11$ and $N=12$ vertices it has been possible with the available computational resources to treat inputs from all structurally different graphs with all existing degrees. For $N\geq14$ this has not been feasible due to the massive growth of the number of potential input graphs $\mathcal{L}_k(N)$~\cite{mer99,rich21,reg_graph}. For nevertheless studying a closed set of graphs covering a complete structural range, we next consider inputs  on $N=\{14,20,26\}$  vertices with degree $k=N-3$. Thus, we have a complete structural variety for the given $N$ and $k$, while the total number of input graphs remains computationally manageable, see Tab. \ref{tab:14_20_graphs} and Fig. \ref{fig:graph_14_26}.

 \begin{table}[htb]
\centering
\caption{\small{Results of Algorithm 1 (approximative, greedy search) for $N=\{14,20,26\}$ and $k=(11,17,23)$. $\mathcal{L}_k(N)$ is the total number of simple, connected, pairwise nonisomorphic $k$-regular graphs on 11 vertices.   $\mathcal{A}_k(N)$ is the proportion of these regular graphs from which transient amplifiers of death-Birth updating are obtained.  $\#_{tot}$ is the total number of transient amplifiers found,   $\#_{noniso}$ is the number of pairwise non-isomorphic transient amplifiers for each $k$, and $\bar{\bar{k}}$ is the mean degree averaged over these non-isomorphic amplifiers for each $k$. 
}}
\label{tab:14_20_graphs}
\center
\begin{tabular}{ccccccc}

  $N$ &
 $k$ & $\mathcal{L}_k$ & $\mathcal{A}_k$  & $\#_{tot}$ & $\#_{noniso}$ & $\bar{\bar{k}}$  \\ 
 \hline 
 14 & 11 & 13 & 4 &  1.597 & 9 & 5.2540 \\
 20 & 17 & 49 & 2 &  4.212 & 43 & 8.4163 \\
 26 & 23 &  130 &   13& 58.355 & 55 & 11.6154  \\

\hline

\end{tabular}
\end{table}

\clearpage

\begin{figure}[htb]
\centering

\includegraphics[trim = 25mm 90mm 40mm 80mm,clip, width=6.6cm, height=5.1cm]{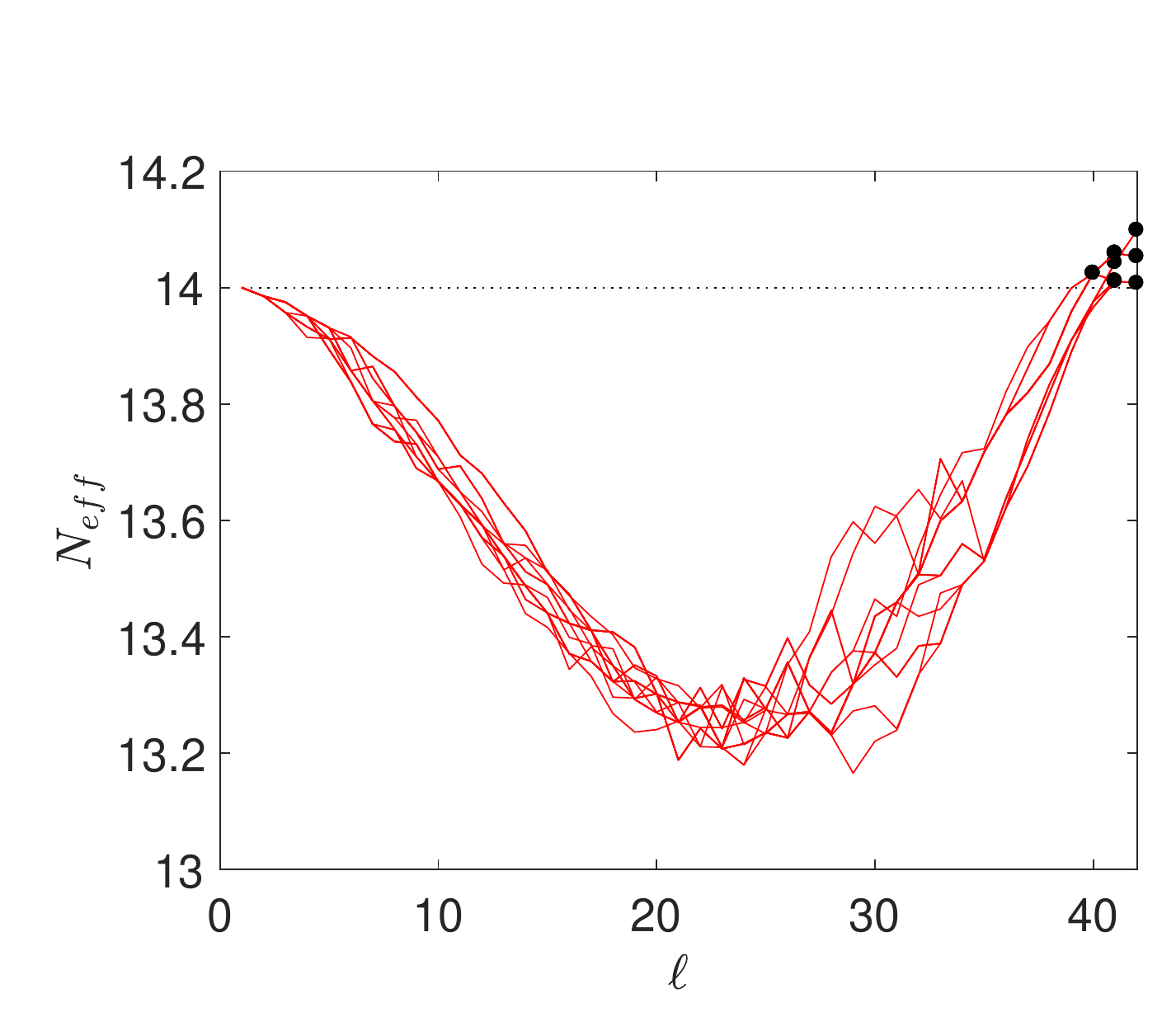}
\includegraphics[trim = 5mm 1mm 6mm 20mm,clip, width=6.6cm, height=5.1cm]{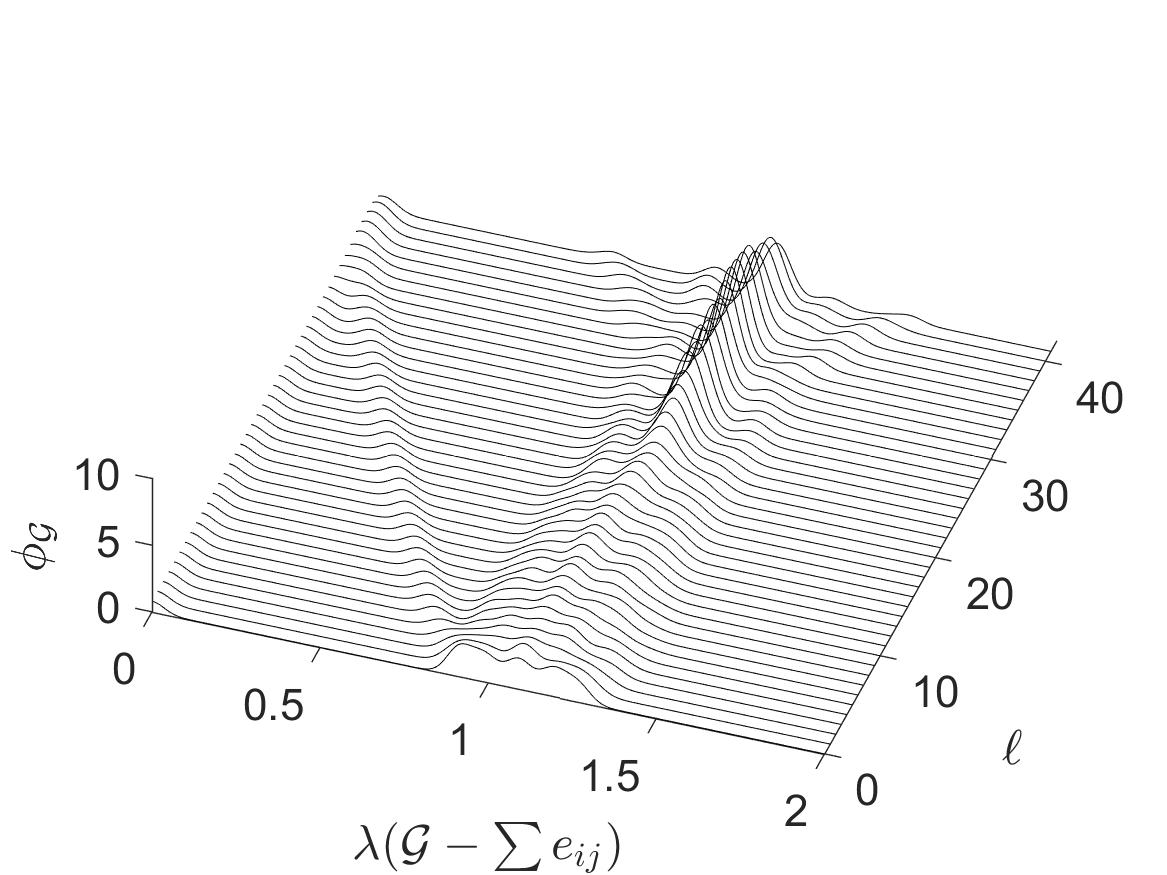}

 (\textbf{a}) \hspace{4.7cm} (\textbf{b})

\includegraphics[trim = 25mm 90mm 40mm 80mm,clip, width=6.6cm, height=5.1cm]{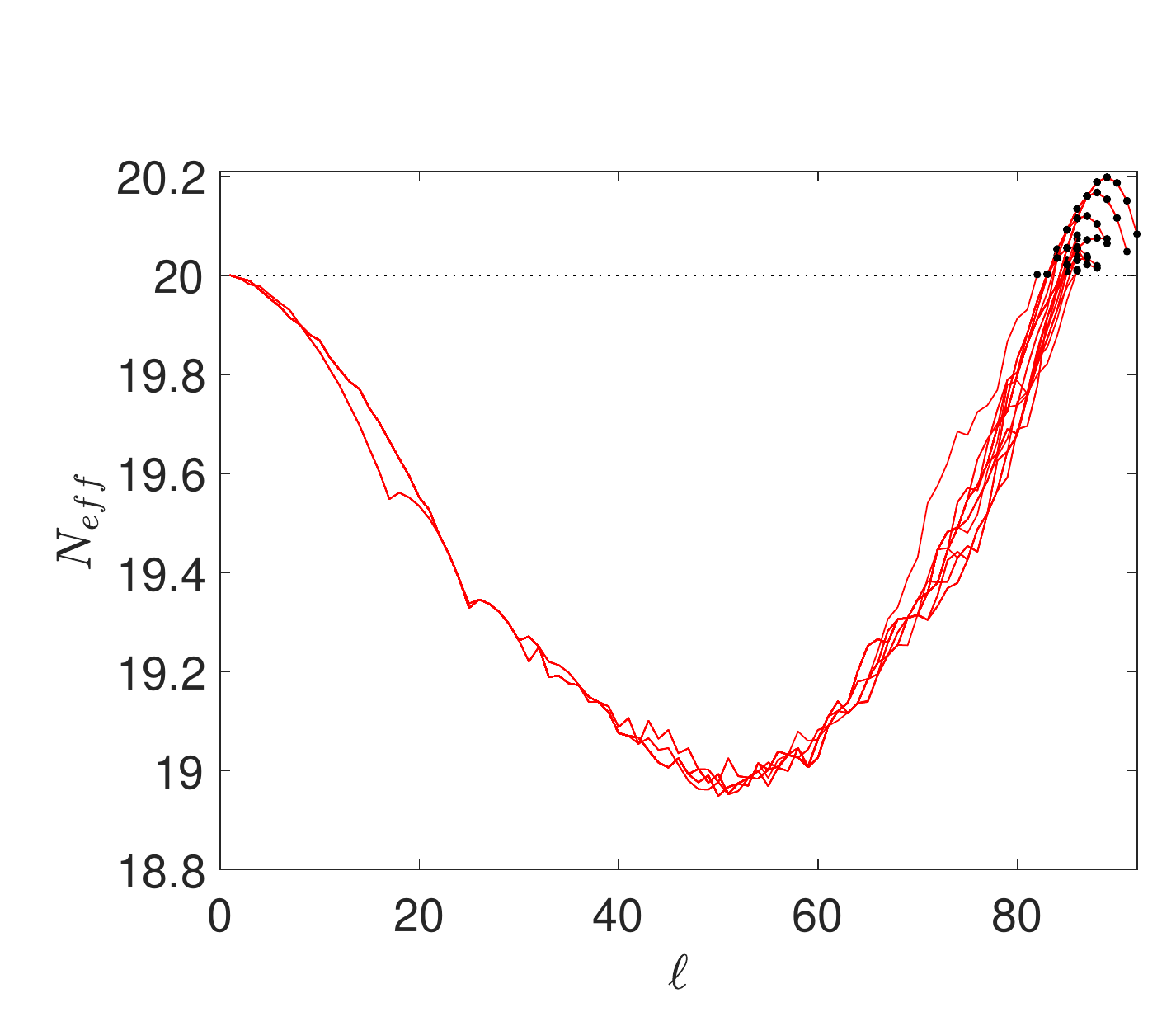}
\includegraphics[trim = 5mm 1mm 6mm 20mm,clip, width=6.6cm, height=5.1cm]{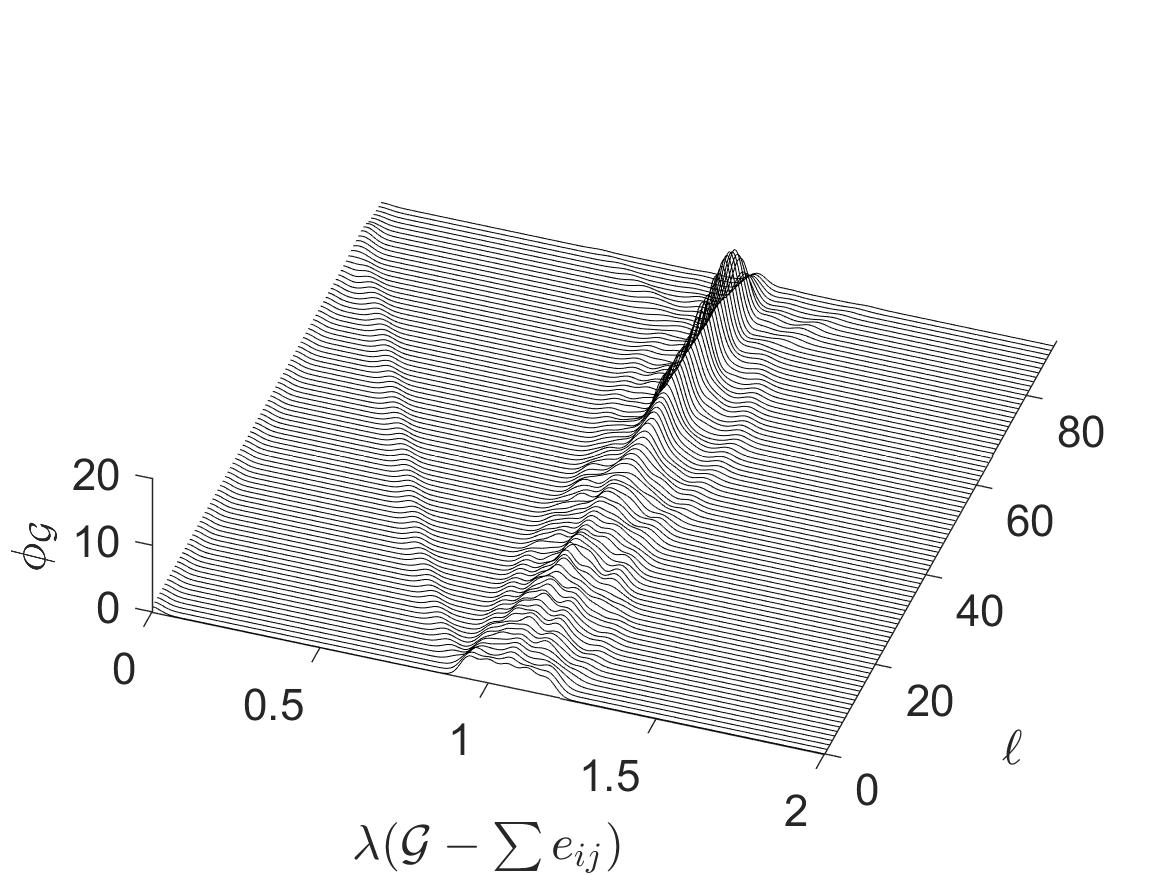}

  (\textbf{c}) \hspace{4.7cm} (\textbf{d})

\includegraphics[trim = 25mm 90mm 40mm 80mm,clip, width=6.6cm, height=5.1cm]{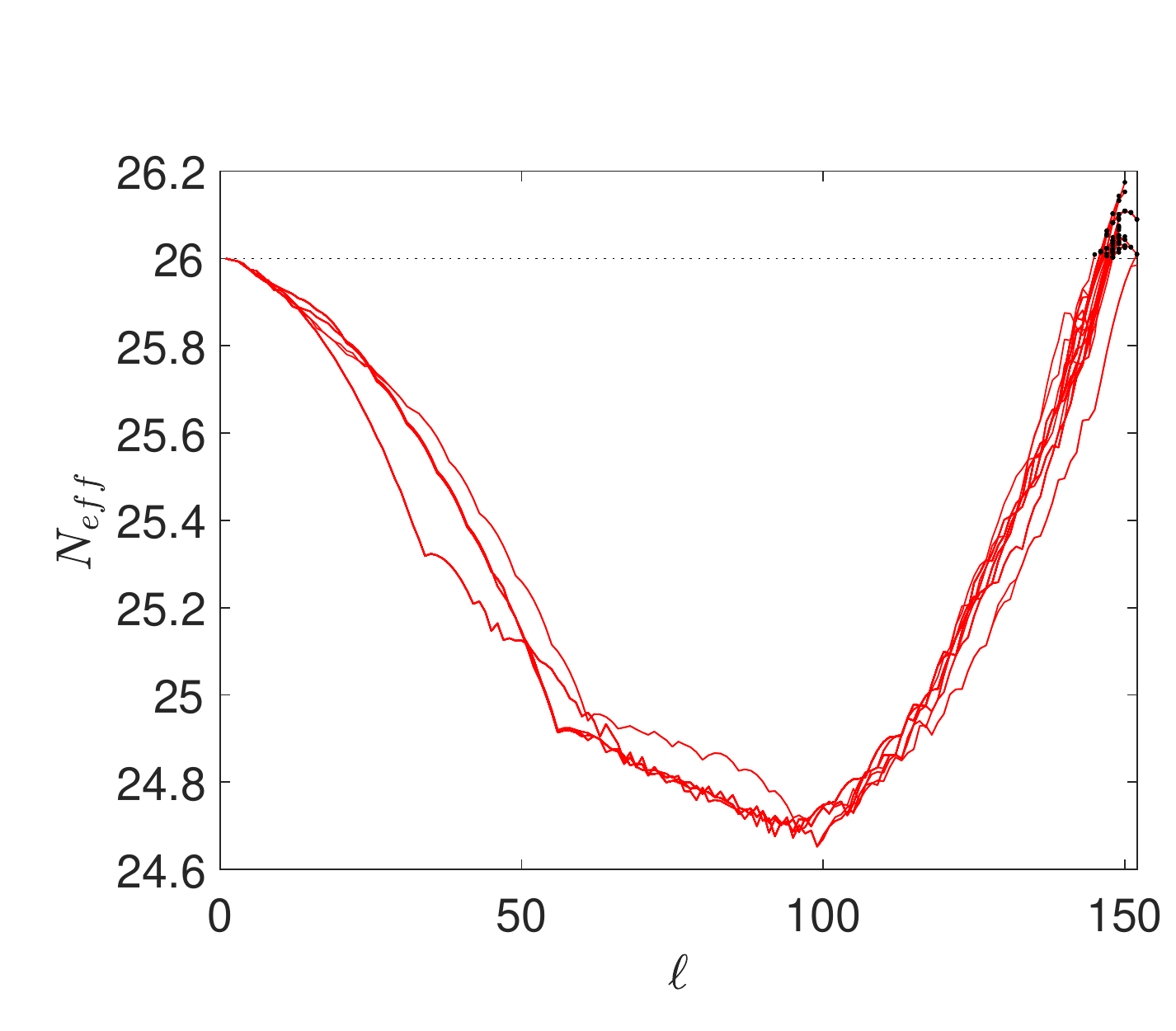}
\includegraphics[trim = 5mm 1mm 6mm 20mm,clip, width=6.6cm, height=5.1cm]{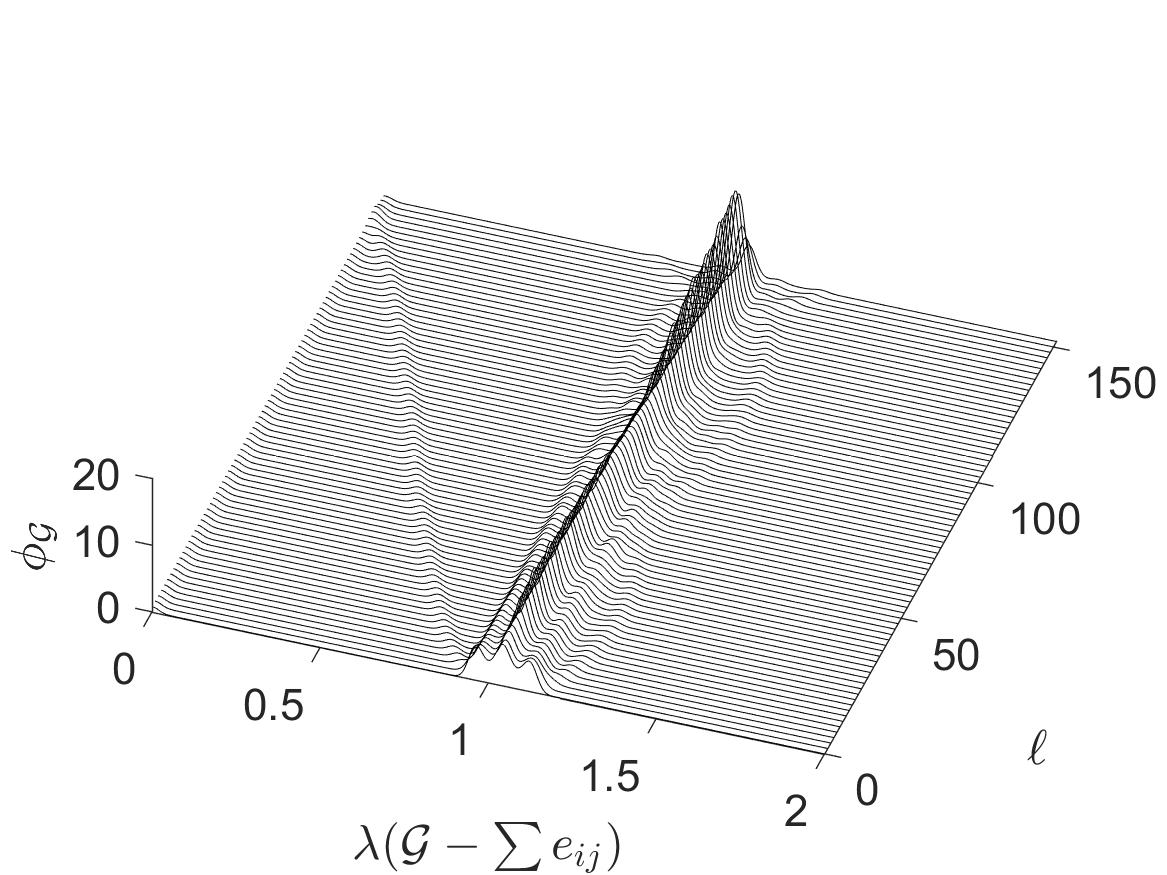}

  (\textbf{e}) \hspace{4.7cm} (\textbf{f})  

\caption{\small{Behaviour of the approximative search of Algorithm 1 for $N=\{14,20,26\}$ and $k=\{11,17,23\}$. The effective population size $N_{eff}$ and the spectral density $\phi_\mathcal{G}$ describing graph evolutions leading to transient amplifiers over edge removal repetitions $\ell$. (a),(b) $N=14$, $k=11$. (c),(d) $N=20$, $k=17$. (e),(f) $N=26$, $k=23$. See Appendix, Fig. \ref{fig:graph_14_20_26_x_dense}, for the spectral density $\phi_\mathcal{G}\prime$ and the difference $|\phi_\mathcal{G}-\phi_\mathcal{G}\prime|$.  }  }
\label{fig:graph_14_X_dense}
\end{figure}

 With increasing order of the considered input graphs also a rising number of transient amplifiers has been identified. Fig. \ref{fig:graph_14_26} gives 4 examples each for $N=\{14,20,26\}$ out of the $\#_{noniso}=\{9,43,55\}$ non-isomorphic amplifiers according to Tab. \ref{tab:14_20_graphs}. The examples are again selected by the largest and smallest values of the effective population size $N_{eff}$ and the mean degree $\bar{k}$. In addition, the maximum and minimum degree, $\Delta(\mathcal{G})$ and $\delta(\mathcal{G})$, as well as the  algebraic connectivity $\lambda_2$ are given.  By comparing the obtained graphs we once more observe characteristic structural features favoring amplification. We find again exclusively graphs consisting of two highly connected cliques. They are mostly joined by a single bridge of one or two edges, but there are also rare instances with two bridges.  Next to these structural similarities there are also differences for a varying number of vertices. For the order of input graphs going up also the maximum, minimum and mean degree ($\Delta(\mathcal{G})$, $\delta(\mathcal{G})$, and $\bar{k}$, respectively) increase with an approximately linear ratio. For transient amplifiers with fixed order $N$ we see no or or very little variance in the maximum degree $\Delta(\mathcal{G})$, which also applies for the mean
  degree $\bar{k}$.  The largest variance can be found for the minimum degree 
$\delta(\mathcal{G})$, which can be as low as $\delta(\mathcal{G})=2$ for bridges with two or more edges, but also as high as $\delta(\mathcal{G})=\Delta(\mathcal{G})-1$  for amplifiers with two bridges, see for instance Fig. \ref{fig:graph_14_26}(f) and (k). Similarly to the results for $N=12$, the algebraic connectivity $\lambda_2$ of transient amplifiers has very small values, while the examples with two bridges have largest. 

\begin{figure}[htb]
\centering

\includegraphics[trim = 25mm 90mm 40mm 80mm,clip, width=6.6cm, height=5.5cm]{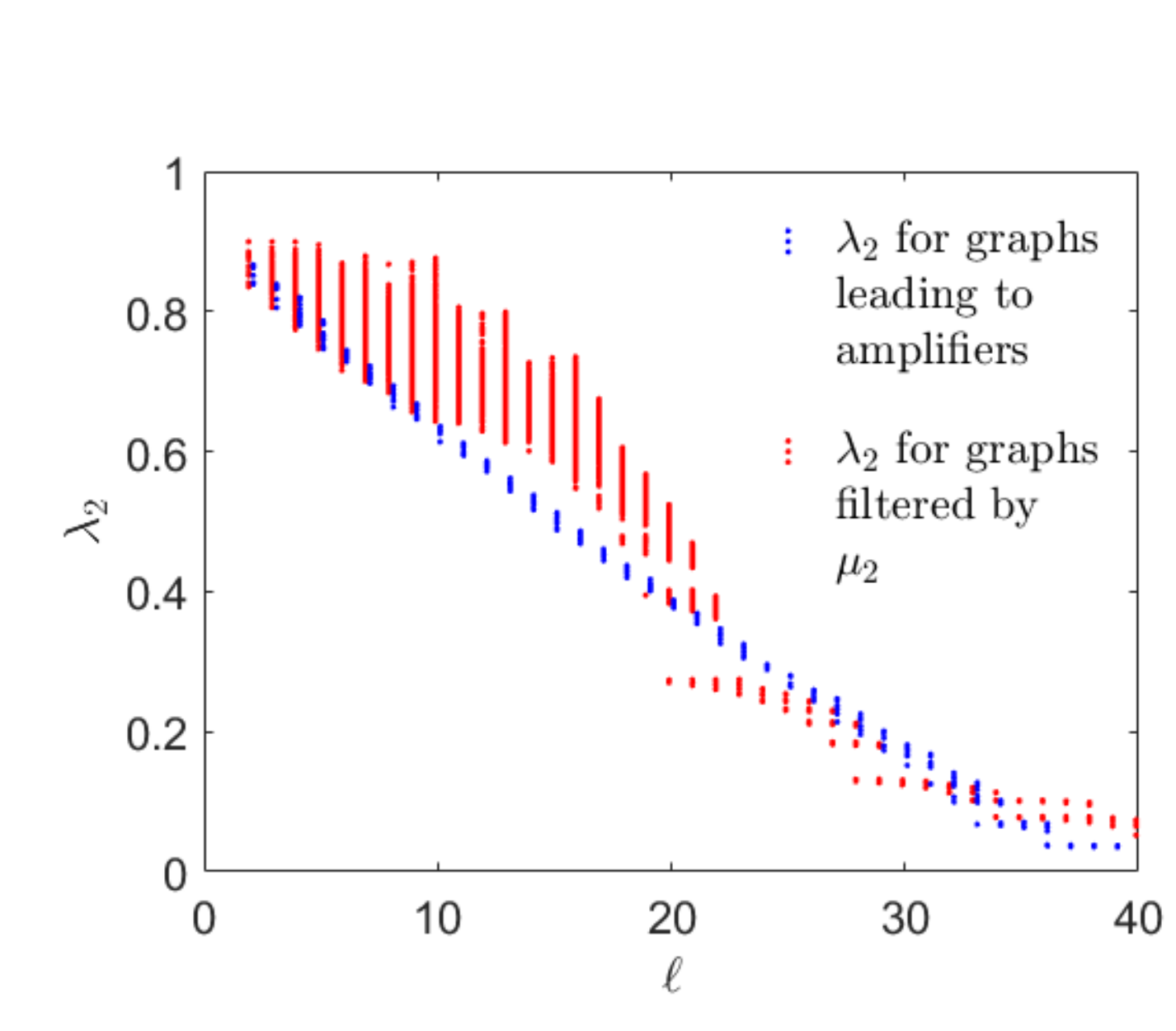}
\includegraphics[trim = 25mm 90mm 45mm 80mm,clip, width=6.6cm, height=5.5cm]{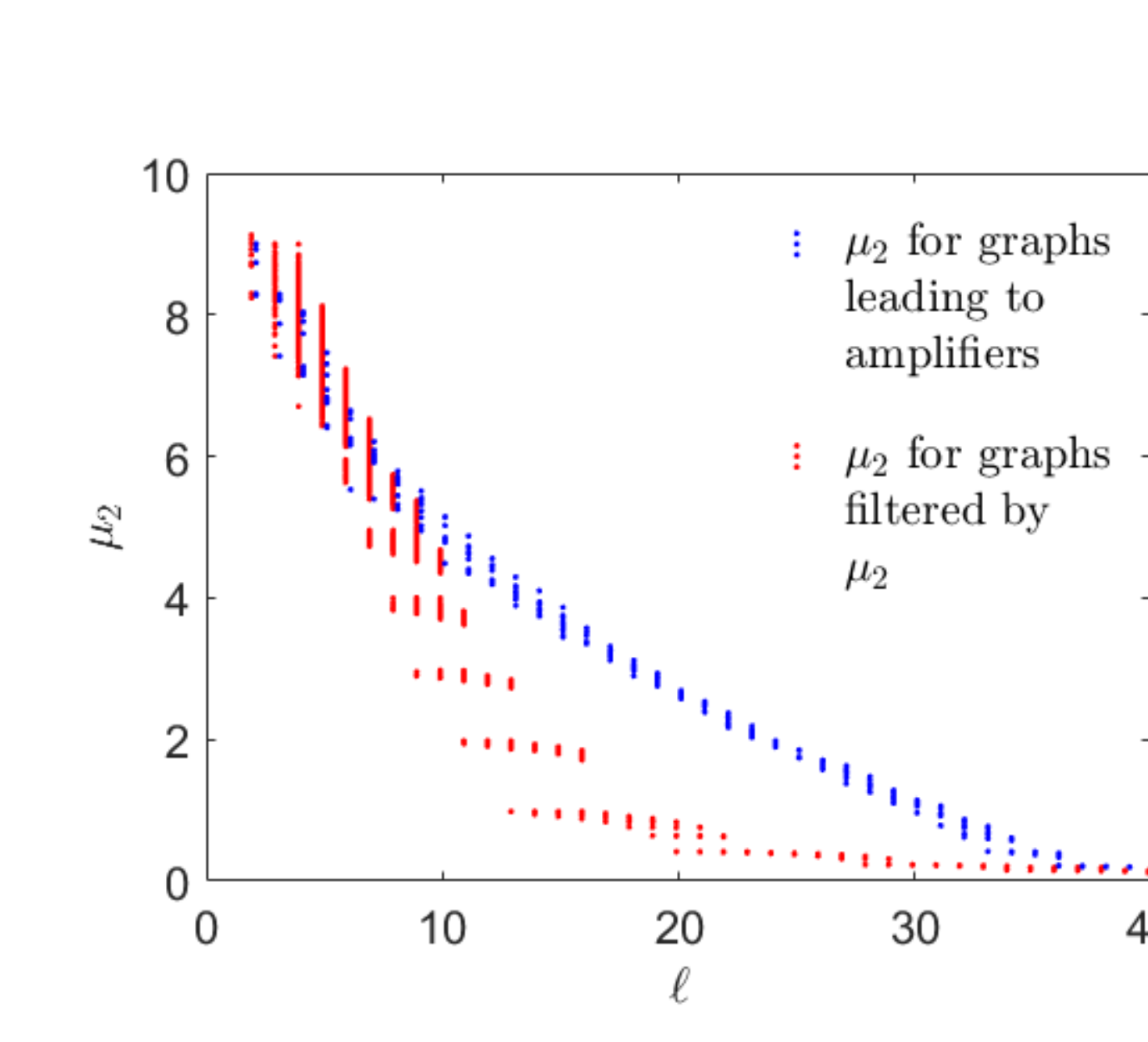}

 (\textbf{a}) \hspace{4.7cm} (\textbf{b})

\caption{\small{Spectral dynamics of guided edge removals. Comparing between graph evolutions leading or not leading to transient amplifiers evaluated and filtered by spectral graph measures. (a)  Algebraic connectivity $\lambda_2$ derived from the normalized Laplacian. (b) Algebraic connectivity $\mu_2$ derived from the standard Laplacian.  }  }
\label{fig:graph_14_compare}
\end{figure}

Fig. \ref{fig:graph_14_X_dense}(a),(c),(e) gives the behavior of the effective population size $N_{eff}$    over edge removal repetitions $\ell$ for $N=\{14,20,26\}$. The results are generally similar to $N=12$, compare to Fig.~\ref{fig:graph_12_X_neff}(a). Though, for $N=20$ and  $N=26$ we frequently find  graph trajectories with consecutive transient amplifiers.  This means for a certain $\ell$ we have a transient amplifier graph and by removing an edge from this graph, we get another transient amplifier. The effective population size $N_{eff}$ may vary for  consecutive transient amplifiers, and we find successively increasing values as well as a parabolic succession. Fig. \ref{fig:graph_14_X_dense}(b),(d),(f) shows the spectral dynamics expressed by spectral densities $\phi_\mathcal{G}$. Again there is a general similarity to $N=12$,  compare to Fig.  \ref{fig:graph_12_X_dense}(a). Particularly, the two geometrical features already discusses, the travelling peak of $\lambda_2$ progressively getting smaller and the standing peak indicating eigenvalue multiplicity can be found in almost the same manner. Thus, it can be concluded that they are features independent of the considered $N$ and $k$.    For the spectral density $\phi_\mathcal{G}\prime$ and the difference $|\phi_\mathcal{G}-\phi_\mathcal{G}\prime|$, see Appendix, Fig. \ref{fig:graph_14_20_26_x_dense}.

The transient amplifiers discussed up to now have been identified using as filter the algebraic connectivity $\lambda_2$ derived from the normalized Laplacian. A noteworthy result is that contrary to using  $\lambda_2$ as filter, taking the algebraic connectivity $\mu_2$ derived from the standard Laplacian does not yield amplifiers, at least not for the tested input graphs on $N=\{11,12,14,20,26\}$ vertices and filter sizes up to $\#_\mathcal{G}=2.500$. In the discussion about using input graphs on $N=11$ vertices and degree $k=6$, it has been observed that the quantities $\lambda_2$ and $\mu_2$ behave differently for edges being removed from a regular graph, see Fig.  \ref{fig:graph_11_6_neff}(d). This is the case for all $N$ and $k$ tested.  Using the example $N=14$  and $k=11$ this behaviour is now analysed by their spectral dynamics. It is furthermore argued that such an analysis offers a possible explanation as to why  $\lambda_2$ as filter leads to amplifiers while $\mu_2$ does not. 

We compare for subsequent edge removal repetitions $\ell$ how graph evolutions leading to transient amplifiers guided by low values of $\lambda_2$ would be evaluated if the filter were using low values of $\mu_2$. The setup of the analysis is this. We take a single input graph from the pool of input graphs finally leading to transient amplifiers. The results given in Fig. \ref{fig:graph_14_compare} are for the graph with $N=14$ and $k=11$ which yields in total $530$ amplifiers of which $9$ are pairwise non-isomorphic. For other input graphs, also with other $N$, similar results have been obtained. Using this input graph we track the values of $\lambda_2$ and $\mu_2$ for these $530$ trajectories leading to amplifiers in the process guided by $\lambda_2$, see the blue dots in Fig. \ref{fig:graph_14_compare}(a) and (b). Then, we rerun the edge removal process taking the same input graph and the graphs on the trajectory towards  transient amplifiers, but  filter and select for each $\ell$ according to $\mu_2$. In other words, we track the values of $\lambda_2$ and $\mu_2$ for the graph evolution leading to amplifiers for each consecutive $\ell$ as if the graphs were to be evaluated and filtered by $\mu_2$,  see the red dots in Fig. \ref{fig:graph_14_compare}(a) and (b). The results show that for  $\lambda_2$, see Fig. \ref{fig:graph_14_compare}(a),  the values of graphs leading to amplifiers are mostly below the values for graphs that would have been taken if they were filtered by $\mu_2$. As the filter selects for small values, graphs leading to amplifiers actually remain in the pool of candidate graphs.  There is an interval in edge removals $ 20<\ell <30$ where the values of $\lambda_2$ for graphs selected according to $\mu_2$ are lower than those on the trajectory towards amplifiers, but if the filter size is large enough this does not lead to exclusion of candidate graphs needed to finally obtain amplifiers.

If we look at  the
spectral dynamics from the 
perspective of $\mu_2$, we get different results,  see Fig. \ref{fig:graph_14_compare}(b). Here the values of $\mu_2$ leading to transient amplifiers are mostly above the values of $\mu_2$, particularly for $\ell >10$. Thus, as the filter selects for small values of $\mu_2$ the graphs which would have led to amplifiers are gradually sorted out of the pool of candidate graphs and thus no transient amplifiers are identified.  It is quite possible that using $\mu_2$ as filter would lead to amplifiers if the filter size  $\#_\mathcal{G}$ is larger than some threshold. However, tests with filter sizes up to $\#_\mathcal{G}=2.500$ have brought no results.

 \subsection{Barbell and dumbbell graphs} \label{sec:bar}
\begin{figure}[htb]
\centering

\includegraphics[angle=5,trim = 58mm 110mm 55mm 110mm,clip, width=3.8cm, height=3.5cm]{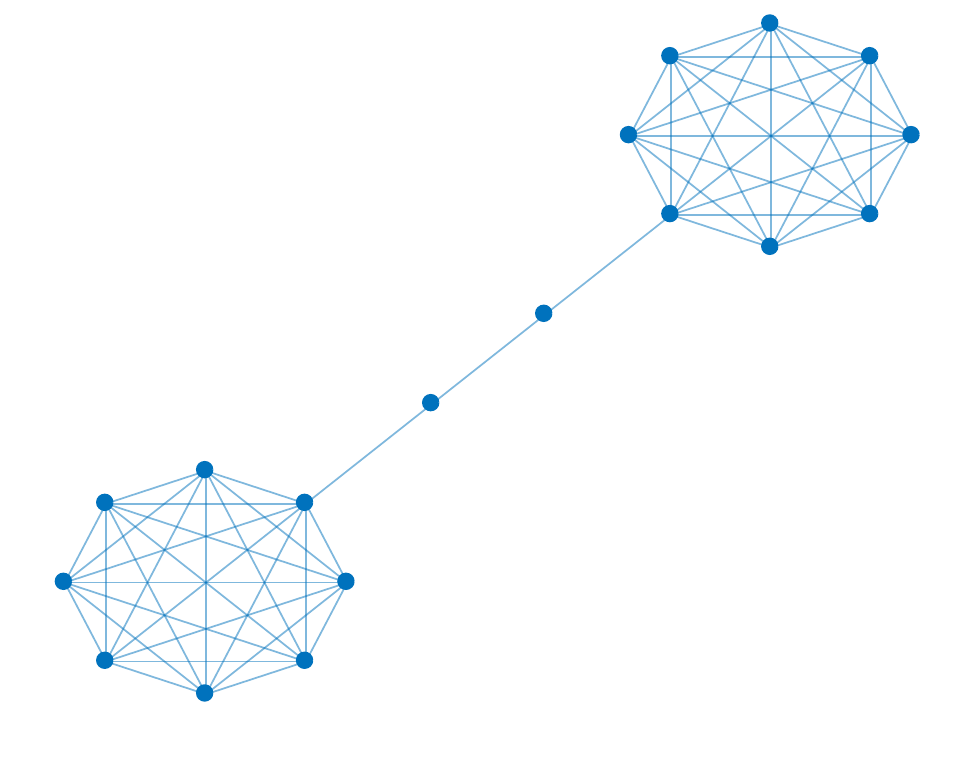}
\includegraphics[angle=5,trim = 58mm 110mm 55mm 110mm,clip, width=3.8cm, height=3.5cm]{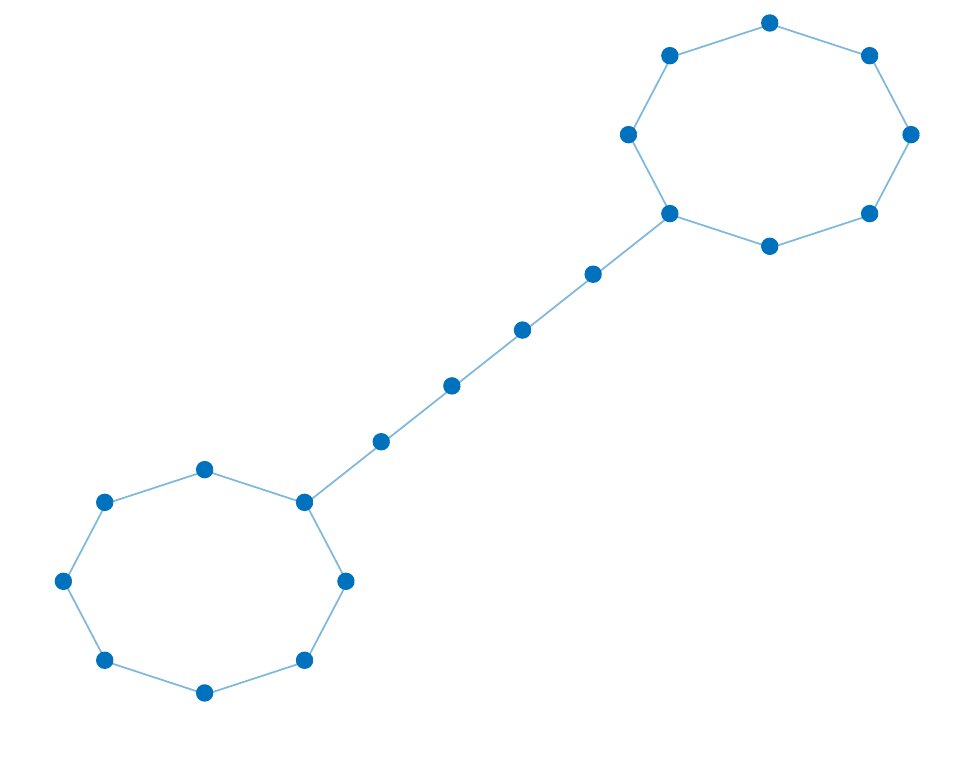}
\includegraphics[angle=5,trim = 58mm 110mm 55mm 110mm,clip, width=3.8cm, height=3.5cm]{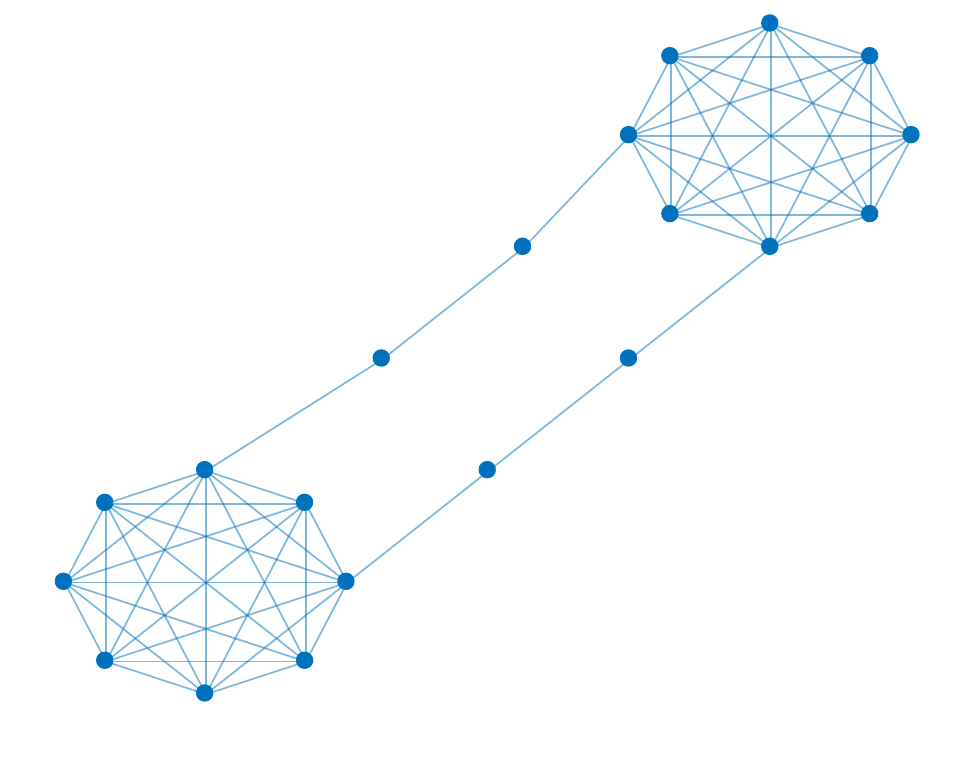}
\includegraphics[angle=5,trim = 58mm 110mm 55mm 110mm,clip, width=3.8cm, height=3.5cm]{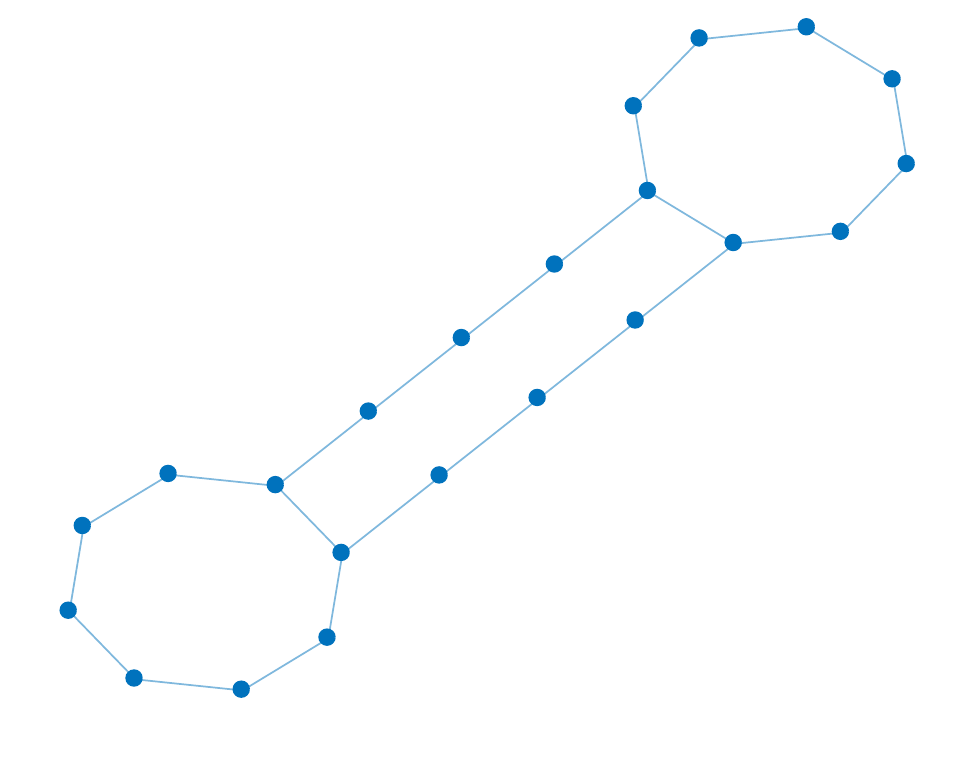}

 (\textbf{a}) \hspace{2.7cm} (\textbf{b})   \hspace{2.7cm} (\textbf{c})  \hspace{2.7cm} (\textbf{d})
 
 \caption{\small{Barbell graphs $\mathbf{B}(a,b)$ (and dumbell graph $\mathbf{D}(a,b)$) consisting of two complete graph (or two circle graph) with $a$ vertices each and a bridge with $b$ edges. (a) $\mathbf{B}(8,3)$. (b) $\mathbf{D}(8,5)$. Barbell graphs $\mathbf{B}_2(a,b)$ (and dumbell graph $\mathbf{D}_2(a,b)$) with two bridges of $b$ edges each. (c) $\mathbf{B}_2(8,3)$. (d) $\mathbf{D}_2(8,4)$. }  }
\label{fig:x_bell_graph}
\end{figure}

Barbell and dumbbell graphs are two families of graphs with a standardized structure~\cite{ghosh08,wang09}. Barbell graphs $\mathbf{B}(a,b)$ consist of two complete graphs with $a$ vertices each which are connected by a bridge with $b$ edges, while dumbbell graphs $\mathbf{D}(a,b)$ consist of two circles with $a$ vertices each which are also connected by a bridge with $b$ edges.  According to such a definition a $\mathbf{B}(a,b)$  barbell graph as well as a $\mathbf{D}(a,b)$ dumbbell graph has $N=2a+b-1$ vertices, see the barbell graph $\mathbf{B}(8,3)$ in Fig. \ref{fig:x_bell_graph}(a) and the dumbbell graph $\mathbf{D}(8,5)$ in Fig. \ref{fig:x_bell_graph}(b).
In view of the findings that most, if not all, amplifier graphs identified using the algorithmic framework discussed in the previous sections resemble barbell and dumbbell graphs in some way or another, we next discuss amplification properties of these graphs.  

\begin{figure}[htb]
\centering
 
 \includegraphics[trim = 25mm 90mm 40mm 80mm,clip, width=6.6cm, height=5.5cm]{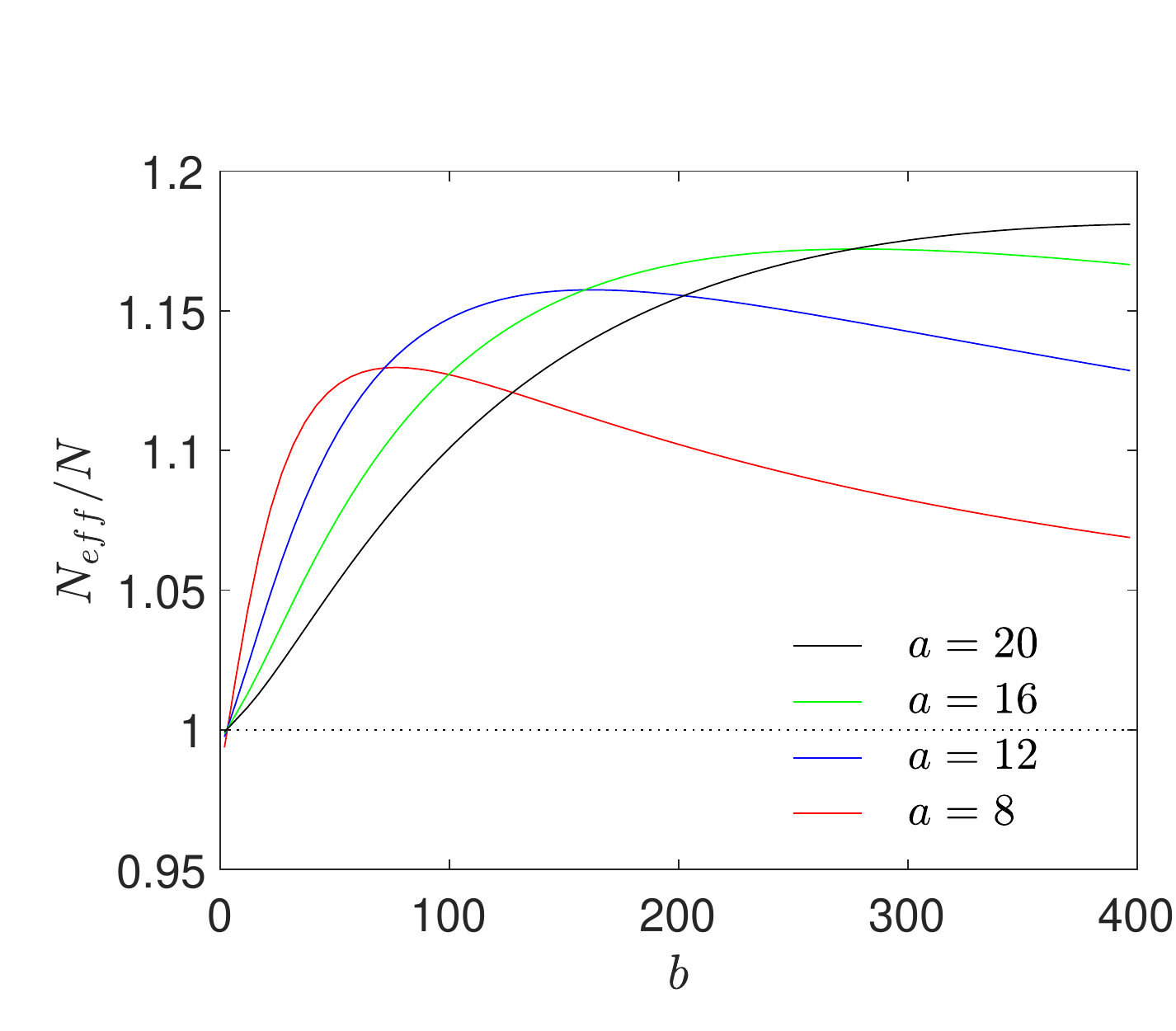}
 \includegraphics[trim = 25mm 90mm 40mm 80mm,clip, width=6.6cm, height=5.5cm]{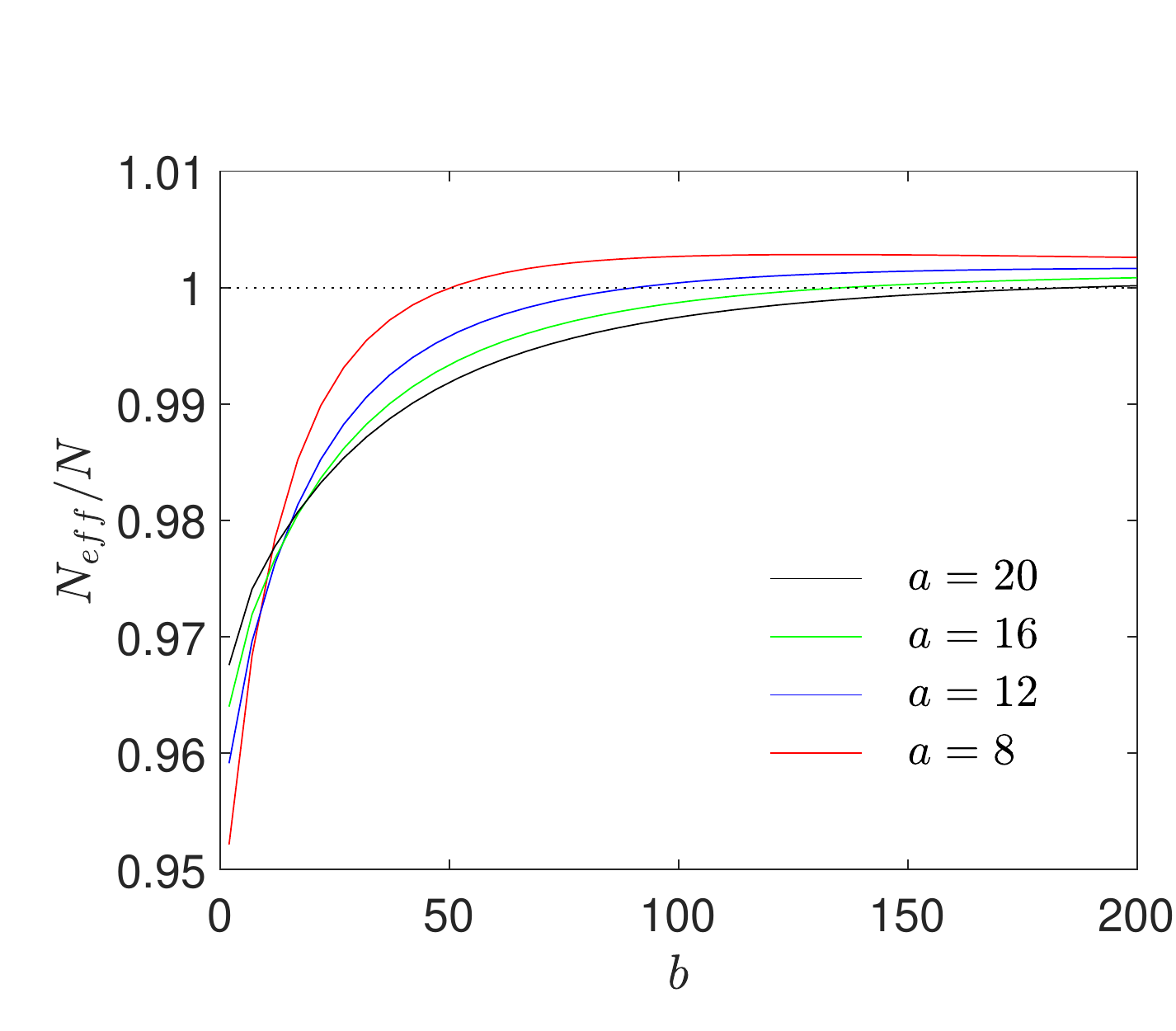}

  (\textbf{a}) \hspace{4.7cm} (\textbf{b})

 \includegraphics[trim = 25mm 90mm 40mm 80mm,clip, width=6.6cm, height=5.5cm]{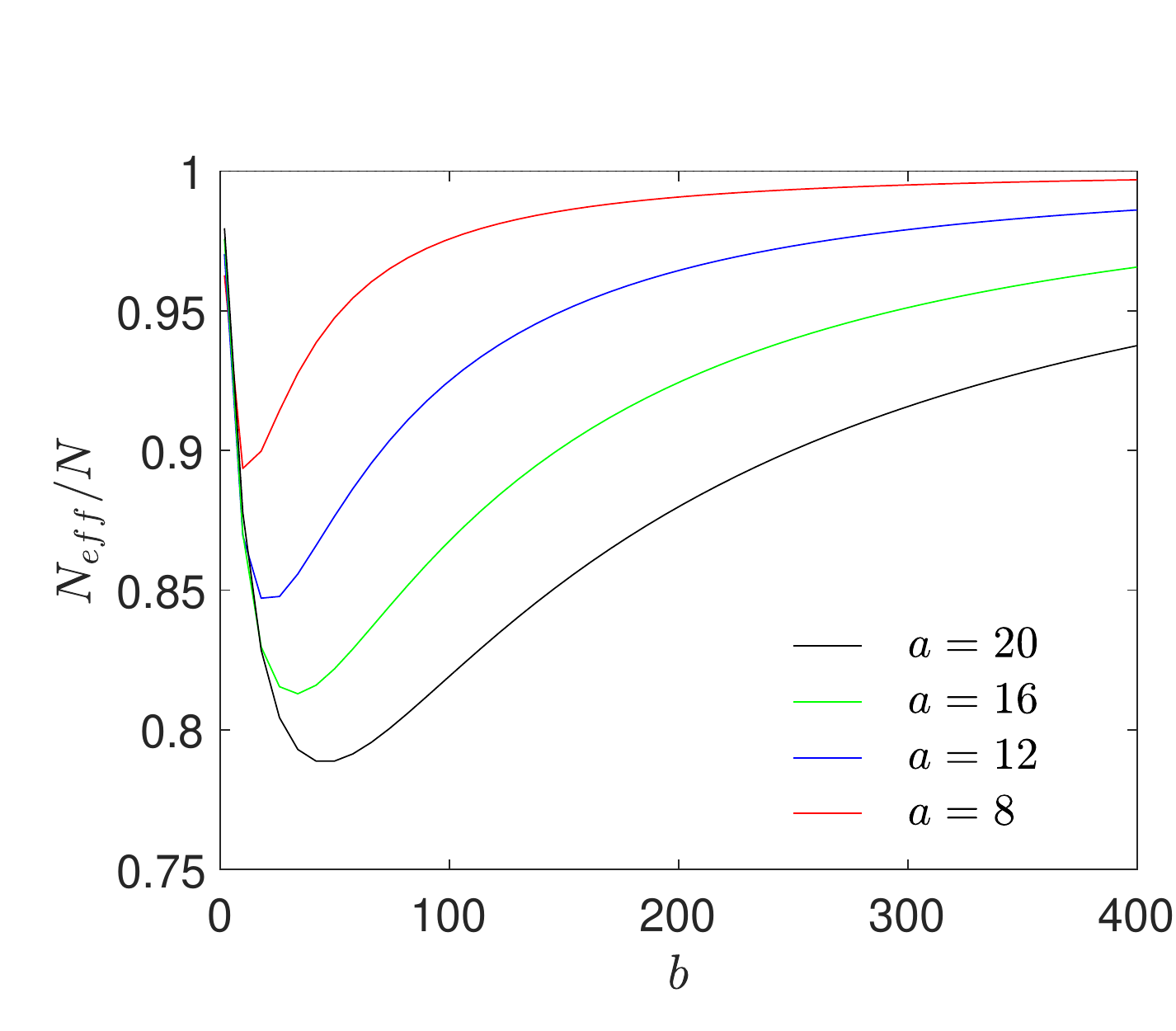}
 \includegraphics[trim = 25mm 90mm 40mm 80mm,clip, width=6.6cm, height=5.5cm]{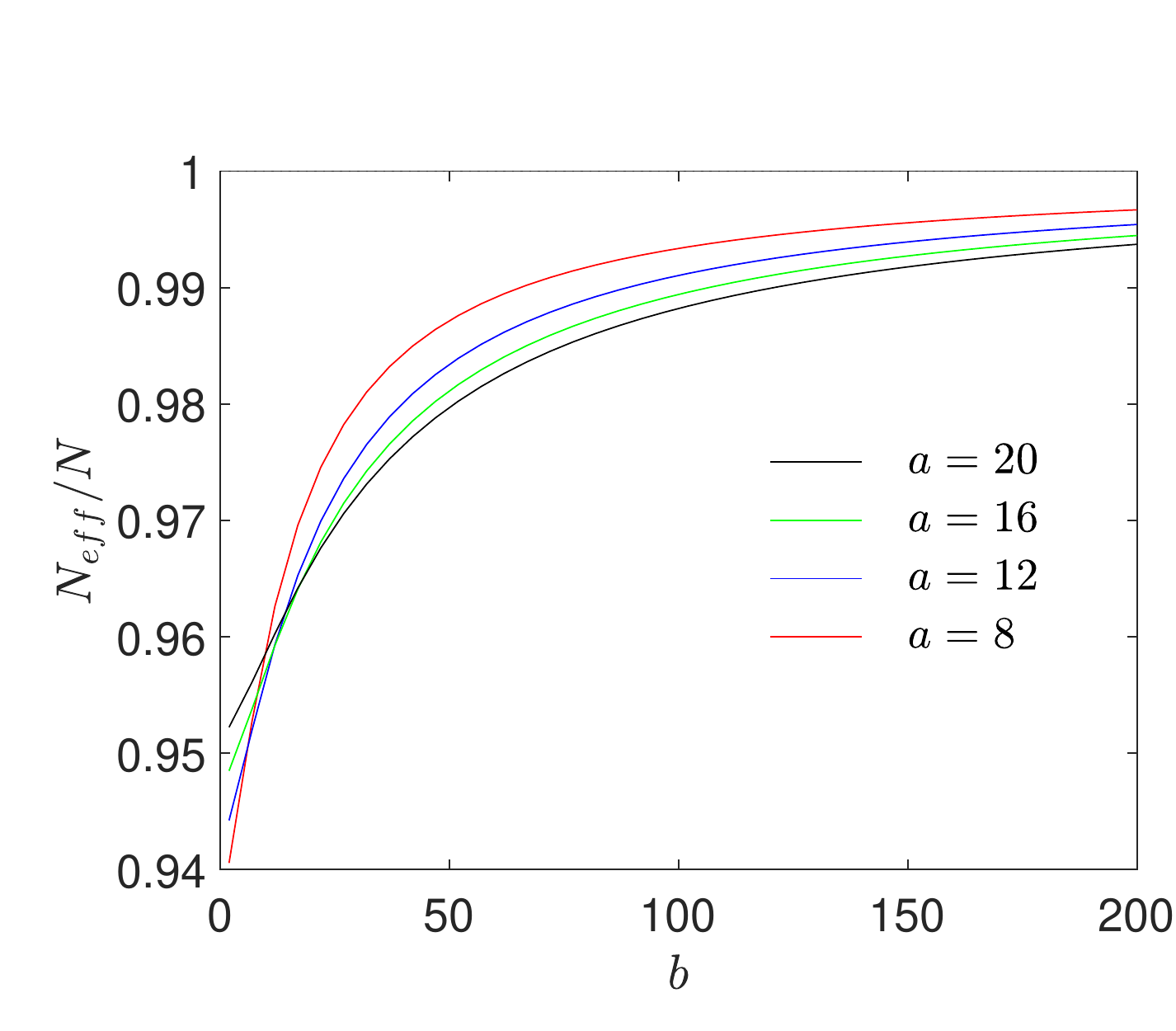}

  (\textbf{c}) \hspace{4.7cm} (\textbf{d})

\caption{\small{The quantity $N_{eff}/N$ for barbell graphs $\mathbf{B}(a,b)$ and $\mathbf{B}_2(a,b)$ as well as dumbbell graphs  $\mathbf{D}(a,b)$ and $\mathbf{D}_2(a,b)$ for different $a$ over $b$. The variable $a$ is the number of vertices in each of the two complete (or circle) graphs, $b$ is the number of edges in the bridge or the two bridges, see also Fig. \ref{fig:x_bell_graph}. Values of  $N_{eff}/N>1$ indicate transient amplification properties.  }}
\label{fig:x_bell_graph_amp}
\end{figure}
Fig. \ref{fig:x_bell_graph_amp}(a) gives the quantity $N_{eff}/N$ for barbell graphs $\mathbf{B}(a,b)$  with $a=\{8,12,16,20\}$ and $1\leq b \leq 400$. A ratio $N_{eff}/N>1$ indicates transient amplification properties.  
We see that for all $a$ and $b>3$ transient amplifiers exist. The ratio $N_{eff}/N$ increases with rising $b$ (and by $N=2a+b-1$ with rising $N$) for a certain interval in $b$, before  reaching a maximum and then slowly falling and finally converging to a value $N_{eff}/N>1$. The larger the  value $a$ is (indicating the number of vertices in the two complete subgraphs) the higher is the maximum $N_{eff}/N$ itself and the larger the associated number of bridges $b$.
The smallest barbell graph with amplification properties is  $\mathbf{B}(4,5)$, which has $N=12$. 
For dumbbell graph the results are qualitatively similar, but some details differ, see Fig. \ref{fig:x_bell_graph_amp}(b) for the same values of $a$ and     $1\leq b \leq 200$. Also for dumbbell graphs amplification properties can be found, but the ratio  $N_{eff}/N$ is much smaller than for barbell graphs and also the number of bridges (and thus the order of the graph) needed is higher. The smallest dumbbell graph with amplification properties is  $\mathbf{D}(3,12)$, which has $N=17$.  Generally speaking,
amplification properties of barbell and dumbbell graphs are universal and for other values of $a$ similar results are obtained. 

As some of the algorithmically identified transient amplifiers have two bridges (see for instance, Fig. \ref{fig:graph_12_}(d) or Fig. \ref{fig:graph_14_26}(f) and (k)), we finally study amplification properties of barbell and dumbbell graphs with two bridges. Therefore, we define  two bridge barbell graphs $\mathbf{B}_2(a,b)$
as two complete graphs with $a$ vertices each connected by two bridges of $b$ edges each. As in a complete graph each vertex is connected to all other vertices (except itself) it makes no difference which two vertices serve as bridgeheads. A two bridge dumbbell graph   $\mathbf{D}_2(a,b)$ is defined likewise, but here  the location matters where the bridges branch off. We define that the two bridgeheads on each side are directly connected, see Figs. \ref{fig:x_bell_graph}(c) and (d) for the examples of $\mathbf{B}_2(8,3)$ and  $\mathbf{D}_2(8,4)$. 
With respect to transient amplification, we see that most likely neither  $\mathbf{B}_2(a,b)$ nor $\mathbf{D}_2(a,b)$ have this properties, see Figs. \ref{fig:x_bell_graph_amp}(c) and (d) which give the ratio $N_{eff}/N$ for different $a$ over $b$. We observe that the curves are always below $N_{eff}/N=1$, become lower for $a$ increasing with  $N_{eff}/N \rightarrow 1$ from below for $b$ getting large. Also, tests with other $a$ and $b$ have not revealed amplification. Eventually, minor modification in the barbell and dumbbell graphs were introduced, for instance deleting the edge between the bridgeheads of the     $\mathbf{B}_2(a,b)$ barbell graphs or varying the edge distance between the bridgeheads of the     $\mathbf{D}_2(a,b)$ dumbbell graphs, or having different amounts of vertices in the two complete and circle graphs, or connecting the bridge to two bridgeheads. However, also these graphs  have not shown amplification properties. It remains to be concluded that although transient amplifier with two bridges have been  found, for instance, Fig. \ref{fig:graph_12_}(d) or Fig. \ref{fig:graph_14_26}(f) and (k), there are more subtle rules as to how barbell and dumbbell graphs must be modified to possess this property.   This may be a topic for future work.

\section{Discussion} \label{sec:dis}

\subsection{Identifying transient amplifiers}  \label{sec:dis_ident}
In the previous section, results about using an iterative algorithmic process for identifying transient amplifier for dB updating have been presented. We next discuss some implications of the results obtained. The algorithm has been tested for all regular graphs on $N=\{11,12\}$ vertices and all degrees, and all regular graphs on $N=\{14,20,26\}$ vertices and degree $k=N-3$.  It has been shown that although  transient amplifiers for dB updating are rather rare, a substantial number of instances has been identified for all tested graph orders. For $N=11$ and $N=12$, most structurally different transient amplifiers are obtained for middle range $k$, that is $k \approx N/2$. It seems to be reasonable to assume that this also applies to $N=\{14,20,26\}$ and the amplifiers obtained for $k=N-3$  are just a small subset of all amplifiers. Unfortunately, a direct test of this assumption was not possible with the available numerical resources due to the massive growth  
in structurally different regular input graphs (for instance, there are $\mathcal{L}_7(14)=21.609.301$ regular graphs for degree $k=7$ and order $N=14$, $\mathcal{L}_{10}(20)$ and $\mathcal{L}_{13}(26)$ are still not exactly known).

Furthermore, all amplifier graph share certain structural characteristics. They are  graphs consisting of two cliques of highly (frequently completely) connected vertices, which are joined by a bridge of one or more edges. Occasionally, structures  with two bridges connecting the cliques have amplification properties. Furthermore, these structures resemble those of barbell and dumbbell graphs, which themselves have amplification properties, see Sec. \ref{sec:bar}. Considering the space of all structurally different graphs with a given order $N$, these structures are rather special and consequently rare.  This is in agreement with a previous work~\cite{rich21} studying the structural and spectral properties connected with removing a single edges from cubic (and quartic) regular graphs up to an order of $N=22$ (and $N=16$).  Also these results showed that transient amplifiers for dB updating exist for all $N$ tested, are rather rare and have certain graph structures. 
Extending these results, in this study we have been interested in the transition process from a regular graphs to a transient amplifier over multiple edge removals. Thus, we obtained a larger variety of transient amplifiers with a stronger perturbation to the regularity of the input graphs. Nevertheless, also this larger variety is subject to similar
structural restrictions. On way of accounting for these restrictions is the degree distribution of  graphs $\mathcal{G}$ expressed by the maximum degree $\Delta(\mathcal{G})$, the minimum degree  $\delta(\mathcal{G})$ and the mean degree $\bar{k}$. If we compare over varying order $N$, we see that generally the  maximum degree $\Delta(\mathcal{G})$ is bounded by $\frac{1}{3}N < \Delta(\mathcal{G}) < \frac{1}{2}N$, while   the mean degree $\bar{k}$ is restricted to $\frac{4}{5}\Delta(\mathcal{G})<\bar{k}<\Delta(\mathcal{G})$. 

Only for the  minimum degree  $\delta(\mathcal{G})$ we find a rather large variety, which can be as low as $\delta(\mathcal{G})=2$ for transient amplifiers with bridges of two and more edges, or as high as $\delta(\mathcal{G})=\Delta(\mathcal{G})-1$ for some amplifiers with two bridges.   In other words, transient amplifiers seems to have an upper and lower bounds of maximal and mean degree.  Such a distribution of  the mean degree $\bar{k}$ differs from random graphs, for instance  Erd{\"o}s-Rényi or Barabási-Albert graphs, which have a binomial and power-law distribution with a much larger spread. Furthermore, this  means that the degree $k$ of the regular input graph plays a role in what structure the transient amplifiers has only insofar as it bounds the maximum degree $\Delta(\mathcal{G})$. This is particularly visible for input graphs on $N=12$ vertices where for all degrees $k=\{3,4,\ldots,9\}$ transient amplifiers have been identified, see Tab. \ref{tab:12_graphs}.   If we compare over varying input degrees $k$, we see that the mean degree $\bar{k}$ slightly increases with increasing $k$ but the transient amplifiers remain in a rather small range of $\bar{k}$ ($2.6666 \leq \bar{k} \leq 4.1666$). In other words, the input degree 
$k$ does not matter very much as if $k$ is large then just more edges need to be removed before a transient amplifiers appears.  Thus, a main result of this study is that many graph structures resembling barbell and dumbbell graphs  with two cliques of highly connected vertices joined by a bridge are transient amplifier of dB updating. These structures expand the collection of graph structures already known to have this property and denoted as fans, separated hubs and stars of islands~\cite{allen20}. They also complement graph structures known as amplifiers of Bd updating and denoted as
lollipop, balloon,  balloon-star graphs~\cite{allen21,moell19}. 

An interesting question is why the iterative algorithmic process does head for graphs structures resembling barbell and dumbbell graphs but not for  structures similar to fans, separated hubs or stars of islands. A main reason most likely is that the approximative search using as filter small values of the algebraic connectivity $\lambda_2$ particularly promotes such structures. A value $\lambda_2=0$ means to disconnect the graph, and low values of $\lambda_2$ imply bottlenecks, clusters, low conductance and path-like graphs which can rather easily be divided into disjointed subgraphs by removing edges or vertices.~\cite{ban08,ban09,hoff19,wills20}. Fans, separated hubs or stars of islands are structurally further away from being close to disconnection than barbell and dumbbell graphs. It could be an  topic of future work if a filter using different spectral or other graph measures apart from (or in addition to) the algebraic connectivity would be suitable to identify also these structures.

\subsection{Spectral dynamics of guided edge removals}
In this study we are equally interested in the performance and the behavior of the iterative process for identifying transient amplifier for dB updating. While Sec.~\ref{sec:dis_ident} mainly focused on algorithmic performance, we next discuss some aspects of algorithmic behaviour.  A main tool in analyzing the algorithmic behaviour is the spectral dynamics of guided edge removals from regular input graphs. The search process is guided by two quantities derived from the graph, the maximal remeeting time $\max(\tau_i)$ and algebraic connectivity $\lambda_2$ of the normalized Laplacian. Both quantities guide the search on different levels. 
The largest remeeting time $\max(\tau_i)$ is suitable to compare vertices and determines from which vertex an edge should be removed. It is not suitable to compare graphs, but the algebraic connectivity $\lambda_2$ is. It determines for the approximative search which graphs remain in the pool of candidate graphs. The decision to use the quantity $\lambda_2$ as filter for candidate graphs is itself a result of preliminary analysis and previous work. On the one hand,  previous results showed that one edge removals yielding transient amplifiers are connected with low $\lambda_2$~\cite{rich21}. 
Moreover, there are applications of graph breeding and graph pruning algorithms in network science which successfully used spectral properties, particularly algebraic connectivity, for guiding the search process~\cite{chan16,ghosh06,ghosh08,li18,shi19,syd13}. 
Finally, a preliminary analysis revealed that local graph measures such as betweenness or closeness centrality and degree distribution, but also motive and cycle count somehow correlate to graph evolutions leading to transient amplifiers, but are generally not promising as filter criteria.  The lack of usefulness of another global graph measure, the algebraic connectivity $\mu_2$ associated with the standard Laplacian, has been discussed in Sec. \ref{sec:14_20_26}, see also Fig. \ref{fig:graph_14_compare}.

Spectral dynamics generally refers to changes in the graph spectra over graph manipulations~\cite{chen17,zhang09}. We here consider the graph manipulations to be repeated edge removals.
In Secs.~\ref{sec:11_12} and~\ref{sec:14_20_26} several instances are given of how the algebraic connectivity $\lambda_2$ as well as the smoothed spectral density $\phi_\mathcal{G}$ changes if we remove edges from a regular input graph and either obtain a transient amplifiers in the end, or not. These results demonstrate that the spectral dynamics towards transient amplifiers subtly differs from the spectral dynamics of graph evolutions not doing so. This is particularly visible if we consider the spectral dynamics of the smoothed spectral density $\phi_\mathcal{G}$  representing the whole normalized Laplacian and focus on the initial and the final  phase of the edge removals.   The spectral dynamics towards amplifiers  furthermore is also substantially different from graph evolutions which are not guided, for instance random graphs and random edge removals. Thus, the results of this paper also expand the applications of spectral analy\-sis of evolutionary graphs~\cite{allen19,rich17,rich19a,rich19b} as they link structural with spectral properties and allow to differentiate between amplifiers and evolutionary graphs in general.


\section{Conclusions}

We have studied  the performance and the behavior of an iterative process for identifying transient amplifier for dB updating. Transient amplifiers are networks representing population structures which shift the balance between natural selection and random drift. They are highly relevant for understanding the relationships between spatial structures and evolutionary dynamics.
The iterative process implies 
dynamic graph structures as we remove edges from regular input graphs. We use the concept of spectral dynamics for analyzing the edge removal process  connected with the algorithmic search for transient amplifiers. Our results particularly showed that the spectral dynamics of edge removals finally leading to transient amplifiers are distinct  and thus  enable differentiation.  Thus, we add to answering the question of what  structural and spectral characteristics transient amplifier have and how these  characteristics can be achieved by edge deletion from a regular graph. 
Moreover, 
the results of analyzing the spectral dynamics flow back to the algorithmic process as structural and spectral properties are usable for informing and guiding the process, particularly as the variety of possibilities to delete edges from a graph grows massively and therefore needs to be pruned due to computational constraints.

As discussed above
the problem of identifying and analyzing transient amplifiers is important for understanding the relationships between spatial structure and evolutionary dynamics and has substantial relevance for real biological processes, as for instance shown for
 cancer initiation and progression, ageing of tissues,   spread of infections and microbial evolution of antibiotic resistance. On the other hand, our topic is also related to a fundamental mathematical question in graph theory which is the relationships between the graph spectra and the graph structure. Thus, the problems discussed in this paper are also interesting from a graph-theoretical point of view.
They contribute to our understanding of how edge manipulations are related to spectral properties and reflect upon similarities and differences between the spectra of the normalized and the standard Laplacian.

\section*{Acknowledgments} I wish to thank Markus Meringer  for making available~\texttt{genreg}~\cite{mer99} used for generating the regular input graphs and for helpful~discussions, and Benjamin Allen for sharing the algorithm for calculating coalescence times.

\clearpage

\section*{Appendix}

The results of this paper are calculated and visualized with MATLAB. The adjacency matrices of the set of all transient amplifier as well as code to produce the results are available at 

\small \url{https://github.com/HendrikRichterLeipzig/IterativeTransientAmplifiers}.

\noindent
\normalsize
Additional graphs are given in the following Appendix.

\begin{figure}[htb]
\centering

 \includegraphics[trim = 5mm 1mm 6mm 20mm,clip, width=5.6cm, height=4.5cm]{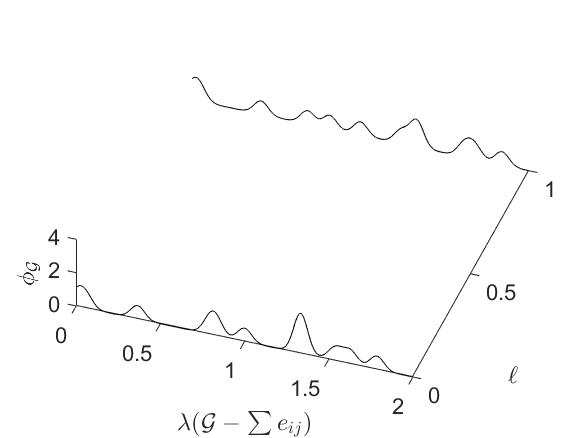}
 \includegraphics[trim = 5mm 1mm 6mm 20mm,clip, width=5.6cm, height=4.5cm]{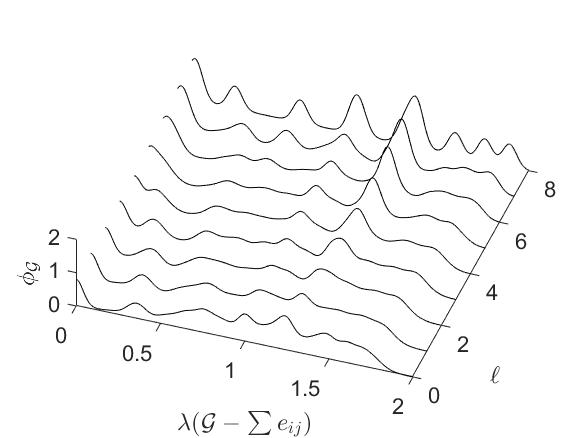}

 (\textbf{a}) $k=3$\hspace{4.7cm} (\textbf{b}) $k=4$

 \includegraphics[trim = 5mm 1mm 6mm 20mm,clip, width=5.6cm, height=4.5cm]{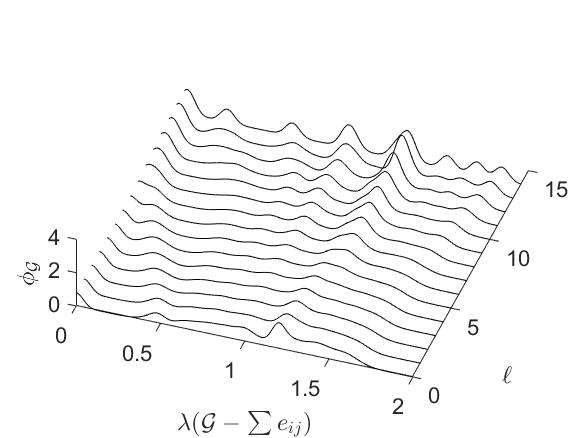}
 \includegraphics[trim = 5mm 1mm 6mm 20mm,clip, width=5.6cm, height=4.5cm]{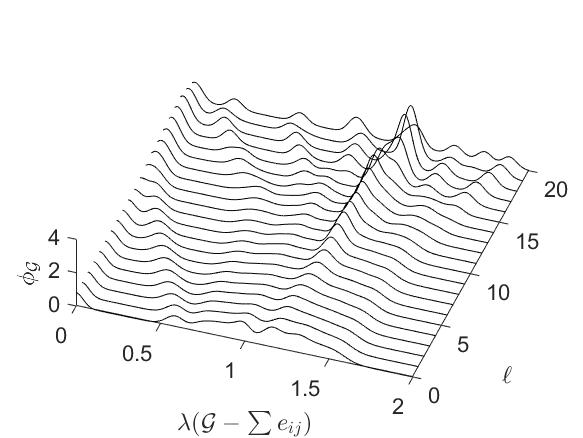}

 (\textbf{c}) $k=5$\hspace{4.7cm} (\textbf{d}) $k=6$

 \includegraphics[trim = 5mm 1mm 6mm 20mm,clip, width=5.6cm, height=4.5cm]{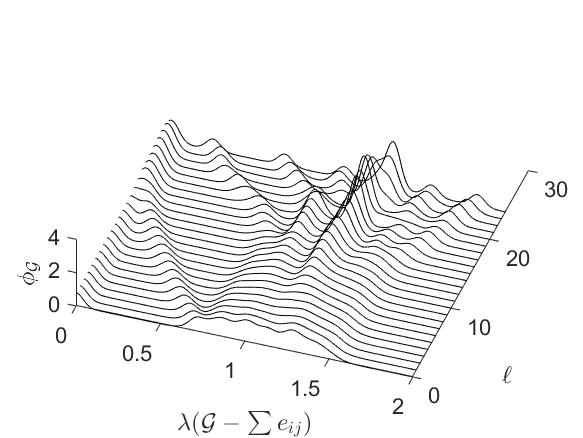}
 \includegraphics[trim = 5mm 1mm 6mm 20mm,clip, width=5.6cm, height=4.5cm]{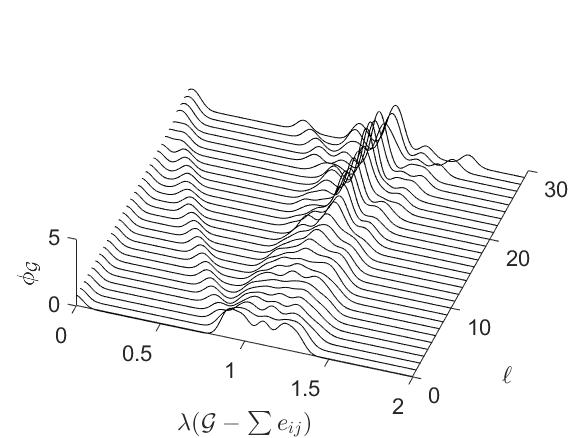}

 (\textbf{e}) $k=7$\hspace{4.7cm} (\textbf{f}) $k=9$  

\caption{\small{Spectral dynamics expressed by the spectral density $\phi_\mathcal{G}$ describing graph evolutions towards transient amplifiers for order $N=12$ and different $k$. Supplement to Fig. \ref{fig:graph_12_X_dense} }  }
\label{fig:graph_12_xx_dense}
\end{figure}

\begin{figure}[htb]
\centering

 \includegraphics[trim = 5mm 1mm 6mm 20mm,clip, width=5.6cm, height=4.5cm]{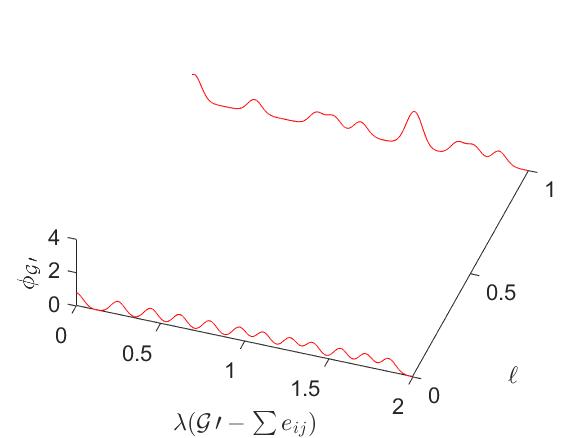}
 \includegraphics[trim = 5mm 1mm 6mm 20mm,clip, width=5.6cm, height=4.5cm]{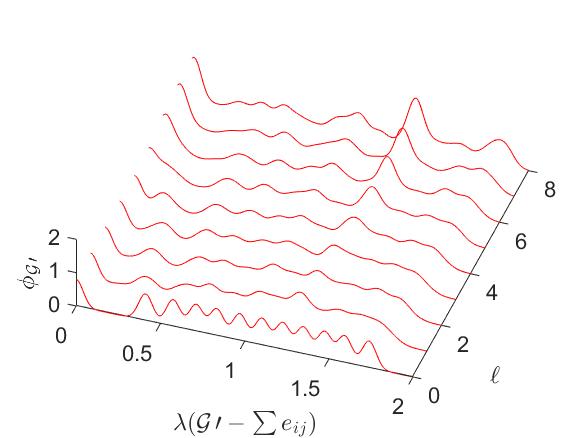}

 (\textbf{a}) $k=3$\hspace{4.7cm} (\textbf{b}) $k=4$

 \includegraphics[trim = 5mm 1mm 6mm 20mm,clip, width=5.6cm, height=4.5cm]{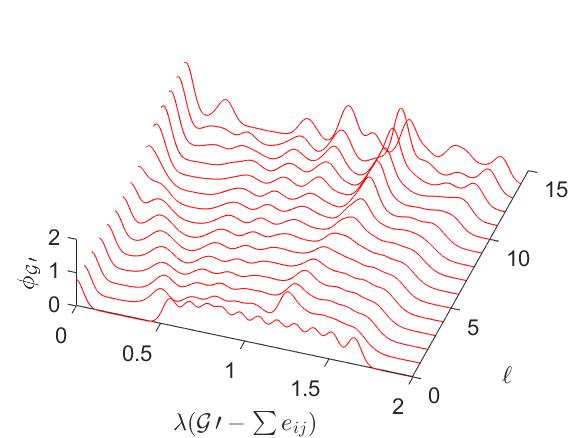}
 \includegraphics[trim = 5mm 1mm 6mm 20mm,clip, width=5.6cm, height=4.5cm]{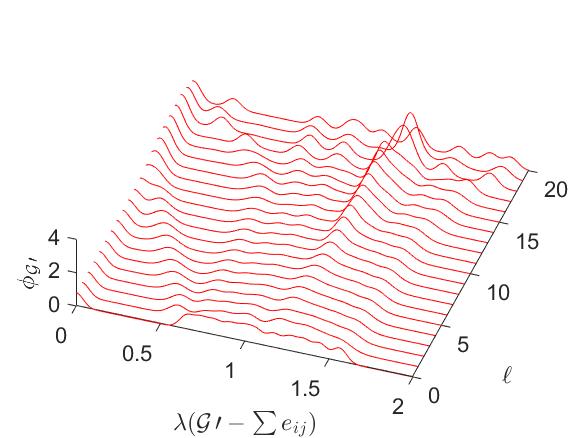}

 (\textbf{c}) $k=5$\hspace{4.7cm} (\textbf{d}) $k=6$

 \includegraphics[trim = 5mm 1mm 6mm 20mm,clip, width=5.6cm, height=4.5cm]{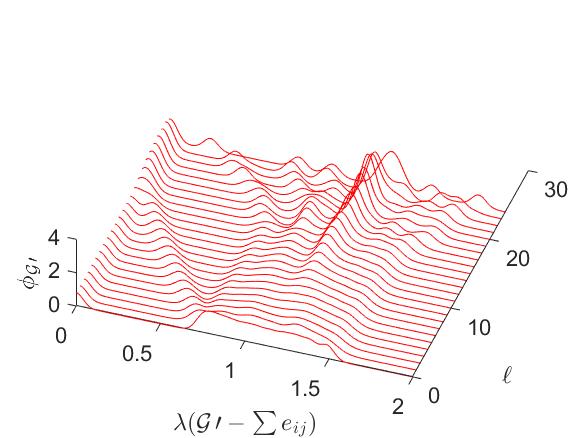}
 \includegraphics[trim = 5mm 1mm 6mm 20mm,clip, width=5.6cm, height=4.5cm]{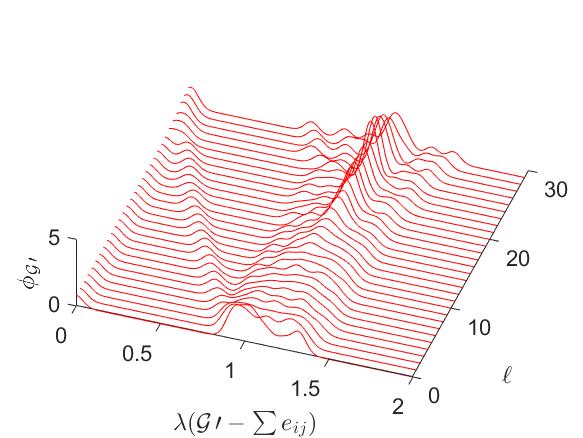}

 (\textbf{e}) $k=7$\hspace{4.7cm} (\textbf{f}) $k=9$  

\caption{\small{Spectral dynamics expressed by the spectral density $\phi_\mathcal{G}\prime$ describing graph evolutions not towards transient amplifiers for order $N=12$ and different $k$. Supplement to Fig. \ref{fig:graph_12_X_dense} }  }
\label{fig:graph_12_xxx_dense}
\end{figure}

\begin{figure}[htb]
\centering

 \includegraphics[trim = 5mm 1mm 6mm 20mm,clip, width=5.6cm, height=4.5cm]{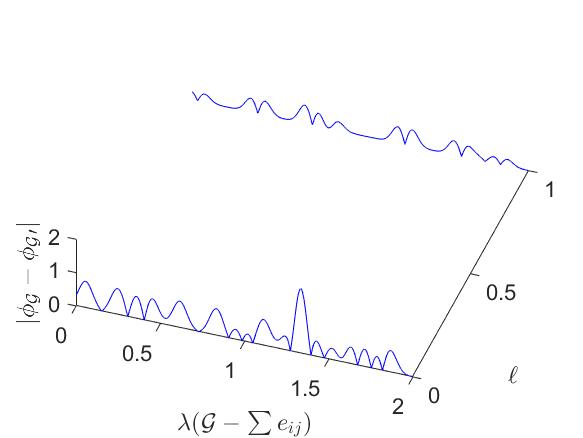}
 \includegraphics[trim = 5mm 1mm 6mm 20mm,clip, width=5.6cm, height=4.5cm]{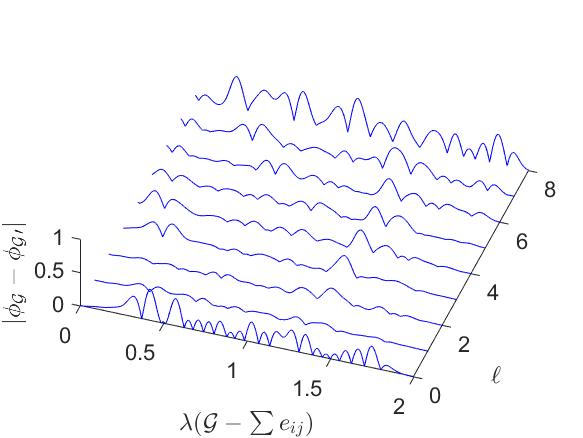}

 (\textbf{a}) $k=3$\hspace{4.7cm} (\textbf{b}) $k=4$

 \includegraphics[trim = 5mm 1mm 6mm 20mm,clip, width=5.6cm, height=4.5cm]{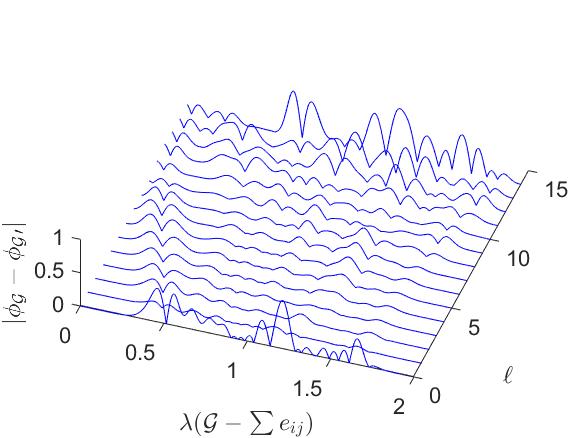}
 \includegraphics[trim = 5mm 1mm 6mm 20mm,clip, width=5.6cm, height=4.5cm]{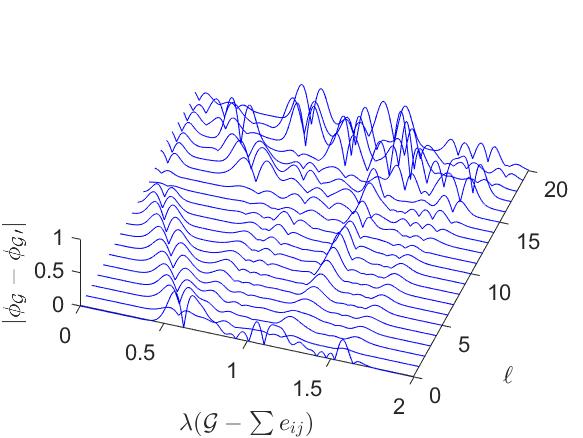}

 (\textbf{a}) $k=5$\hspace{4.7cm} (\textbf{b}) $k=6$

 \includegraphics[trim = 5mm 1mm 6mm 20mm,clip, width=5.6cm, height=4.5cm]{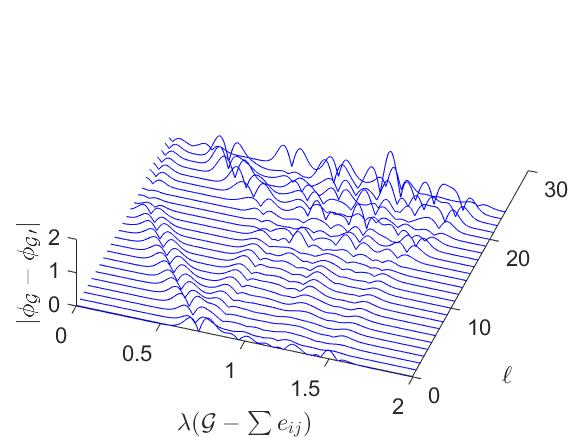}
 \includegraphics[trim = 5mm 1mm 6mm 20mm,clip, width=5.6cm, height=4.5cm]{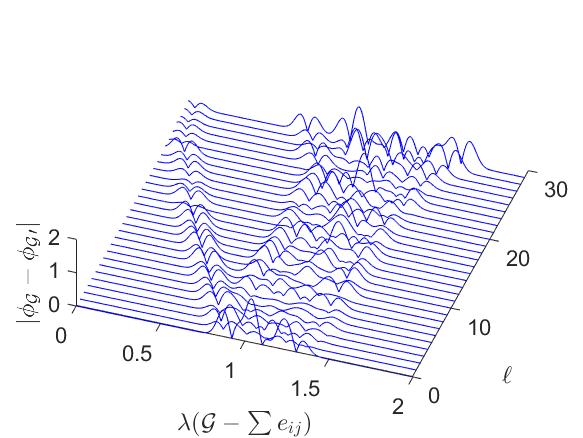}

 (\textbf{a}) $k=7$\hspace{4.7cm} (\textbf{b}) $k=9$  

\caption{\small{Spectral dynamics expressed by $|\phi_\mathcal{G}-\phi_\mathcal{G}\prime|$ describing the difference between $\phi_\mathcal{G}$ and $\phi_\mathcal{G}\prime$ for order $N=12$ and different $k$. Supplement to Fig. \ref{fig:graph_12_X_dense} }  }
\label{fig:graph_12_xxxx_dense}
\end{figure}

\begin{figure}[htb]
\centering

 \includegraphics[trim = 5mm 1mm 6mm 20mm,clip, width=5.6cm, height=4.5cm]{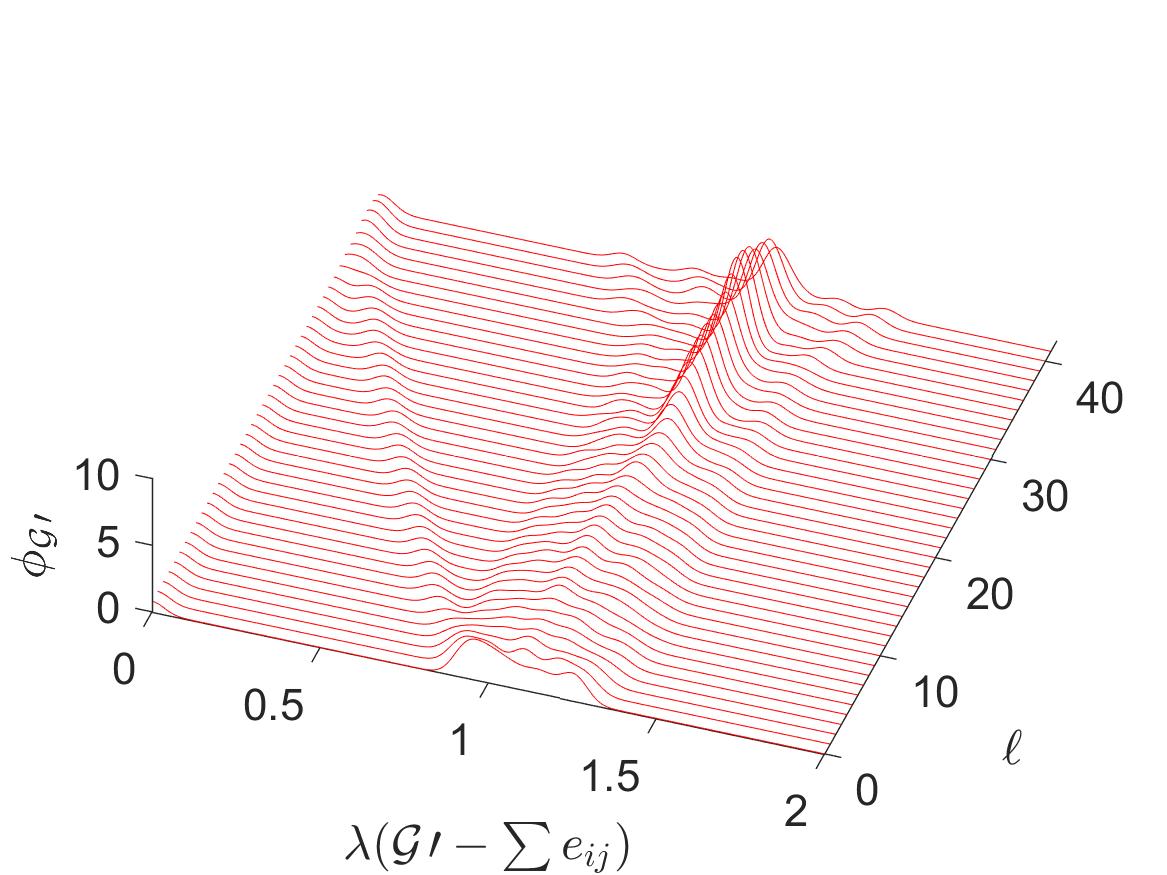}
 \includegraphics[trim = 5mm 1mm 6mm 20mm,clip, width=5.6cm, height=4.5cm]{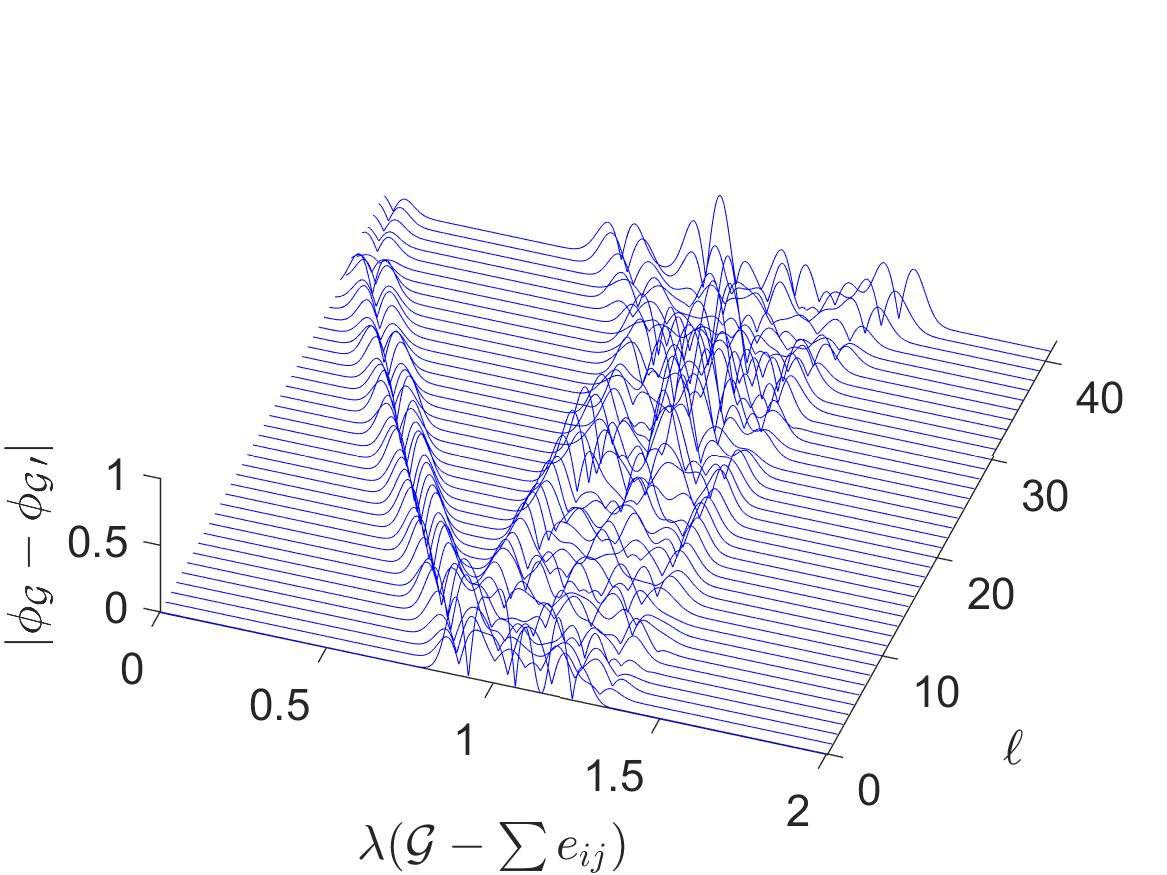}

 (\textbf{a}) \hspace{4.7cm} (\textbf{b})

 \includegraphics[trim = 5mm 1mm 6mm 20mm,clip, width=5.6cm, height=4.5cm]{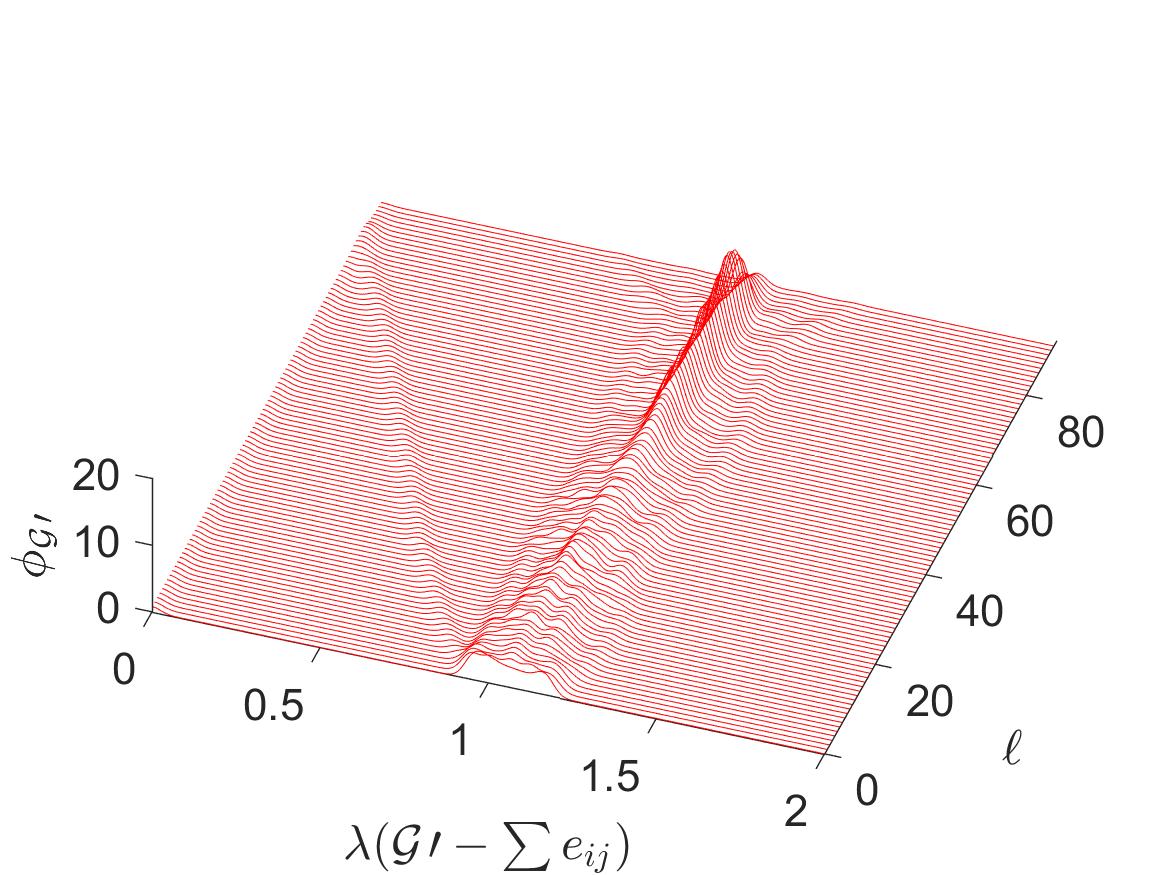}
 \includegraphics[trim = 5mm 1mm 6mm 20mm,clip, width=5.6cm, height=4.5cm]{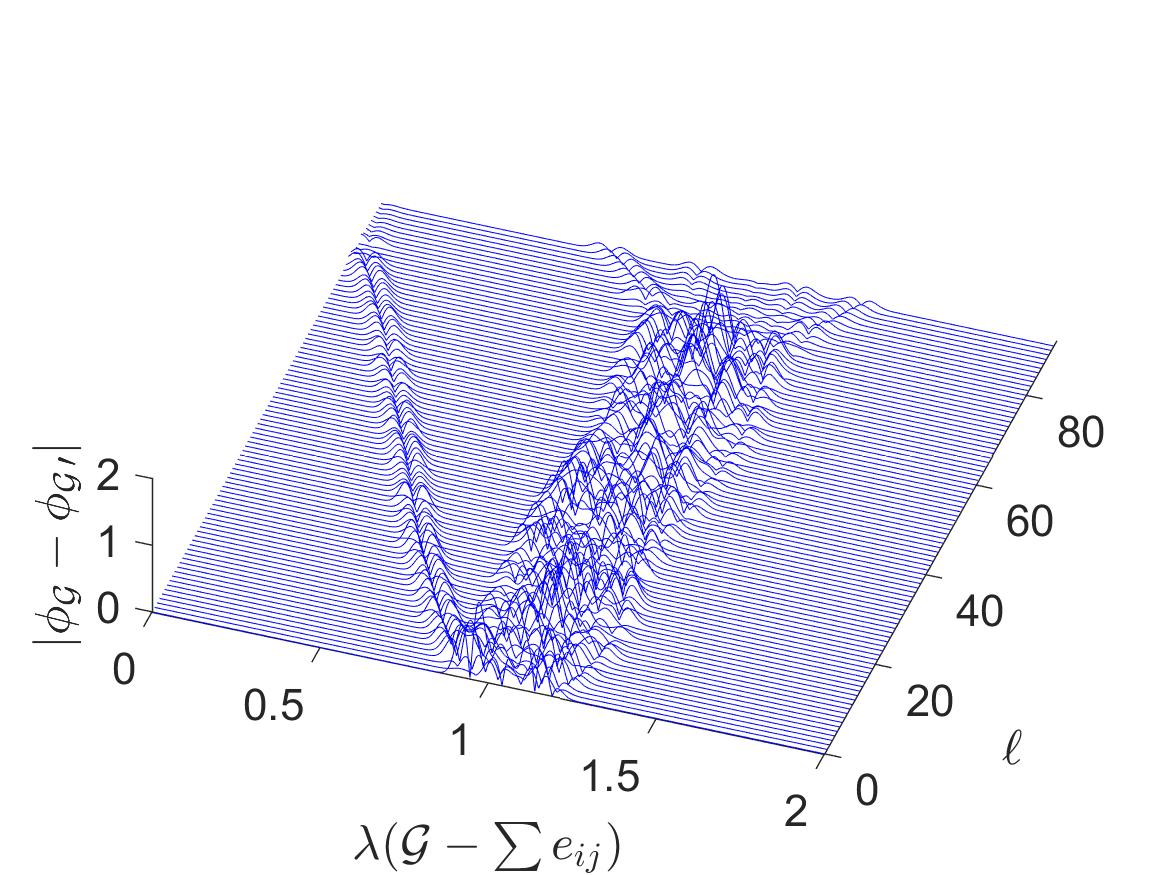}

  (\textbf{c}) \hspace{4.7cm} (\textbf{d})

 \includegraphics[trim = 5mm 1mm 6mm 20mm,clip, width=5.6cm, height=4.5cm]{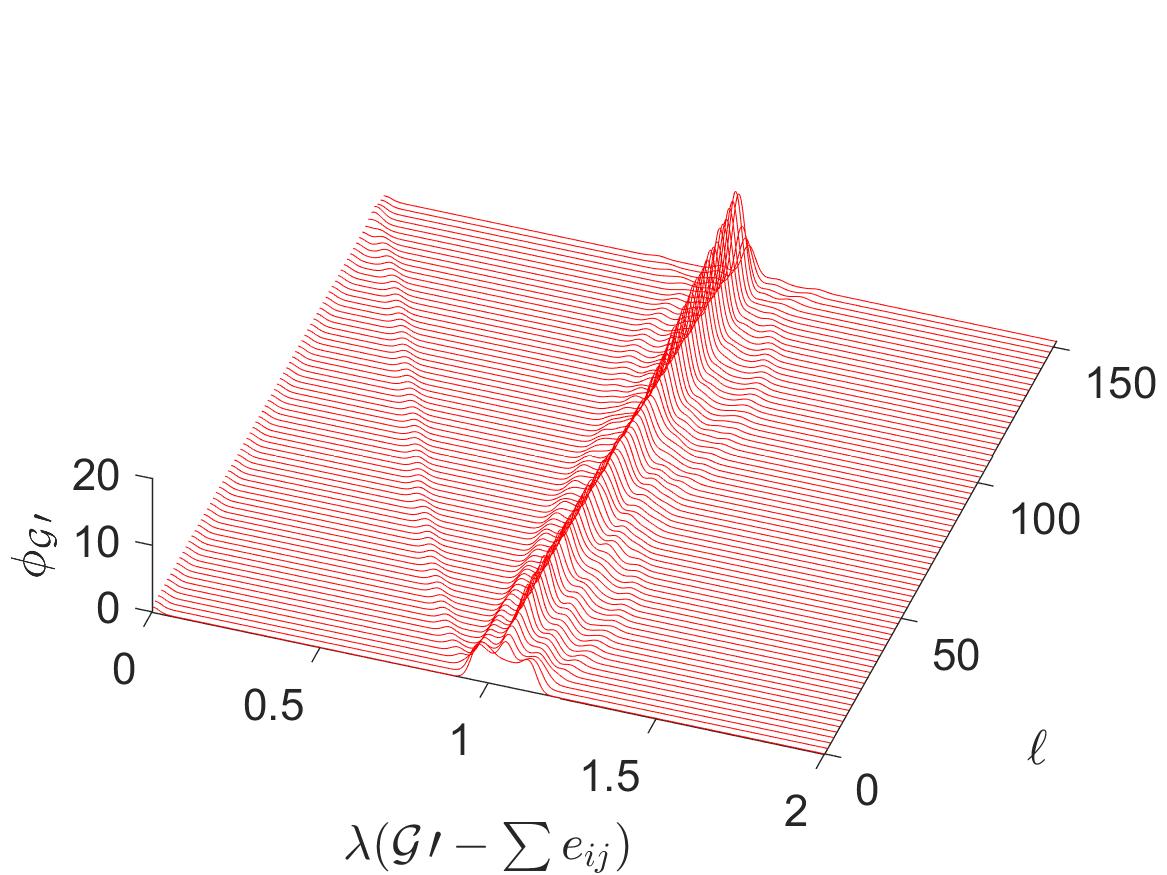}
 \includegraphics[trim = 5mm 1mm 6mm 20mm,clip, width=5.6cm, height=4.5cm]{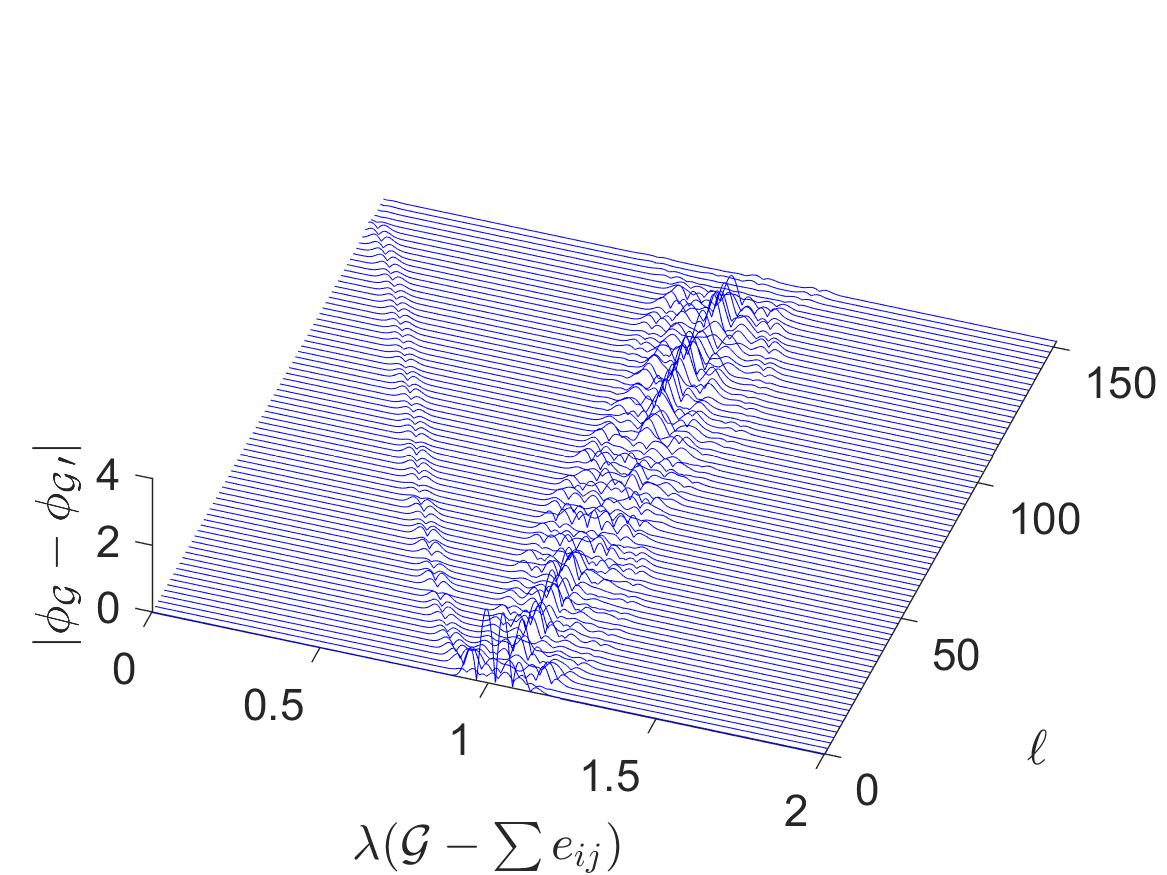}

 (\textbf{e}) \hspace{4.7cm} (\textbf{f})  

\caption{\small{ Behaviour of the approximative search of Algorithm 1 for $N=\{14,20,26\}$ and $k=\{11,17,23\}$.  The spectral density $\phi_\mathcal{G}\prime$ describing graph evolutions not leading to transient amplifiers and the quantity $|\phi_\mathcal{G}-\phi_\mathcal{G}\prime|$ describing the difference between $\phi_\mathcal{G}$ and $\phi_\mathcal{G}\prime$. Supplement to Fig. \ref{fig:graph_14_X_dense}.      }  }
\label{fig:graph_14_20_26_x_dense}
\end{figure}

\end{document}